\documentclass[british]{scrartcl}
\usepackage{threeparttable}
\usepackage{caption}
\usepackage{subcaption}

\usepackage{float}
\usepackage{bm}
\usepackage{amsthm}
\usepackage{amsmath}
\usepackage{amssymb}
\usepackage{graphicx}
\usepackage{esint}
\usepackage[authoryear]{natbib}
\usepackage{slashbox}
\usepackage{url}
\usepackage{afterpage}
\usepackage{multirow}

\makeatletter

\floatstyle{ruled}
\newfloat{algorithm}{tbp}{loa}
\providecommand{\algorithmname}{Algorithm}
\floatname{algorithm}{\protect\algorithmname}

\theoremstyle{plain}

  \theoremstyle{plain}
  
  \theoremstyle{plain}

  \theoremstyle{plain}
\newtheorem{definition}{\protect\defname}
  \theoremstyle{plain}
\newtheorem{prop}{\protect\propname}

\usepackage{dsfont} 

\usepackage{algpseudocode}
\usepackage{mathtools}

\renewcommand{\P}{\mathbb{P}}
\newcommand{\E}{\mathbb{E}}

\newcommand{\bx}{\mathbf{x}}

\newcommand{\bz}{\mathbf{z}}

\newcommand{\dd}{\mathrm{d}}

\newcommand{\comment}[1]{ \ifthenelse{ \equal{\showcomment}{true} }{ {\bf #1} }{} }
\newcommand{\showcomment}{true}
\newcommand{\bs}{\boldsymbol}

\makeatother

\usepackage{babel}
  \providecommand{\lemmaname}{Lemma}
\providecommand{\theoremname}{Theorem}
\providecommand{\corollaryname}{Corollary}
\providecommand{\defname}{Definition}
\providecommand{\propname}{Proposition}

\title{Bayesian Inference for the Multivariate Extended-Skew Normal Distribution}

\begin{document}
\allowdisplaybreaks

\author{Mathieu Gerber\thanks{Present address: Harvard University, Department of Statistics. Email: mathieugerber@fas.harvard.edu} \\
\emph{{\footnotesize {Faculty of Business and Economics (HEC)}}}\\
\emph{{\footnotesize {University of Lausanne, Switzerland}}} \\
\emph{CREST}
\and Florian Pelgrin\thanks{Email: florian.pelgrin@edhec.edu} \\
\emph{{\footnotesize {EDHEC Business School, France}}} \\
\emph{CIRANO}
}
\date{}
\maketitle
\setlength{\baselineskip}{18pt}

\begin{abstract}

The multivariate  extended skew-normal distribution allows for accommodating raw data which are skewed and heavy tailed, and has at least three appealing statistical properties, namely closure under conditioning, affine transformations, and marginalization. In this paper we propose a  Bayesian computational approach based on  a sequential Monte Carlo (SMC) sampler to estimate such distributions. The practical implementation of each step of the algorithm is discussed and the elicitation of prior distributions takes into consideration some unusual behaviour of the likelihood function and the corresponding Fisher information matrix.  Using Monte Carlo simulations, we provide strong evidence regarding the  performances of the SMC sampler  as well as some new insights regarding the parametrizations of the extended skew-normal distribution.  A generalization to the extended skew-normal sample selection model is also presented. Finally we proceed with the analysis of two real datasets.

\end{abstract}

\section{Introduction}

Recent years have seen a growing interest for flexible parametric families of multivariate distributions that can accommodate both the skewness and the kurtosis often observed on data. This is especially important, e.g., in health, finance and environmental data, which are often skewed and heavy tailed. For instance, these two features of health (-care) expenditures have fundamental implications in topics related to risk adjustments, program and treatment evaluations, or insurance choices \citep{Manning2005}. In this respect, the application of the skew-elliptical family of distributions (i.e., all non-symmetric distributions obtained from an elliptical distribution) has been put forward in the literature, and for good reasons. On the one hand, this family (or certain distributions of this family) has at least three appealing properties: closure under conditioning, affine transformations, and marginalization. 
On the other hand, these parametric distributions appear in the natural and important context of selection models \citep{Heckman1976, CopasLi1997, Arnold2002}. This last feature is particularly relevant in various research topics (e.g., economics, environmetrics or political sciences). 

Within the class of skew-elliptical distributions, the extended skew-normal (ESN) distribution appears in different areas of statistical theory, e.g., Bayesian statistics \citep{Ohagan1976}, regression analysis \citep{CopasLi1997} or graphical models \citep{Capitanio2003}. This generalization of the skew normal distribution, studied  in the seminal paper of \citet{Azzalini1985},  has  been developed and studied by \citet{Arnold1993,Arnold2000,Capitanio2003}. Such a family of distributions may be obtained  as a convolution between a multivariate normal random variable, $\widetilde{\mathbf{Y}}$,  and a truncated standard normal random variable, $Z$, say $\mathbf{Y}{\buildrel d \over =} \widetilde{\mathbf{Y}}+\boldsymbol{d} Z$. The derivation and statistical properties of ESN distributions make it a natural candidate to model skewed and (leptokurtic, platikurtic) mesokurtic data generating processes. However, as pointed out by \citet{Arellano2006}, statistical inference of  skew-elliptical distributions, and thus of the  extended skew-normal family of distributions, is still mostly unsolved (even in the univariate case).

Indeed, if Bayesian analysis of some families of skew-elliptical distributions has been proposed in the literature, they mainly focus on the skew-normal (or skew-t) distribution. In addition, Bayesian analysis of these distributions  usually rely  on objective prior-based methods, Gibbs sampling or population Monte Carlo. In particular, \citet{Liseo2006} consider a Bayesian estimation of the univariate skew-normal distribution based on objective priors whereas \citet{Wiper2008} analyse the half-normal and half-t cases, and \citet{Branco2013} focus on the skew-t distribution. 
\citet{Cabral2012} propose a full Bayesian estimation of a mixture of skew-normal densities while \citet{Fruhwirth2010} provide a Gibbs sampler to estimate a mixture of skew-normal and skew-t densities. As an alternative to the Gibbs sampler, \citet{Liseo2013} advocate the application of a Population Monte Carlo (PMC) algorithm for missing data \citep{Celeux2006} in order to sample from the posterior distribution of the  skew-normal model.

In this paper we propose a Bayesian computational method to estimate the extended (multivariate) skew-normal distribution. The Bayesian approach for this family of distributions is motivated by  some severe anomalies of the likelihood function and some identification issues. Notably, and as shown in this paper, the maximum likelihood estimator may not be uniquely defined for univariate ESN distributions.  In a Bayesian approach, these anomalies can be tackled to some extent by using a suitable elicitation of prior distributions. In the light of the properties of the likelihood function, we make use of a (tempered) sequential Monte Carlo  sampler \citep{DelMoral2006} rather than a Monte Carlo Makov chain (MCMC) algorithm. Briefly speaking, sequential Monte Carlo (SMC) iterates importance sampling steps, resampling steps and Markov kernel transitions in order to recursively approximate a sequence of distributions by making use of a sequence of weighted particle systems. Relative to MCMC algorithms, SMC might be called for at least three arguments. On the one hand, the great generality of SMC allows to build  efficient algorithms in order to sample from complicated distributions and thus to overcome some distortions of the likelihood functions of the  skew-normal distribution. 
On the other hand, compared to MCMC methods, it is easier to make SMC algorithms adaptive in the sense that they can be adjusted sequentially and automatically to the problem at hand, and the evidence or marginal likelihood of data can be derived formally. Finally, due to the convolution representation of the ESN distribution presented above, a natural idea to sample from the posterior distribution would be  to implement a Gibbs sampler in which the hidden random variable $Z$ is an extra parameter. However, in the case of ESN models, the support of the hidden variable $Z$  depends on the parameters of interest and thus the posterior distribution in the augmented space does not satisfy the positivity condition \citep[see][chapter 9]{Robert2004}.

The rest of the paper is organized as follows. Section \ref{section:ESN} defines two equivalent representations of extended skew-normal random vectors, review some useful properties of this class of distributions, and discuss some unpleasant features of the maximum likelihood function. Section \ref{sec:Bayes} proposes some prior distributions which take into consideration theses anomalies of the likelihood function and describes the proposed tempered sequential Monte Carlo algorithm. Section \ref{sec:num2} presents some Monte Carlo simulations regarding the inference of univariate ESN distributions and of some regressions with missing data. Moreover we discuss the testing and model selection problems. Section \ref{sec:app} deals with two applications, namely the distribution of transfer fees of soccer players in major European leagues and the bivariate distribution of two financial returns \citep{Liseo2013}. The last section provides some concluding remarks.

\section{The extended skew-normal distribution}\label{section:ESN}

In this section we first define the extended skew-normal (ESN) distribution using two different parametrizations. Then, we review some appealing properties of this class of distributions, especially in the light of the subsequent derivations of this paper. Finally, we provide a new theoretical justification for the unsatisfactory behaviour of the maximum likelihood estimator of the ESN distribution.

\subsection{Definition and main properties}

We consider  two parametrizations of the ESN distribution. The first parametrization, denoted P1, is based on hidden truncation (and/or selective reporting) using normal component densities whereas the second parametrization, denoted P2, rests on the convolution of a multivariate normal distribution with a truncated standard normal variable.

\begin{definition}\label{def:P1}
A random vector $\mathbf{Y}$ is said to have a $d$-dimensional extended skew-normal distribution, denoted  $\mathbf{Y}\sim \mathcal{ESN}^{(P1)}_d(\boldsymbol{\xi},\Sigma,\boldsymbol{\alpha},\lambda)$, with covariance (correlation) matrix $\Sigma$, shape parameter $\bs{\alpha}$, and shift parameter $\lambda$, if
\begin{equation}\label{eq:stoch}
\mathbf{Y}{\buildrel d \over =}\big(\boldsymbol{\xi}+\widetilde{\mathbf{Y}}_1|\lambda+\boldsymbol{\alpha}'\widetilde{\mathbf{Y}}_1> Z_1\big),\qquad
\begin{pmatrix}
\widetilde{\mathbf{Y}}_1\\
Z_1
\end{pmatrix}\sim \mathcal{N}_{d+1}\left(
\begin{pmatrix}
\mathbf{0}\\
0
\end{pmatrix},
\begin{pmatrix}
\Sigma& \mathbf{0}\\
\mathbf{0}&1
\end{pmatrix}
\right)
\end{equation}
where $\mathcal{N}_{d+1}(\bs{\mu},B)$ denotes the $(d+1)$-dimensional Gaussian distribution with mean $\bs{\mu}$ and variance-covariance matrix $B$. Its density function is defined to be:
\begin{equation}\label{eq:ESN_Density}
f_Y(\mathbf{y})=\phi_d\left(\mathbf{y},\boldsymbol{\xi},\Sigma\right)\frac{\Phi(\lambda
+\boldsymbol{\alpha}'(\mathbf{y}-\boldsymbol{\xi}))}
{\Phi\left(\lambda/c_0\right)},\quad c_0=\sqrt{1+\boldsymbol{\alpha}'\Sigma
\boldsymbol{\alpha}}
\end{equation}
where  $\phi_d\left(\cdot{ },\boldsymbol{\mu},B\right)$ is the density of the $d$-dimensional normal distribution with mean $\boldsymbol{\mu}$ and covariance (correlation) matrix $B$  and $\Phi(\cdot{ })$ is the cumulative density function (cdf) of the $\mathcal{N}_1(0,1)$ distribution.
\end{definition}

On the other hand, the ESN distribution can be defined from a convolution.

\begin{definition}\label{def:P2}
A random vector $\bs{Y}$ is said to have a $d$-dimensional extended skew-normal distribution, denoted  $\mathbf{Y}\sim \mathcal{ESN}^{(P2)}_d(\boldsymbol{\xi},\Omega,\boldsymbol{d},c)$, if
\begin{equation*}
\mathbf{Y}{\buildrel d \over =} \widetilde{\mathbf{Y}}_3+\boldsymbol{d} Z_3
\end{equation*}
where $-Z_3\sim\mathcal{TN}_{c}(0,1)$, the $\mathcal{N}_1(0,1)$ distribution truncated to $(-\infty, c]$, and $\widetilde{\mathbf{Y}}_3\sim \mathcal{N}_d(\mathbf{0},\Omega)$.
Its probability density function is defined to be:
\begin{equation*}
f_Y(\mathbf{y})=\phi_d\left(\mathbf{y},\boldsymbol{\xi},\Omega+\mathbf{d}\mathbf{d}'\right)\frac{\Phi\left(c_0\left\{c
+\mathbf{d}'\left[\Omega+\mathbf{d}\mathbf{d}'\right]^{-1}(\mathbf{y}-\boldsymbol{\xi})\right\}\right)}
{\Phi\left(c\right)}.
\end{equation*}
\end{definition}

Several points are worth commenting. First, the ESN distribution belongs to the families of  skew-elliptical distributions proposed by \citet{Arnold2002}, \citet{Dominguez2003}, \citet{Fang2003}, and \citet{Arellano2010a}. 
Alternatively, using P2, the ESN distribution belongs to the family of distributions proposed by \citet{Sahu2003}. Irrespective of the parametrization, the ESN distribution generalizes the multivariate skew-normal  distribution \citep{Azzalini1996} and thus the Gaussian distribution. More specifically, when the shift parameter $\lambda$ is set to zero, one obtains the (multivariate) SN distribution. On the other hand, the standard normal distribution results from the nullity of the shape  parameter vector $\boldsymbol{\alpha}$. 
As explained in Section \ref{subsec:lik}, this constraint on the shape parameter vector has some key implications on inference. Indeed, the Fisher information matrix of the ESN (and of the SN) distribution is singular, preventing a straightforward application of standard likelihood-based methods to test the null hypothesis of normality. The problem is even made worse by the parameter $\lambda$, which indexes the distribution in the case of non-normality (nuisance parameter).

Second, the choice of the parametrization might be critical for the estimation of ESN distributions since, in a Bayesian perspective, different parametrizations lead to alternative choices of prior distributions and thus different models (see Section \ref{sec:Bayes}). 
Third, one key feature of the  ESN distribution over the SN distribution is that the former has an extra parameter that allows for a larger range of values for  skewness and kurtosis and thus for more flexibility to accommodate real data.  
 For instance, using the moment generating function of \citet{Dominguez2003}, one can provide evidence with a numerical analysis of the univariate ESN that the skewness coefficient is bounded by 2 (in absolute value) while the kurtosis coefficient varies roughly between 2.75 and 7. In contrast, \citet{Azzalini1985} points out that the skewness is smaller (in absolute value) than 0.995 and that the kurtosis lies  between 3 and $3.87$ in the case of the skew-normal distribution. 

Fourth, the  ESN distribution has three familiar and useful properties, especially for regression-type models. It is closed under affine transformations, conditioning and marginalization. On the one hand, ESN random vectors share the affine transformation of normal random vectors. 
 In particular, let $A$ denote an $d\times d$  non-singular  matrix and $\tilde{\boldsymbol{\xi}}\in\mathbb{R}^d$. Then, taking  \eqref{eq:stoch}, one obtains
$
\tilde{\boldsymbol{\xi}}+A'\mathbf{Y}\sim \mathcal{ESN}^{(P1)}_d(\tilde{\boldsymbol{\xi}}+A'\boldsymbol{\xi},A'\Sigma A,A^{-1}\boldsymbol{\alpha},\lambda).
$
On the other hand, if an ESN vector is partitioned into two components, the conditional distribution of one component given the other is extended skew-normal and each component is marginally extended skew-normal. 
For sake of completeness, Proposition \ref{prop:closure} due to \citet{Fang2003} and \citet{Dominguez2003} reports the closure of the ESN distribution under conditioning and marginalization.

\begin{prop}\label{prop:closure}
Assume that $\mathbf{Y}\sim \mathcal{ESN}^{(P1)}_d(\boldsymbol{\xi},\Sigma,\boldsymbol{\alpha},\lambda)$. Partition $\mathbf{Y}$, $\boldsymbol{\xi}$, $\boldsymbol{\alpha}$ and $\Sigma$ as $\mathbf{Y}=(\mathbf{Y}_1,\mathbf{Y}_2)'$, $\bs{\epsilon}=(\bs{\epsilon}_1,\bs{\epsilon}_2)'$, $\bs{\alpha}=(\bs{\alpha}_1,\bs{\alpha}_2)'$ and
$
\Sigma=\left(
\begin{smallmatrix}
\Sigma_{11}&\Sigma_{12}\\
\Sigma_{21}&\Sigma_{22}
\end{smallmatrix}\right)
$
where $\mathbf{Y}_i$, $\boldsymbol{\xi}_i$ and $\bs{\alpha}_i
$ are $m_i\times 1$ and $\Sigma_{ii}$ is $m_i\times m_i$. Then,
$$
\mathbf{Y}_i\sim \mathcal{ESN}^{(P1)}_{m_i}(\boldsymbol{\xi}_i,\Sigma_{ii},c_i\tilde{\boldsymbol{\alpha}}_i,c_i
\lambda),\qquad
\left(\mathbf{Y}_i|\mathbf{Y}_j=\mathbf{y}_j\right)\sim \mathcal{ESN}^{(P1)}_{m_i}(\boldsymbol{\xi}_i^c, \Sigma_{ii.1},\boldsymbol{\alpha}_i,\lambda_i)
$$
where
$c_i=(1+\boldsymbol{\alpha}_j'\Sigma_{i.1}
\boldsymbol{\alpha}_j)^{-1/2}$,  $\boldsymbol{\xi}_i^c=\boldsymbol{\xi}_i
+\Sigma_{ij}\Sigma_{jj}^{-1}(\mathbf{y}_j-\boldsymbol{\xi}_j)$, $ \Sigma_{ii.1}=\Sigma_{ii}-\Sigma_{ij}
\Sigma_{jj}^{-1}\Sigma_{ji}$,
$
\tilde{\boldsymbol{\alpha}}_i
=\boldsymbol{\alpha}_i+\Sigma_{ii}^{-1}\Sigma_{ij}
\boldsymbol{\alpha}_j$ and $\lambda_i=\lambda+\tilde{\boldsymbol{\alpha}}_j
(\mathbf{y}_j-\boldsymbol{\xi}_j)$.
\end{prop}

Finally, the stochastic representation \eqref{eq:stoch} of ESN random vectors leads to the following expression of the cumulative density function (henceforth, cdf):
$$
\P(\mathbf{Y}\leq \mathbf{y})=\frac{\Phi_{d+1}(\mathbf{y}-\boldsymbol{\xi},\Sigma,\boldsymbol{\alpha},\lambda)}{\Phi(\lambda/c_0)}
$$
with $\Phi_{d+1}(\mathbf{a},\Sigma,\boldsymbol{\alpha},\lambda)=\P\big(\widetilde{\mathbf{Y}}_2\leq \mathbf{a}, Z_2\leq \lambda\big)$ and where
$$
(\widetilde{\mathbf{Y}}_2,Z_2)\sim \mathcal{N}_{d+1}\left(
\begin{pmatrix}
\mathbf{0}\\
0
\end{pmatrix},
\begin{pmatrix}
\Sigma&-\Sigma\boldsymbol{\alpha}'\\
-\boldsymbol{\alpha}'\Sigma&c_0^2
\end{pmatrix}
\right).
$$
Notably, the evaluation of the cdf of the $d$-dimensional ESN distribution  has the same complexity as the computation of the cdf of the $(d+1)$-dimensional Gaussian distribution, for which  efficient methods are available \citep[e.g., see][]{huguenin2014}. It turns to be very useful in practice since, for instance, the cdf of the ESN distribution arises naturally when deriving the expression of the likelihood function in the presence of missing data (see Section \ref{subsect:numESN}).

\subsection{Log-likelihood function}\label{subsec:lik}

Since our methodology rests on Bayesian estimation and thus on the posterior distribution associated to the ESN-based model, it is fundamental to study the statistical properties of the likelihood function. This might provide some useful insights in order to determine the prior distribution and thus challenge some identified anomalies regarding the likelihood function (i.e., to correct at least partially the odd behaviour of the likelihood function with external information). 
For sake of exposition, we concentrate on the univariate ESN distribution.

Maximum likelihood estimation of ESN distributions is challenging and quite difficult to manage. More specifically, it is widely acknowledged that (i) there are no closed form expressions for the maximum likelihood estimator (MLE), (ii) the MLE of $\alpha$ can be infinite even in very simple settings, (iii) the multimodality  of the log-likelihood profile (and thus local solutions) can not be ruled out and (iv) there exists an inflexion point at $\alpha=0$.  
In particular, the Fisher information matrix tends to be singular as $\alpha$ goes toward zero irrespective of the $\lambda$ parameter. Note that, in this case, ESN distributions are no longer indexed by the normal cumulative density functions and, consequently, the rank of the information matrix might be at least two less than its full rank. On the other hand, the presence of a stationary point (e.g., using the profile log-likelihood for the $\alpha$ parameter) and of multiple modes generally cause numerical issues. 

While these issues have been outlined in the literature, to the best of our knowledge, there is not yet a formal proof of the near unidentifiability of the log-likelihood function and the $\lambda$ parameter. Therefore, we show in Proposition \ref{prop:lik} that the presence of the shift parameter $\lambda$ in P1 might lead to local maxima for the maximum likelihood estimator of the univariate extended skew-normal distribution. 
Indeed, irrespective of the data and for all $l\in\mathbb{R}$, the ESN distribution admits a stationary point at $\theta_{n,G}^l:=(\boldsymbol{\xi}_{n,G},\Sigma_{n,G},\mathbf{0}_d,l)$, where $\boldsymbol{\xi}_{n,G}$ and $\Sigma_{n,G}$ are the MLE of $\boldsymbol{\xi}$ and $\Sigma$ under the Gaussian assumption. In so doing, if this stationary point is an inflexion point when we impose the $\lambda$ parameter to be zero \citep{Azzalini1998}, the problem becomes even more severe when $\lambda$ is a free parameter as stated in Proposition \ref{prop:lik}. 

\begin{prop}\label{prop:lik}
Let $Y_1,...,Y_n$ be $n$ i.i.d. random variables, $Y_1 \sim \mathcal{ESN}^{(P1)}_1(\xi,\sigma^2,\alpha,\lambda)$ with $\alpha\neq 0$. Let $\theta_{n,G}^l=(\xi_{n,G},\sigma_{n,G}^2,0,l)$ with $\xi_{n,G}=\frac{1}{n}\sum_{i=1}^n Y_i$ and $\sigma_{n,G}^2=\frac{1}{n}\sum_{i=1}^n Y_i^2-\xi_{n,G}^2$. Let $L_n(\theta)$ denote the log-likelihood function. Then,
\begin{enumerate}
\item With probability one, there exists a $l^*\in\mathbb{R}$ such that $L_n(\theta_{n,G}^{l})$ is a local maximum of $L_n(\cdot)$ for all $l\leq l^*$;
\item With strictly positive probability,  $L_n(\theta^l_{n,G})=L_n(\theta_n)$, $l\in\mathbb{R}$, where $\theta_n\neq \theta^l_{n,G}$ is a global maximizer of $L_n(\cdot)$.
\end{enumerate}
\end{prop}
\noindent See Appendix \ref{app:lik} for a proof.

The first result of Proposition \ref{prop:lik}  has an intuitive interpretation. When $\alpha=0$, the value of the log-likelihood function is insensitive to any change of the $\lambda$ parameter and thus any small deviation of $\alpha$ leads to large deviations from the true log-likelihood value (since the estimate $\lambda$ was initially far from its true unknown value). Consequently, a small deviation from $\theta_{n,G}^l$ in any direction reduces the value of the likelihood. The second part of Proposition \ref{prop:lik} is more puzzling because it implies that, with a positive (but decreasing with $n$) probability,  the likelihood function does not allow to discriminate between the Gaussian and the ESN model. This is a particularly severe anomaly of the likelihood function because it implies that the MLE might be not uniquely defined. 

\section{Bayesian analysis of the ESN distribution}\label{sec:Bayes}

In this section we first discuss the elicitation of prior distributions and then explain how to estimate the parameters of the two parameterizations of ESN distributions using Sequential Monte Carlo \citep{DelMoral2006}.

\subsection{A default Prior specification}\label{subsec:Prior}

In contrast to the standard approach of default prior distributions, and in the spirit of \citet{Gelman2008}, we propose a prior specification that embeds enough information to circumvent the anomalies of the log-likelihood function listed in Section \ref{subsec:lik}.

On the one hand, the $(\boldsymbol{\xi},\Sigma,\boldsymbol{\alpha},\boldsymbol{\lambda})$-parametrization (P1) of ESN random vectors must tackle two issues, namely the potential existence of multiple modes and the identification (estimation) of the truncation point, that are related to the identification of the $\lambda$ parameter. First, the multi-modality of the log-likelihood function might be attenuated by setting a prior that assigns less weight on very negative values of $\lambda$. Second, as argued in Section \ref{subsec:lik}, values of $\lambda$ such that the truncation point exceeds a certain threshold, say $|\lambda|/c_0>2$, are difficult to identify and therefore, both to avoid extreme estimates of $\lambda$ and to facilitate its identification in this region of the parameter space, it is important to choose a prior $\pi(\dd\lambda|\Sigma,\bs{\alpha})$ that puts small weights on $\{l\in\mathbb{R}: |l|/c_0>2\}$. In so doing, we propose to consider a conditional normal prior distribution with mean zero and variance $c_0^2$, i.e.  $\lambda|(\Sigma,\boldsymbol{\alpha})\sim \mathcal{N}_1(0,c_0^2)$. This naturally leads to a normal-inverse Wishart distribution as a prior for $(\boldsymbol{\xi},\Sigma)$, which is the conjugate prior for Gaussian models \citep[e.g., see][]{Gelman2004}. Note that it turns to ease the Bayesian model selection procedure (see Section \ref{sec:discussion}). Hence,
\begin{equation}\label{eq:prior_iid}
\pi(\boldsymbol{\xi},\Sigma|\bs{\alpha})\propto \exp\Big(-\frac{1}{2}\mathrm{tr}(V\Sigma^{-1})-\frac{\kappa}{2}(\boldsymbol{\xi}-\boldsymbol{\xi}_0)'\Sigma^{-1}(\boldsymbol{\xi}-\boldsymbol{\xi}_0)\Big)|\Sigma|^{-\frac{\nu+d+2}{2}}
\end{equation}
where $V$ is a $d\times d$ positive definite matrix, $\kappa$ and $\nu$ are real such that $\nu>d+3$. This last condition ensures that the mean of the prior distribution of $\Sigma$ exists and that all  its components has finite variance. Finally, one can choose  a vague prior for $\bs{\alpha}$, e.g. $\boldsymbol{\alpha}\sim \mathcal{N}_d(\bs{\mu}_{\bs{\alpha}},\sigma_{\bs{\alpha}}^2I_d)$ with $\sigma_{\bs{\alpha}}^2$ large and $I_d$ the $d\times d$ identity matrix. In practice, it is likely to have information on the sign of $\alpha_i$, $i=1,...,d$, through information about the asymmetry of the full conditional distribution  of $Y_i$ (see Proposition \ref{prop:closure}).  This prior knowledge can be incorporated in the Bayesian analysis by taking $\bs{\mu}_{\bs{\alpha}}\neq\mathbf{0}_d$.

 On the other hand, the $(\boldsymbol{\xi},\Omega,\boldsymbol{d},c)$-parametrization shares the same issues as the P1-parametrization since  $c=\lambda/c_0$. In addition, since an ESN random vector  $\mathbf{Y}$ is defined by $\mathbf{Y}{\buildrel d \over =}\boldsymbol{\xi}+ \mathbf{d}Z_3+\Omega^{1/2} \widetilde{\mathbf{Y}}_3$, where $-Z_3\sim \mathcal{TN}_c(0,1)$ and $\widetilde{\mathbf{Y}}_3\sim \mathcal{N}_d(\mathbf{0}_d,I_d)$, the convolution representation of ESN random vectors leads to an additional identification problem that arises when $\Omega$ is ``large'' or ``small'' relative to $\mathbf{d}$---more variability of one of the convolution-based density is obtained at the expense of weak identification of the other. In this respect, we assume that $\mathbf{d}|(\bs{\xi},\Omega)\sim \mathcal{N}_d(\boldsymbol{\mu}_{\mathbf{d}},\kappa^{-1}_{\mathbf{d}}\Omega)$, with $\kappa_{\bs{d}}=2(\sigma_{\alpha}^2(\tilde{\nu}-d-1))^{-1}$, yielding, on average, the same variance for both $\mathbf{d}$ and $\boldsymbol{\alpha}$.  The choice of a Gaussian distribution for $(\mathbf{d}|\Omega)$ is motivated by the fact that, together with the assumption that the prior distribution of $(\boldsymbol{\xi},\Omega)$ is the normal-inverse Wishart distribution $\pi(\boldsymbol{\xi},\Sigma|\tilde{\boldsymbol{\xi}}_0,\tilde{\kappa},\tilde{\nu},\tilde{V})$, we can easily implement a Gibbs sampler when $c$ is known \citep{Sahu2003}. Say differently, the Gaussian prior for $\mathbf{d}$ and the normal-inverse Wishart prior for $(\boldsymbol{\xi},\Omega)$ are some natural candidates for the SN distribution of \citet{Sahu2003}. Since $\Sigma-\Omega=\Sigma\boldsymbol{\alpha}
\boldsymbol{\alpha}'\Sigma/c_0^{2}$, which is a positive definite matrix, we choose  $(\tilde{\nu},\tilde{V})$ such that the inverse Wishart distribution $\mathcal{W}^{-1}(\tilde{V},\tilde{\nu})$ gives more weight to ``small'' values than  the $\mathcal{W}^{-1}(V,\nu)$ distribution. This can be done by taking $\tilde{V}$ such that $V-\tilde{V}$ is positive definite and $\tilde{\nu}\geq \nu$. In this case, the difference between the mode under $(\tilde{\nu},\tilde{V})$ and the mode under $(\nu,V)$ is negative definite. This also holds for the mean.

\subsection{A  SMC sampler for multivariate ESN distributions}\label{subsec:SMC}

\subsubsection{General description}

Let $\theta$ be the vector whose components are the parameters of the model (either under P1 or under P2), $f(\mathbf{z}_{1:n}|\theta)$ be the likelihood function, where $\mathbf{z}_{1:n}=(\mathbf{z}_1,...,\mathbf{z}_n)$ is the set of observations, and $\pi(\theta)$ be the prior distribution of the parameters, which is either $\pi_{\text{P1}}(\theta)$ under P1 or $\pi_{\text{P2}}(\theta)$ under P2. Using these notations, the posterior distribution we want to sample from is given by:
$$
 \pi(\theta|\mathbf{z}_{1:n})\propto f(\mathbf{z}_{1:n}|\theta)\pi(\theta).
$$
As pointed out by \citet{DelMoral2006}, sequential Monte Carlo samplers are  relevant when there is no fully-eligible proposal distribution, say $\eta_1(\theta)$, in order to implement the importance sampler. The SMC sampler requires to define a sequence of distributions $\{\pi_t(\theta)\}_{t=1}^T$ such that (i) $\pi_T(\theta)=\pi(\theta|\mathbf{z}_{1:n})$
and (ii) $\pi_1(\theta)=\eta_1(\theta)$, with $\eta_1$ a distribution we can easily sample from.  This sequence of intermediary distributions is purely instrumental and could be defined by making use of an appropriate real sequence of so-called temperatures $\{\rho_t\}_{t=1}^T$, increasing from zero to one. Following \citet{Gelman1998}, \citet{Neal2001} and  \citet{DelMoral2006}, we consider the geometric bridge:
\begin{equation*}
\pi_t(\theta):\propto\eta_1(\theta)^{1-\rho_t}\pi(\theta|\bz_{1:n})^{\rho_t}.
\end{equation*}

The basic idea of a SMC algorithm  is first to sample $N\geq 1$ particles $\theta_1^m$ from the initial  distribution $\pi_1$ in order to obtain a Monte Carlo approximation $\pi_1^N=N^{-1}\sum_{m=1}^N\delta_{\theta_1^m}$ of $\pi_1$. Then, using resampling and propagation steps, SMC  uses the approximation $\pi_1^N$ of $\pi_1$ to construct $\pi_2^N=\sum_{m=1}^NW_2^m\delta_{\theta_2^m}$, a Monte Carlo approximation of  $\pi_2$. Informally, if  $\pi^N_1$ is a good approximation of $\pi_1$ and if $\pi_2$ is close to $\pi_1$, then one may expect  $\pi^N_2$ to be close to $\pi_2$, and so on.

More precisely, suppose  that, for a $t\in\{2,\dots,T\}$,   one has at hand a  sample  $\left\{ \theta_{t}^{m}\right\}_{m=1}^N$  such that:
$$
\frac{1}{N}\sum_{m=1}^N \delta_{\theta_{t}^m}(\dd\theta)\approx \pi_{t}(\theta)\dd\theta.
$$
Then, we can approximate $\pi_{t+1}$ by the empirical distribution
\begin{equation*}
\sum_{m=1}^N W_{t+1}^m(\rho_{t+1})\delta_{\theta_{t}^m}(\dd\theta)
\end{equation*}
where the corresponding importance functions $W^m_{t+1}(\cdot)$ are defined to be:
\begin{equation*}
W_{t+1}^m(\rho) = \frac{w_{t+1}(\theta_{t}^m,\rho)}{\sum_{j=1}^N w_{t+1}(\theta_{t}^j, \rho)},\qquad w_{t+1}(\theta, \rho) 
= \left[\frac{\pi(\theta|\bz_{1:n})}{\eta_1(\theta)}\right]^{\rho-\rho_{t}}.
\end{equation*}
Note that $\rho_{t+1} - \rho_{t}$ measures the step length at time $t+1$ so that, the larger the difference, the more the accuracy of the importance weighting worsens. To control such a degeneracy, we consider a procedure to determine a suitable sequence of $\left\{\rho_t\right\}_{t=1}^T$ through the effective sample size  criterion. More specifically, instead of regarding $T$ and the set $\{\rho_t\}_{t=1}^T$ as parameters of the algorithm, we view them as self-tuning parameters using the method proposed by \citet{Schafer2012}. Given a value of $\rho_t$ and a sample $\{\theta_t^m\}_{m=1}^N$ that approximates $\pi_t$, we compute the largest value of  $\rho\in (\rho_t,1]$ such that the particle system $\{\theta_t^m\}_{m=1}^N$, once being properly weighted, allows to approximate ``reasonably well'' the probability distribution $\pi_{\rho}\propto \eta_1^{1-\rho}\pi^{\rho}$ through the effective sample size  criterion \citep{Liu1995}:
$$
\mathrm{ESS}_{t}(\rho)=\left[\sum_{m=1}^N W^m_{t}(\rho)^2\right]^{-1}
$$
where, by definition,  $W^m_{t}(\rho)$ is the weight assigned to $\theta_t^m$ to target $\pi_{\rho}$. If the effective sample size equals $N$, the interpretation is that the weights are equally balanced and that all $N$ particles are equally contributing to the estimation. Then, $\rho_{t+1}$ is defined as the minimum  between 1 and $\rho_{t+1}^\star$ with:
\begin{equation*}
\rho_{t+1}^\star =  \sup\left\{\rho > \rho_{t}: \mathrm{ESS}_t(\rho)\geq \beta \right\}
\end{equation*}
where $\beta$ is a pre-specified threshold, say $\beta = N/2$. The fixed value $\rho_{t+1}^\star$ can be obtained by solving the equation $\mathrm{ESS}_t(\rho) = \beta$ using the bi-sectional search algorithm of \citet{Schafer2012} (see Algorithm \ref{alg:rho}).

At every $\rho_t$, a resampling step, using the systematic resampling method of \citet{Carpenter1999}, is first performed in order to suppress particles that are in the region of the parameter space that receives very little mass from $\pi_t$. Say differently, the particles with the largest weights have multiplied whereas those with the smallest weights have vanished after the resampling step. Then, to restore  particle diversity, new particles are generated from a Markov Kernel $K^N_t(\theta',\dd\theta)$ with invariant distribution $\pi_t$ (see further).

The complete procedure  is summarized in Algorithm \ref{algo:SMC}. Any operation involving the superscript $m$ (respectively, subscript $t$)  must be understood as performed for $m \in 1:N$ (respectively, $t \in 0:T$) where $N$ (respectively, $T$) is the total number of particles (respectively, number of iterations). Note that $n$ denotes the sample size.
In addition, the procedure to find the step length is described in Algorithm \ref{alg:rho}.

\begin{algorithm}
\caption{Tempering Sequential Monte Carlo Sampler }\label{algo:SMC}
\begin{algorithmic}
\State Operations must be performed for all $m=1,\dots, N$.
\State\underline{Initialization}
\State Set $t=2$ and $\rho_1=0$.
\State Generate $\theta_1^{m}\sim \eta_1(\dd\theta)$ and compute $W^m(\rho_1)$.

\While{$\rho_{t-1}<1$}
\State Compute $\rho_t$ using Algorithm \ref{alg:rho} with inputs $\rho_{t-1}$ and $\{\theta_{t-1}^m\}_{m=1}^N$.
\State \underline{Resampling}: Generate $a_{t-1}^m=F^{-1}_{t,N}(u_t^m)$ where $u_t^m=\frac{m-1+u_t}{N}$, $u_t \sim\mathcal{U}\left((0,1)\right)$ and
$$
F_{t,N}(i)=\sum_{m=1}^NW_t^m(\rho_t)\mathbb{I}(m\leq i).
$$

\State  \underline{Propagation}: Generate  $\theta_t^m\sim K_t^N(\theta^{a_{t-1}^m}_{t-1},\dd \theta)$.
\State Set $t\leftarrow t+1$.
\EndWhile
\end{algorithmic}
\end{algorithm}

\begin{algorithm}
\caption{Find step length using \citet{Schafer2012}}\label{alg:rho}
\begin{algorithmic}
\State \textbf{Input}: $\epsilon$, $\rho$, $\{\theta^m\}_{m=1}^N$.
\State $l\gets 0$, $u\gets1.05$, $\delta\gets0.05$.
\While{$|u-l|\geq \epsilon$ and $l\leq 1-\rho$}
\If{$\left[\sum_{m=1}^N W^m{}(\rho+\delta)^2\right]^{-1}<N/2$}
\State $u\gets \delta$, $\delta\gets(\delta+l)/2$
\Else
\State $l\gets \delta$, $\delta\gets(\delta+u)/2$
\EndIf
\EndWhile

\State \textbf{Return} $\min (\rho+a,1)$.
\end{algorithmic}
\end{algorithm}

\subsubsection{Implementation}

In our implementation we follow the usual approach and take for $K^N_t(\theta',\dd\theta)$ the Markov kernel that corresponds to $\tau$ steps of the Gaussian random-walk Metropolis-Hastings algorithm with variance-covariance matrix given by $c_s\Omega^N_{t}$. The constant $c_s>0$ is a scale factor such that the acceptance rate of the kernel lies in the range $[0.2,0.6]$ while $\Omega^N_t$ is a particle-based  estimation of the variance-covariance matrix that corresponds to the distribution $\pi_t$.

The initial distribution $\eta_1$ is another critical element for the speed of convergence of the algorithm and for the precision of the estimates. The first obvious option is to take the prior distribution, so that the SMC sampler moves simulations from the prior to simulations from the posterior distribution. Nevertheless, starting the SMC sampler with simulations from the prior can lead to a very low convergence rate of the algorithm and some large Monte-Carlo errors since there is no reason for the prior to be close to the posterior distribution.  A better approach consists in  initializing the sampler with an approximation of the target distribution from which we can easily sample. When one can maximize the posterior distribution, this is effectively done by a Laplace approximation. In this case, $\eta_1$ would be a normal distribution with mean $\mathbf{m}_1$ and covariance matrix $\Sigma_1$, where $\mathbf{m}_1$ is set to the posterior mode and $\Sigma_1$ is equal to minus the inverse of the Hessian matrix evaluated at the posterior mode. In some settings (see Section \ref{sec:num2}), the numerical maximization of the posterior distribution might be particularly  troublesome. In this case, we use a pilot run of a Gaussian random walk Metropolis-Hastings  algorithm to get an estimate $\hat{\mathbf{m}}$ of the posterior mean and an estimate $\hat{\Sigma}$ of the posterior covariance matrix, and we define $\eta_1$ as the a normal distribution with mean $\hat{\mathbf{m}}$ and covariance matrix $\hat{\Sigma}$.

\subsubsection{Discussion} \label{sec:discussion}

Using Algorithm \ref{algo:SMC}, one can obtain estimates of the target distributions and the normalizing constants directly from the variables generated by the sampler. Indeed, at the end of iteration $T$,  an approximation of the target distribution $\pi(\theta|\bz_{1:n})$ is given by:
\begin{eqnarray*}
\pi_T^N(\dd\theta)=\frac{1}{N}\sum_{m=1}^N \delta_{\theta_T^m}(\dd\theta).
\end{eqnarray*}
Moreover an estimate of the normalizing constant $Z_T$ of the posterior distribution $\pi_T(\theta)$ can be obtained as follows. Let $Z_t$ be the normalizing constant of $\pi_t$. Then, we
can estimate of $Z_T/Z_{1}$ by \citep{DelMoral2006}:
$$
\widehat{\frac{Z_T}{Z_{1}}}=\prod_{t=2}^T\widehat{\frac{Z_t}{Z_{t-1}}},\quad \widehat{\frac{Z_t}{Z_{t-1}}}=\sum_{m=1}^NW_{t-1}^m(\rho_{t-1})\left[\frac{\pi(\theta_{t-1}^m|\bz_{1:n})}{\eta_1(\theta_{t-1}^m)}\right]^{\rho_t-\rho_{t-1}},\quad t\geq 2.
$$

A question of particular interest is whether the SN or the Gaussian distributions are more appropriate than the ESN distribution. In a Bayesian framework, the answer to this question is obtained by comparing the evidence, or marginal likelihood of the data, between the competing models.  More specifically, consider the general test $H_0: \theta = \theta^0$ against $H_1: \theta \neq \theta^0$, where $\theta^0$ is the vector of parameters under the null hypothesis. In this respect, we make use of the Bayes factor defined by:
$$
B_{10}=\frac{m_1(\mathbf{z}_{1:n})}{m_0(\mathbf{z}_{1:n})}
$$
where $\mathbf{z}_{1:n}$ is the observations and where $m_i(\mathbf{z}_{1:n})=\int f_i(\mathbf{z}_{1:n}|\theta)\pi_i(\dd\theta)$ is the evidence of model $i\in\{0,1\}$, with $f_i(\mathbf{z}_{1:n}|\theta)$ and $\pi_i(\dd\theta)$ the corresponding likelihood and the prior distribution.

It is well known \citep{Morin2013} that, if the competing models $i\in\{0,1\}$ are regular, then the Bayes factor is a consistent criterion to discriminate between $H_1$ and $H_0$. However, and  as discussed in Section \ref{subsec:lik}, if we wrongly assume that data are generated by some ESN distributions when the true underlying model is Gaussian, then the Fisher information matrix is singular and therefore there is no theoretical guarantee that the  Bayes factor selects asymptotically the true model.  We leave this issue for further research and rather assess the Bayes factor reliability through Monte Carlo simulations in Section \ref{sec:num2}.

At this stage it is worth noting that testing the ESN distribution against the Gaussian model is straightforward. Indeed, the evidence under the ESN distribution can be directly obtained as a by-product of Algorithm \ref{algo:SMC}, as explained above, while that under the Gaussian distribution can be computed explicitly thanks to the Gaussian conjugate prior \eqref{eq:prior_iid} for $\boldsymbol{\xi}$ and $\Sigma$ \citep[see e.g. ][]{Gelman2004}:
$$
m_0(\mathbf{z}_{1:n})=\frac{1}{\pi^{nd/2}}\frac{\Gamma_d(\nu_n/2)}{\Gamma_d(\nu/2)}\frac{|V|^{\nu/2}}{|V_n|^{\nu_n/2}}\left(\frac{\kappa}{\kappa_n}\right)^{d/2}
$$
where
$$
\kappa_n=\kappa+n,\quad \nu_n=\nu+n,\quad V_n=V+\sum_{i=1}^n(\mathbf{z}_i-\bar{\mathbf{z}}_n)(\mathbf{z}_i-\bar{\mathbf{z}}_n)^t+\frac{\kappa n}{\kappa+n}(\bar{\mathbf{z}}_n-\boldsymbol{\xi}^0)(\bar{\mathbf{z}}_n-\boldsymbol{\xi}^0)'.
$$

Finally, note that one key feature of  SMC algorithms is their flexibility. Indeed, the implementation of  Algorithm \ref{algo:SMC} only requires to be able to evaluate the likelihood function. The Bayesian methodology  developed in this section can therefore be easily modified to carry out parameter  inference in (complicated) parametric models based on the ESN distribution. This point is illustrated in  Section \ref{subsect:numESN} where we apply the proposed methodology  on an ESN sample selection model.



\section{Numerical study}\label{sec:num2}

In this section we provide some Monte Carlo simulations in order to assess the performances of the proposed Bayesian approach and the behaviour of the posterior distribution. We consider two main data generating processes: (1) IID univariate extended skew-normal random variables, and (2) an extended skew-normal sample selection model (ESNSM).  For the IID setting, the SMC sampler is initialized with a Laplace approximation of the posterior distribution while, for the ESNSM, the maximization of the posterior distribution turns out to be too sensitive to the choice of initial values in order to be  useful in the construction of a good approximation of the posterior distribution. In that case, and as described above, we calibrate the initial distribution of the sampler using 10\,000 iterations of a pilot Metropolis-Hastings algorithm. Finally, in all of the simulations presented below, the propagation step of the tempered sequential Monte Carlo algorithm is based on $\tau=3$ iterations of the Gaussian random walk Metropolis-Hastings kernel described in Section \ref{subsec:SMC}.

\subsection{Example 1: IID univariate ESN random variables}\label{subsect:iidNum}

We first consider a sample of IID  ESN random variables. To study the implications of the parametrization of ESN distributions, we use two data generating processes:
\begin{equation}\label{sim:eq:ESN1}
Z_1,\cdots,Z_n \sim \mathcal{ESN}^{(P1)}_1(2,6,5,-2)
\end{equation}
and
\begin{equation}\label{sim:eq:ESN2}
Z_1,\cdots,Z_n \sim \mathcal{ESN}^{(P2)}_1(2,1,5,-0.8)
\end{equation}
where  the sample size $n$ is successively 1\,000, 5\,000, and 10\,000. The variance, skewness and kurtosis of the first ESN distribution \eqref{sim:eq:ESN1} are respectively given by 2,  1, 4 whereas those of  the second ESN distribution \eqref{sim:eq:ESN2} are respectively given by 6.60, 0.99, and  4.28. For all of the simulations, the parameters for the prior distributions,  defined in Section \ref{subsec:Prior}, are set as follows: $\kappa=0.1$, $\bs{\mu}_{\bs{\alpha}}=\bs{\mu}_{\mathbf{d}}=\boldsymbol{\xi}_0=\mathbf{0}$, $\nu=\max(6,d+4)$, $V=12 I_d$, $\tilde{V}=2I_d$, $\tilde{\nu}=\nu$, $\tilde{\boldsymbol{\xi}}_0=\boldsymbol{\xi}_0$, $\tilde{\kappa}=\kappa$ and $\sigma_{\bs{\alpha}}^2=10$. Finally, the tempered sequential Monte Carlo algorithm makes use of 10\,000 particles.


\subsubsection{Parameters estimation}

Tables \ref{Table:runiv}  and \ref{Table:runiv2} report respectively  the results for  the two $\mathcal{ESN}_1$ distributions \eqref{sim:eq:ESN1} and  \eqref{sim:eq:ESN2}  when the sample size is 1\,000 and 5\,000. Several points are worth commenting. First, as to be expected, the parametrization matters irrespective of the posterior statistics criteria used to compare the overall fitting (posterior mean, posterior median or posterior mode) and of the sample size. More specifically, when the true model is defined from the hidden truncation-based representation (P1), the posterior mean, median and mode using the second parametrization have a larger bias than in the case of the first parametrization. Unsurprisingly, turning to the Bayes factor, we do observe a clear evidence in favor of the results obtained under P1. In contrast, when the true distribution is defined from the convolution-based representation, results in Table \ref{Table:runiv2}, and especially the Bayes factor, clearly provide support for the P2-based estimates. This parametric dependence is further illustrated in Figure \ref{Fig:UnivPost} which displays the marginal posterior distributions using P1 and  P2 when the sample size is 1\,000.

\begin{center}
[Tables \ref{Table:runiv}-\ref{Table:runiv2} and Figure \ref{Fig:UnivPost} here]
\end{center}

Second, comparing Tables \ref{Table:runiv} and  \ref{Table:runiv2}, it is worth noting that the results obtained using P1 are less sensitive to the parametrization of the underlying model than those obtained under P2. To understand this point, note that the parameter values of the $\mathcal{ESN}$ distribution \eqref{sim:eq:ESN1} are such that $\omega^2$ is close to the boundary of the parameter space ($\omega^2\approx  0.038$) and, consequently,  inference for this parameter is very sensitive to the choice of the prior distribution \citep[see e.g.][]{Newton1994, Gelman2006}. In particular, the prior we choose for $\omega^2$ puts a very small weight to values close to zero and therefore tends to overestimate $\omega^2$.
This nearly boundary problem is critical in the sense that even "non informative" prior distributions can have a substantial effect on  inference \citep[see e.g.][]{Gelman2008}. For that reason, and contrary to the current practise  \citep[see e.g.][]{Adock2004,Liseo2013}, we advocate for the use of the $(\bs{\xi},\Sigma,\bs{\alpha},\lambda)$-parametrization to carry out parameter inference in the ESN (and in the SN) distribution.

However, and this is our third observation, when the sample size gets larger and larger, posterior modes converge toward the true parameter values irrespective of the chosen parametrization. In particular, the middle panel of Figure \ref{Fig:UnivPost} provides strong support for the convergence of the marginal posterior modes when the sample size is 10\,000.

Finally, taking the low number of particles ($N=10\,000$), the Monte Carlo error is rather small in all cases and for all parameters, especially as the sample size increases. However, it is at the expense of a somehow large computing time which is, for both parametrization, around 90 seconds for $n=1\,000$ and around 460 seconds for $n=5\,000$.

\subsubsection{Model selection}

As explained in Section \ref{sec:discussion}, it is critical to assess the robustness of ESN distributions with respect to Gaussian distributions. Therefore we conduct some simulation experiments regarding the Bayes factor  to test the null hypothesis of normality against the alternative hypothesis of an extended skew-normal distribution, $\mathcal{ESN}^{(P_1)}_1(2,6,\alpha,\lambda)$, for different $(\alpha,\lambda)$ pairs. The results are reported in Table  \ref{Table:Bayes}, which describes the percentage of samples where the evidence in favor of the ESN hypothesis is poor ($\log_{10} B_{10}\leq 0.5$), substantial ($0.5<\log_{10} B_{10}\leq 1$), strong ($1<\log_{10} B_{10}\leq 2$) and decisive ($\log_{10} B_{10}>2$).

\begin{center}
[Table \ref{Table:Bayes} here]
\end{center}

The results presented in the first three lines of Table \ref{Table:Bayes} are obtained for a sample size of $n=100$. We observe that, despite the small number of observations, the Bayes factor yields very good results for $(\alpha,\lambda)=(0,-)$ (i.e., Gaussian model) and $(\alpha,\lambda)=(5,-2)$. Indeed, in both cases and in all samples, the Bayes factor selects the correct model with a strong confidence. For $(\alpha,\lambda)=(0.5,0)$, estimations are in favour of the Gaussian distribution  although the underlying model is ESN. This  results is intuitive. Indeed, the Bayes factor penalizes for the number of parameters. Therefore, since $\lambda$ is useless when the underlying model is Gaussian, it is natural that the Bayes factor is biased toward the Gaussian distribution when $\alpha$ is close to zero. In contrast, when the sample size increases (from $n= 100$ to  5\,000), the Bayes factor selects the correct model with a strong confidence. These results suggest that the Bayes factor is convergent even if no formal proof for this specific test is yet available in the literature (see Section \ref{sec:discussion}).


\subsection{Example 2: Extended skew-normal sample selection model}\label{subsect:numESN}

One key feature of the proposed methodology is its adaptability since SMC samplers can be used, at least from a theoretical point of view, as soon as one can evaluate efficiently the likelihood function. To illustrate this point in a more complicated set-up than in the previous subsection, we consider the estimation of a sample selection model based on ESN error terms. 

\subsubsection{Model description}

Thanks to Definition \ref{def:P1},  the application of ESN distributions in sample selection models or Tobit-type models \citep{Amemiya1986, Maddala1976} is a natural choice since any hidden truncation of normal component densities leads to such a distribution \citep[see][]{Arnold2002}. In this respect, starting from the Gaussian sample selection model \citep{Heckman1976}, a (multivariate) extended skew-normal sample selection model (ESNSM) can be defined by:
\begin{equation}\label{introduction:eq:model}
\begin{cases}
\mathbf{Y}_{i}^*=B\bx_{i}+\boldsymbol{\epsilon}_{1i}\\
S_{i}^*=\bs{\beta}_2'\bx_{i}+\epsilon_{2i},\quad i=1,...,n
\end{cases}
\end{equation}
where  $B\in\mathbb{R}^{d\times k_1}$, $\bs{\beta}_2\in\mathbb{R}^{k_1}$, and
\begin{equation}\label{eq:law}
\boldsymbol{\epsilon}_i\sim\mathcal{ESN}^{(P_1)}_{d+1}
\left(\boldsymbol{\xi}=(\boldsymbol{\xi}_1,\xi_2),\Sigma=
\begin{pmatrix}
\Sigma_1&\Sigma_{12}\\
\Sigma_{21}&1
\end{pmatrix}
, \boldsymbol{\alpha}=(\boldsymbol{\alpha}_1,\alpha_2),\lambda\right)
\end{equation}
with $\boldsymbol{\xi}=-\frac{\Sigma\boldsymbol{\alpha}}{c_0}\frac{\phi(\lambda/c_0)}{\Phi(\lambda/c_0)}$  such that $\E[\boldsymbol{\epsilon}_i]=\mathbf{0}_d$. We assume that we observe $S_i=\mathbb{I}_{\mathbb{R}_+}(S_{i}^*)$ and $\mathbf{Y}_i=\mathbf{Y}_{i}^*S_{i}$, with $\mathbb{I}_{A}(\cdot{ })$ the indicator function of $A\subseteq\mathbb{R}$. The likelihood function of the model, which is required to compute the importance weights of the SMC sampler (Algorithm \ref{algo:SMC}), follows from a direct application of Proposition \ref{prop:closure} (closure under conditioning and marginalization of the extended skew-normal family of distributions):
\begin{equation*}
\begin{split}
&L_n(\theta,\bs{\beta}_2, B)=\prod_{i=1}^n\left[\frac{\Phi_2\left(-\bs{\beta}'_2\bx_i-\xi_2,1
,c_2\tilde{\alpha}_2,c_2\lambda\right)}{\Phi\left(\frac{c_2\lambda}{\sqrt{1+c_2^2\tilde{\alpha}_2^2}}\right)}\right]^{1-s_{i}}\\
&\times \left[\phi_d\left(\mathbf{y}_i,B\bx_i+\boldsymbol{\xi}_1,\Sigma_1\right)
\frac{\Phi_2(m_i,\sigma_{22.2}^2,-\alpha_2,\lambda+\tilde{\boldsymbol{\alpha}}'_1(\mathbf{y}_{i}-
B\bx_{i}-\boldsymbol{\xi}_1))}
{\Phi\left(\frac{c_1\lambda}{\sqrt{1+
\tilde{\boldsymbol{\alpha}}'_1\Sigma_1
\tilde{\boldsymbol{\alpha}}_1c_1^2}}\right)}\right]^{s_{i}}
\end{split}
\end{equation*}
where $m_i=\bs{\xi}_1+\bs{\beta}'_2\bx_i+
\Sigma_{12}\Sigma^{-1}_1(\mathbf{y}_i-B\bx_i
-\boldsymbol{\xi}_1)$ and with $\tilde{\alpha}_1$ and $\tilde{\bs{\alpha}}_2$ defined as in Proposition \ref{prop:closure}. The prior distributions for the $\bs{\alpha}$ and  $\lambda$ parameters are the same as the ones defined in Section \ref{subsec:Prior}, while those for $\Sigma$, $B$ and $\bs{\beta}_2$ are discussed in Appendix \ref{appendix:ESNSM}.

\subsubsection{Simulation set-up}

The numerical study is conducted for the univariate extended skew-normal sample selection model:
\begin{equation}\label{sim:eq:model}
\begin{cases}
Y^*_i=\beta_{10}+\beta_{11}x_{1i}+\epsilon_{1i}\\
S_i^*=\beta_{20}+\beta_{22}x_{2i}+\epsilon_{2i},
\end{cases}
\end{equation}
and
\begin{equation}\label{sim:eq:resESN}
(\epsilon_{1i},\epsilon_{2i})\sim\mathcal{ESN}^{(P_1)}_2\left(\bs{\xi},\bigl(\begin{smallmatrix}
6&\rho\sqrt{6}\\
\rho\sqrt{6}&1\\
\end{smallmatrix} \bigr),(2,1),-2\right)
\end{equation}
where  $\rho\in\{-0.9,0.3,-0.9\}$. The parameter value of $\rho$ is a key issue in sample selection models. Notably when $\rho = 0$, there is no selection effect. On the other hand, it can be shown that the correlation between $\epsilon_{1i}$ and $\epsilon_{2i}$ increases with this parameter, as shown  in Figure \ref{sim:fig:shape}. The parameter values for the  $\beta$'s parameters are respectively given by $\beta_{10}=3$, $\beta_{11}=-2$, $\beta_{20}=1.5$ and $\beta_{22}=2$ while the covariates  $x_{1i}$ and $x_{2i}$ are assumed to be independent $\mathcal{N}_1(0,2)$ random variables (without loss of generality). This setup implies that $S_i=0$ for about $30\%-35\%$ of the $n=1\,000$ observations.

\begin{center}
[Figure \ref{sim:fig:shape} here]
\end{center}

We  discuss below some Monte Carlo simulations results for the extended skew-normal sample selection model \eqref{sim:eq:model}-\eqref{sim:eq:resESN}. The purpose of this numerical study is to  compare it with the  standard Tobit-type 2 model \citep[i.e., the sample selection model with Gaussian errors, see][] {Amemiya1986} regarding the estimation of the parameters and of the marginal effects.

\subsubsection{Parameters estimation}

Table \ref{sim:table1} provides the posterior mean and the standard deviation of 50 independent estimates of the parameters of the model \eqref{sim:eq:model}-\eqref{sim:eq:resESN}  under the two parametric assumptions (i.e., the bivariate extended skew-normal and the Gaussian distribution of the error terms). Results are reported for the three different values of $\rho$.

\begin{center}
[Table \ref{sim:table1} here]
\end{center}

Regarding the estimation of the constant and slope parameters of the regression equation, $\beta_{10}$ and $\beta_{11}$, we observe that the distributional assumption has a limited effect on the estimated values in all  scenarios. A similar result is obtained for the Student selection model in \citet{Marchenko2012} and for the skew-normal model in \citet{Ogundimu2012}. On the other hand, the estimation of the corresponding parameters in the selection equation, $\beta_{20}$ and $\beta_{22}$, are more sensitive to the choice of the error terms distribution. Indeed, if the Gaussian assumption leads to a small bias for these parameters when the correlation between the variable of interest and the selection variable is low (i.e. when $\rho=0.3$, implying a correlation  between -0.02 and -0.03), the results obtained with the Tobit 2 model for these parameters are significantly biased for larger values of $|\rho|$.

\begin{center}
[Figures \ref{sim:fig:Biais}-\ref{sim:ME1} here]
\end{center}

To illustrate the importance of the bias for $\beta_{20}$ and $\beta_{22}$, Figure \ref{sim:fig:Biais} reports, when  $\rho=0.9$, the individual estimates of these parameters over 50 simulations in  the presence of misspecified error terms---they are wrongly assumed to be normally distributed. Taking that the true parameter vector is given by $(\beta_{20}, \beta_{22}) = (1.5, 2)$, we observe that all of the estimates of $\beta_{20}$ and $\beta_{22}$ are much larger than the true underlying parameter values. 
To some extent, this result is consistent with standard results relative to the misspecification issues of the maximum likelihood estimator of Tobit-type models \citep{Amemiya1986}.

In contrast, when the model is correctly specified, the constant and slope parameters of the selection process are well-estimated irrespective of the correlation parameter $\rho$. Notably, the posterior mean of each parameter is close to the true parameter value and the estimation error is small. Regarding other parameters, we obtain very good estimations of $\sigma_1^2$ and $\rho$ for which we observe both a small bias and a small standard deviation. The estimation of  $\alpha_2$ turns out to be more challenging due to the loss of information engendered by the censorship mechanism through $S_i^*$.  Moreover, the posterior mean of $\lambda$ is close to the true value at the expense of a relatively large standard deviation (precision), especially with respect to other parameters. 

Further evidence is provided by Figure \ref{sim:fig:ME1}, which displays the marginal posterior distributions of the parameters in the case of one realization of the model \eqref{sim:eq:model}-\eqref{sim:eq:resESN} with $\rho=0.3$. In addition to previous results, three points are worth commenting. First, the posterior modes are close to the true parameter values. Second, the marginal distribution for the $\beta$'s parameters are very concentrated around the mode. Third, the sign of the $\alpha$'s parameters, and hence of the skewness of the data, is well-identified since the posterior mass on $\{\alpha_i<0, i=1,2\}$ is close to zero. In contrast, there is a small but significant posterior probability for the event $\{\lambda>0\}$ suggesting that more observations are needed to identify more precisely this parameter.

\begin{center}
[Figures \ref{sim:fig:posterior} here]
\end{center}

\subsubsection{Marginal effects}
 For ease of interpretation, it is arguably better to consider the (average) marginal effects \citep{Cameron2005} since only the sign (but not the magnitude) of the coefficients can be readily interpreted in Tobit-type models. In this respect, we compute the marginal effects (see Proposition \ref{prop:me} in Appendix \ref{appendix:ESNSM}) and Figure \ref{sim:fig:posterior} displays marginal effect estimates of $\beta_{22}$ on $\E[Y_i^*|S_i=1,\bx_i]$ for a realization of the above model with $\rho=0.3$. The main result is that the Gaussian model is not able to account for substantial heterogeneity in marginal effects. Indeed, a visual inspection shows that the distribution of Gaussian estimates is much more concentrated than the distribution of the true values. In addition, the marginal effects obtained from the Tobit type-2 models are in all cases larger than -10 although for a very significant proportion of individuals the marginal effect of $\beta_{22}$ is indeed smaller than this threshold (with a minimum nearby -60). The average marginal effect estimate under the Gaussian assumption is around -0.14 while the true value is about 15 times larger (around -2.08). In contrast, this estimate under the ESN assumption is -2.22. Some marginal effect estimates under the correct parametric assumption are also reported and are, as to be expected, very close to the true values.

\section{Applications}\label{sec:app}

In this section, we proceed with two real applications.  In both cases, estimations are performed using the P1-parametrization (Definition \ref{def:P1}), with the prior distribution as in Section \ref{subsect:iidNum}. The absence of a constraint on $\lambda$ contrasts with most of the applications of skew-elliptical distributions in the literature that set the $\lambda$ parameter to zero (or consider some arbitrary value of $c$). Finally, the SMC sampler is initialized using a pilot run of a Gaussian random walk Metropolis-Hastings and the propagation step is based on $\tau=3$ iterations of a Gaussian  Metropolis-Hastings kernel (see Section \ref{subsec:SMC}).

\subsection{Transfer fees of soccer players}

As an application of the univariate ESN,  we consider a data set with 1\,062 observations on (log-) transfer fees in major European soccer leagues. Data, which have been collected from various sources and are available upon request, cover the period 2008-2012 for the first league (England and France), Bundesliga (Germany), Calcio (Italy) and Liga (Spain).

\begin{center}
[Figures \ref{Fig:UnivPostFootESN} and \ref{Fig:UnivPostFootSN} here]
\end{center}

Figure \ref{Fig:UnivPostFootESN} (resp. Figure \ref{Fig:UnivPostFootSN}) presents the marginal posterior distributions when data are assumed to be randomly generated from an ESN distribution (resp. a SN distribution). Visual inspection of the marginal posterior distribution indicates that the ESN-based marginal posterior distribution of $\lambda$ has most of its mass on the  interval $(-\infty,-2]$. Paired with the fact that the posterior distribution for $\alpha$ has most of its mass on  $[1.2,\infty)$, this suggests that the ESN-based specification fits better the overall distribution of data than the SN-based specification  of \citet{Azzalini1985}. Further evidence can then be provided by comparing the marginal likelihood values of the two models. Notably, using the output of the SMC sampler, the (log-) evidence of the ESN-based model is -685.0374 whereas it is only -797.1437 in the case of the SN-based model. This means that the evidence in favor the ESN-based model can be considered as being "decisive" in the sense of \cite{Jeffreys1939}. Finally, to assess the robustness of our results, Figure \ref{Fig:densityFoot} compares the ESN estimate of the density function of the data with a non-parametric estimate: one can observe that both provide very similar results.

\begin{center}
[Figure \ref{Fig:densityFoot} here]
\end{center}

\subsection{Bivariate ESN: Financial Data}

As a final illustration of the proposed algorithm, we proceed with a real financial data set as in \citet{Liseo2013}. There is an impressive literature in finance that has witnessed the fact that (high-frequency) financial returns are skewed and display leptokurtic tails \citep[e.g., see][]{Jondeau2006, Genton2004} and may have strong implications in portfolio selection, asset pricing models or risk measurement (among others). In this respect, we consider a simple \texttt{i.i.d.} bivariate sampling model. 
More specifically, we analyse the weekly returns (in percentage) of two US stocks, namely ``ABM Industries Incorporated'' (ABM) and ``The Boeing Company'' (BA). The sample size covers the period Jul 19, 1984 to  Jul 28, 2014 (1\,566 observations).\footnote{We also perform estimation with daily and monthly returns. Our main results remain unchanged.}

\begin{center}
[Figures \ref{Fig:UnivPostReal} and \ref{Fig:UnivPostRealSN} here]
\end{center}
Figures \ref{Fig:UnivPostReal} and \ref{Fig:UnivPostRealSN} depict, respectively, the marginal posterior distribution of each parameter under the ESN and the SN assumption. Two points are worth commenting. First the contour plot of the density of the estimated $\mathcal{ESN}_2$ model suggests that raw data, which are skewed and fat-tailed, can be reasonably well-captured by this specification. Second, the marginal posterior modes of the shape ($\alpha_1$ and $\alpha_2$) and the shift parameter ($\lambda$) are roughly given by 0.13, 0.20 and -3, respectively. Combined with the fact that the (marginal) posterior of each of these parameters has a negligible mass with positive (for $\lambda$) or negative (for $\alpha_1$, $\alpha_2$) values, the estimation provides strong support for the application of an extended skew-normal distribution in order to jointly model ABM and BA. Moreover, according to standard stylized financial facts of weekly returns, the location parameters, $\xi_1$ and $\xi_2$, are negative (large negative returns are more important than large positive returns) and the marginal posterior modes of the unconditional variance-covariance parameters ($\sigma_1^2$, $\sigma_2^2$, $\sigma_{12}^2$) support large volatility and co-volatility.

Finally we proceed with model selection. Using the SMC estimate of the Bayes factor, we find that the evidence in favor of the skew-normal bivariate distribution proposed by \citet{Liseo2013}  is poor \citep[in the sense of][]{Jeffreys1939}.

\section{Conclusion}

In this paper, we propose a new Bayesian computational approach, which rests on a tempered sequential Monte Carlo sampler, to estimate (multivariate) extended skew-normal distributions. Among others, the proposed approach have several advantages. First, it overcomes some issues encountered in standard maximum likelihood estimation. Second, in contrast to MCMC methods, it is easy to build a SMC algorithm that is adaptive in the sense that it can  adjust sequentially and automatically its sampling distribution to the problem at hand provided some well-defined prior distributions. Especially, it can implemented for a large class of (multivariate) skew-elliptical distributions. Third, it allows to compute easily as a by-product the marginal posterior distributions, the normalizing constant and thus the Bayes factor. Fourth, it embeds as a special case the population algorithm provided by \citet{Liseo2013}.

 Monte Carlo simulations provide evidence regarding the robustness of the proposed algorithm with different data generating processes. Irrespective of the model considered (sampling models, extended skew-normal sample selection models), posterior statistics  are rather precise (with a low standard deviation) in a tractable computing time. Moreover, results suggest that the hidden truncation-based parametrization is more robust for estimation than the convolution-based parametrization. 
 Directions for future research include more comprehensive empirical applications \citep{Gerber2014} and the derivation of more general models with hidden truncation, censoring or selective report with the (multivariate) extended skew-normal family of distributions or some unified skew-elliptical distributions.

\section*{Acknowledgements}
Financial support from the Swiss
National Science Foundation (research module ``Modelling
simultaneous equation models with skewed and heavy-tailed
distributions'') is gratefully acknowledged.

\appendix

\section{Proof of Proposition \ref{prop:lik}}\label{app:lik}

Let $l_n(\theta)=\log L_n(\theta)$ and $\theta\in\Theta_{\epsilon}^{l^*}$ where $\Theta_{\epsilon}(l^*)=\{\theta:\|\theta-\theta_{n,G}^{l^*}\|\leq \epsilon\}$. Then, $l_n(\theta^{l^*}_{n,G})-l_n(\theta)>0$ means that
\begin{align*}
l_n^G(\theta_G^{l^*})-l_n^G(\theta)-\frac{1}{N}\sum_{i=1}^n\log \Phi\left(l+a(z_i-m)\right)+\log \Phi(l/c_0)\geq 0
\end{align*}
where $l_n^G$ is the log-likelihood corresponding to the Gaussian model. A sufficient condition for the above inequality to hold is
$$
\log\Phi\left(\frac{l^*-\epsilon}{\sqrt{1+(\sigma_{n,G}^2+\epsilon)\epsilon^2}}\right)\geq \log \Phi\left(l^*+\epsilon(1+\bar{z}_n-\xi_{n,G}+\epsilon)\right)
$$
where $\bar{z}_n=\max \{|z_i|\}$. This is equivalent to
$$
l^*\leq l_{n,\epsilon}^*:= \frac{\epsilon+\sqrt{1^*+(\sigma_{n,G}^2+\epsilon)\epsilon^2}
\left(\epsilon(1+\bar{z}_n-\xi_{n,G}+\epsilon)\right)}{1-
\sqrt{1^*+(\sigma_{n,G}^2+\epsilon)\epsilon^2}}.
$$
Hence, for all $\epsilon>0$, there exists a $l^*_{n,\epsilon}$ such that
$$
l^G(\theta_{n,G}^{l^*_{n,\epsilon}})-l(\theta)-\frac{1}{N}\sum_{i=1}^n\log \Phi\left(l+a(z_i-m)\right)+\log \Phi(l/c_0)\geq 0\quad \forall \theta\in \Theta_{\epsilon}(l^*_{n,\epsilon}).
$$
To prove part 2., let $\epsilon$ and $M\geq 1$ be such that $
c_n:=\|\tilde{\theta}_n-\tilde{\theta}_G\|=\frac{\epsilon}{M}$ where $\theta=(\tilde{\theta},l)$. Then, if
$$
l_{n,\epsilon}^*-\epsilon\left[1-\frac{c_n^2}{\epsilon^2}\right]^{1/2}\leq l_n\leq l_{n,\epsilon}^*+\epsilon\left[1-\frac{c_n^2}{\epsilon^2}\right]^{1/2}
$$
we have $\|\theta_n-\theta_G^{l^*_{n,\epsilon}}\|\leq \epsilon$ so that $l_n(\theta_G^{l^*_{n,\epsilon}})\geq l_n(\theta_n)$.

\section{Extended skew-normal sample selection models\label{appendix:ESNSM}}

\subsection{Prior distributions for $\Sigma$ $B$ and $\bs{\beta}_2$}

When available, the conjugate prior distribution is frequently used in bayesian analysis. Under Gaussian error terms and no selection effect,  the conjugate prior distribution for $\beta_1$  and $\Sigma$ is the normal-inverse Wishart distribution:
\begin{align*}
\pi(\bs{\beta}_1,\Sigma|\bs{\mu}_{\bs{\beta}_1},\kappa,\nu,V)&\propto \exp\left(-\frac{1}{2}\text{tr}(V\Sigma^{-1})-\frac{\kappa}{2}(\bs{\beta}_1-\bs{\mu}_{\bs{\beta}_1})'(\Sigma^{-1}\otimes c_{\bs{\beta}_1}X'X)(\bs{\beta}_1-\bs{\mu}_{\bs{\beta}_1})\right)\\
&\times |\Sigma|^{-\frac{\nu+|\bs{\beta}_1|+2}{2}}
\end{align*}
where $c_{\bs{\beta}_1}$ is a scale factor, $V$ is a $d\times d$ positive definite matrix, $\kappa$ and $\nu$ are real such that $\nu>|\bs{\beta}_1|+3$\footnote{This last condition is not necessary but ensures that  all
the components of $\Sigma$ has a finite variance.}. Since the ESN distribution generalizes the Gaussian distribution, and because the presence of selection effect does not modify our prior knowledge, we choose
this prior distribution for $\bs{\beta}_1$ and $\Sigma$.

Using a similar argument, a possible choice of prior distribution for the parameters of the selection equation is $\bs{\beta}_2\sim \mathcal{N}_{|\bs{\beta}_2|}(\bs{\mu}_{\bs{\beta}_2}, c_{\bs{\beta}_2}(X'X)^{-1})$ where $c_{\bs{\beta}_2}$ is a scale factor. This choice or prior distribution for $(\beta,\Sigma)$ is particularly convenient for model selection because under Gaussian error terms and no selection effect the
posterior mean of $\bs{\beta}_1$ (respectively, $\Sigma$) has a closed form expression provided that $\bs{\beta}_1$ and $\bs{\beta}_2$ are \emph{a priori} independent. In the numerical study (Section \ref{subsect:numESN}), parameters of prior distributions are given by $\mu_{\bs{\beta}_1}=\mu_{\bs{\beta}_2}=0$ and $c_{\bs{\beta}_1}=c_{\bs{\beta}_2}=5n$, with $n$ the sample size.

\subsection{Determination of the marginal effects}

\begin{prop}\label{prop:me}
Consider the univariate extended skew-normal sample selection model defined by \eqref{introduction:eq:model} and \eqref{eq:law}. Let
$$
\tau(a,\alpha,\lambda)=\frac{\phi(a)\Phi(\lambda+\alpha a)}{\Phi_2(a,1,\alpha,\lambda)},\quad \delta(a,\alpha,\lambda)=\frac{\phi(\lambda/c_0)\Phi\left(ac_0+\frac{\alpha\lambda}{c_0}\right)}{\Phi_2(a,1,\alpha,\lambda)}.
$$
Then,
\begin{align*}
&\E\left[S_i^*|S_i=1,\bx_i\right]=\beta_2'\bx_i+\tau_{2i}+\frac{c_2\tilde{\alpha}_2}{c_{02}}\delta_{2i}\\
&\E[Y^*_{i}|S_i=1,\bx_i]=\xi_{1i}+\bx_i'\beta_1+\sigma_{12}\tau_{2i}
+\sigma_1 v_{2}\delta_{2i}
\end{align*}
where $
\tau_{2i}=\tau\left(\xi_2+\bx_i'\beta_2,-c_2\tilde{\alpha}_2,c_2\lambda\right)$, $ \delta_{2i}=\delta\left(\xi_2+\bx_i'\beta_2,
-c_2\tilde{\alpha}_2,c_2\lambda\right)$ and  $v_2=\frac{\rho c_2\tilde{\alpha}_2+c_2(1-\rho^2)\alpha_1}{c_{02}}
$.
\end{prop}
Proof: See \citet{Gerber2014}.

\newpage

\section{Tables}

\begin{table}[H]
\caption{Estimation of univariate $\mathcal{ESN}_1$ distributions \eqref{sim:eq:ESN1}}
\centering
\scalebox{0.75}{
\begin{tabular}{lccc|ccc}
\hline\hline
\multirow{2}{*}{}&\multicolumn{3}{c}{Estimation under P1}&\multicolumn{3}{c}{Estimation under P2}\\
&Mean&Median&Mode&Mean&Median&Mode\\
\hline
\multirow{4}{*}{$\xi=2$}
&-15\%&-13\%&-11.5\%&-73.5\%&-73\%&-72.5\%\\
&(0.010)&(0.009)&(0.012)&(0.008)&(0.009)&(0.012)\\
&-8\%&-7.5\%&-7\%&-31\%&-30.5\%&-30\%\\
&(0.002)&(0.003)&(0.005)&(0.003)&(0.004)&(0.006)\\
\hline
\multirow{4}{*}{$\sigma^2=6$}
&-2.3\%&-3.5\%&-4.3\%&25.3\%&24.7\%&24.2\%\\
&(0.017)&(0.016)&(0.019)&(0.011)&(0.013)&(0.016)\\
&3.7\%&3.4\%&3.2\%&15.5\%&15.3\%&15.2\%\\
&(0.004)&(0.005)&(0.009)&(0.005)&(0.006)&(0.010)\\
\hline
\multirow{4}{*}{$\alpha=5$}
&-19.8\%&-20.2\%&-20.4\%&-39.2\%&-39.8\%&-40.0\%\\
&(0.006)&(0.006)&(0.010)&(0.006)&(0.005)&(0.007)\\
&-5.8\%&-6.0\%&-6.2\%&-19.4\%&-19.6\%&-19.6\%\\
&(0.003)&(0.004)&(0.006)&(0.002)&(0.003)&(0.005)\\
\hline
\multirow{4}{*}{$\lambda=-2$}
&55\%&47\%&42\%&212\%&207.5\%&204\%\\
&(0.040)&(0.036)&(0.041)&(0.023)&(0.024)&(0.029)\\
&38\%&35.5\%&34\%&118\%&116.5\%&115\%\\
&(0.011)&(0.014)&(0.021)&(0.011)&(0.013)&(0.022)\\
\hline
\multirow{4}{*}{$\log m(z_{1:n})$}&\multicolumn{3}{c}{-1\,473.84

}\vline&\multicolumn{3}{c}{-1\,532.98

}\\
&\multicolumn{3}{c}{(38.37)}\vline&\multicolumn{3}{c}{(34.00)}\\
&\multicolumn{3}{c}{-8\,065.57

}\vline&\multicolumn{3}{c}{-8\,074.10

}\\
&\multicolumn{3}{c}{(13.01)}\vline&\multicolumn{3}{c}{(18.22)}\\
\hline
\multirow{2}{*}{Time (in seconds)}&\multicolumn{3}{c}{60.22}\vline&\multicolumn{3}{c}{34.05}\\
&\multicolumn{3}{c}{120.44
}\vline&\multicolumn{3}{c}{124.26}\\
\hline \hline
\end{tabular}
}
\caption*{\footnotesize{Notes: The results are obtained from 50 estimations of the model with $N=10\,000$ particles. Mean estimates  are reported as percentage deviation of the true parameter value, and  standard deviations are given in brackets. For each parameter, the first (respectively, last) two rows correspond to $n=$1\,000 (respectively, $n=$5\,000).}} \label{Table:runiv}
\end{table}

\begin{table}[H]
\caption{Estimation of univariate $\mathcal{ESN}_1$ distributions \eqref{sim:eq:ESN2}}
\centering
\scalebox{0.75}{
\begin{tabular}{lccc|ccc}
\hline \hline
&\multicolumn{3}{c}{Estimation under P1}&\multicolumn{3}{c}{Estimation under P2}\\
&Mean&Median&Mode&Mean&Median&Mode\\
\hline
\multirow{4}{*}{$\xi=2$}
&76\%&83\%&88.5\%&-10.5\%&-9\%&-7\%\\
&(0.031)&(0.030)&(0.035)&(0.017)&(0.019)&(0.029)\\
&50.5\%&53.5\%&55.5\%&26\%&27\%&28\%\\
&(0.010)&(0.012)&(0.022)&(0.010)&(0.010)&(0.014)\\
\hline
\multirow{4}{*}{$\sigma^2=26$}
&-12.7\%&-14.1\%&-15.2\%&6.8\%&6\%&5.4\%\\
&(0.092)&(0.082)&(0.094)&(0.045)&(0.051)&(0.071)\\
&-4.4\%&-5\%&-5.5\%&1.1\%&0.9\%&0.6\%\\
&(0.029)&(0.037)&(0.055)&(0.024)&(0.030)&(0.046)\\
\hline
\multirow{4}{*}{$\alpha=0.98$}
&2\%&2\%&2\%&2\%&-2\%&0\%\\
&(0.001)&(0.001)&(0.001)&(0.001)&(0.001)&(0.002)\\
&3.1\%&4.1\%&4.1\%&4.1\%&4.1\%&4.1\%\\
&(0.000)&(0.000)&(0.001)&(0.001)&(0.000)&(0.001)\\
\hline
\multirow{4}{*}{$\lambda=-4.08$}
&-40.2\%&-43.1\%&45.8\%&6.1\%&4.4\%&3.2\%\\
&(0.032)&(0.031)&(0.034)&(0.019)&(0.019)&(0.033)\\
&-25.2\%&-26.5\%&-27.5\%&-12.5\%&-13\%&-13.2\%\\
&(0.010)&(0.013)&(0.021)&(0.009)&(0.010)&(0.018)\\
\hline
\multirow{4}{*}{$\log m(z_{1:n})$}&\multicolumn{3}{c}{-2\,244.49}\vline&\multicolumn{3}{c}{-2\,184.73
}\\
&\multicolumn{3}{c}{(41.69)}\vline&\multicolumn{3}{c}{(30.01)}\\
&\multicolumn{3}{c}{-11\,500.13
}\vline&\multicolumn{3}{c}{-11\,476.46

}\\
&\multicolumn{3}{c}{(25.22)}\vline&\multicolumn{3}{c}{(41.65)}\\
\hline
\multirow{2}{*}{Time (in seconds)}&\multicolumn{3}{c}{38.52}\vline&\multicolumn{3}{c}{61.40}\\
&\multicolumn{3}{c}{125.98}\vline&\multicolumn{3}{c}{165.89}\\
\hline \hline
\end{tabular}
}
\caption*{\footnotesize{Notes: The results are obtained from 50 estimations of the model with $N=10\,000$ particles. Mean estimates  are reported as percentage deviation of the true parameter value, and standard deviations are given in brackets. For each parameter, the first (respectively, last) two rows correspond to $n=$1\,000 (respectively, $n=$5\,000).}} \label{Table:runiv2}
\end{table}

\begin{table}[H]
\caption{Bayes factors}
\centering
\scalebox{0.9}{%
\begin{tabular}{llcccc}
\hline \hline
&$(\alpha,\lambda)$& $\log_{10}B_{10}\leq 0.5$&$0.5<\log_{10}B_{10}\leq 1$&$1<\log_{10}B_{10}\leq 2$&$\log_{10}B_{10}>2$\\
\hline
n=100&(0,-)&100\%&0\%&0\%&0\%\\
&(5,-2)&1\%&1\%&4\%&96\%\\
&(0.5,1)&100\%&0\%&0\%&0\%\\
\hline
n=5\,000&(0.5,1)&0\%&0\%&0\%&100\%\\
\hline \hline
\end{tabular}
}
\caption*{\footnotesize{Notes: The results  are obtained from 100 samples. The number of particles is 10\,000 and  $B_{10}$ denotes the Bayes factor to test the normality hypothesis.}}\label{Table:Bayes}
\end{table}

\begin{table}[H]
\caption{Estimation of sample selection model \eqref{sim:eq:model}-\eqref{sim:eq:resESN}}
\centering
\scalebox{0.95}{
\begin{tabular}{l c c cc c cc}
\hline \hline
\multirow{2}{*}{Parameter} &\multirow{2}{*}{$\rho$}&\multicolumn{2}{c}{Tobit 2} &\multicolumn{2}{c}{ESNM}&\multirow{2}{*}{True value}\\
&&Mean& Standard deviation &Mean& Standard deviation\\
\hline
\multirow{3}{*}{$\beta_{10}$}&0.3&2.92&0.0008&2.94&0.0006&\multirow{3}{*}{3}\\
&0.9&2.98&0.0006&2.99&0.0005\\
&-0.9&2.97&0.0005&2.98&0.0006\\
\hline
\multirow{3}{*}{$\beta_{11}$}&0.3&-1.98&0.0004&-1.96&0.0004&\multirow{3}{*}{-2}\\
&0.9&-1.99&0.0004&-1.99&0.0003\\
&-0.9&-1.99&0.0003&-1.990&0.0352\\
\hline
\multirow{3}{*}{$\beta_{20}$}&0.3&1.58&0.0010&1.37&0.0015&\multirow{3}{*}{1.5}\\
&0.9&2.57&0.0020&1.78&0.0021\\
&-0.9&2.10&0.0015&1.43&0.0020\\
\hline
\multirow{3}{*}{$\beta_{22}$}&0.3&2.04&0.0013&1.77&0.0020&\multirow{3}{*}{2}\\
&0.9&3.32&0.0026&2.30&0.0026\\
&-0.9&2.77&0.0020&1.87&0.0028\\
\hline
\multirow{3}{*}{$\sigma_1^2$}&0.3&2.22&0.0012&6.04&0.0101&\multirow{3}{*}{(6)}\\
&0.9&2.10&0.0011&5.74&0.0074\\
&-0.9&1.60&0.0008&6.08&0.0113\\
\hline
\multirow{3}{*}{$\rho$}&0.3&0.06&0.0010&0.39&0.0015&\\
&0.9&0.63&0.0010&0.82&0.0006\\
&-0.9&-0.76&0.0008&-0.90&0.0004\\
\hline
\multirow{3}{*}{$\alpha_1$}&0.3&-&-&3.04& 0.0154&\multirow{3}{*}{(2)}\\
&0.9&-&-&3.27&0.0165\\
&-0.9&-&-&1.86&0.0066\\
\hline
\multirow{3}{*}{$\alpha_2$}&0.3&-&-&2.17&0.0145&\multirow{3}{*}{(1)}\\
&0.9&-&-&2.15&0.0251\\
&-0.9&-&-&0.55&0.0134\\
\hline
\multirow{3}{*}{$\lambda$}&0.3&-&-&-2.54&0.0234&\multirow{3}{*}{(-2)}\\
&0.9&-&-&-1.85&0.0207\\
&-0.9&-&-&-3.38&0.0164\\
\hline
\multirow{3}{*}{$\log m(z_{1:n})$}&0.3&-1\,448.91&0.00251& -1\,313.91&0.0269\\
&0.9&-1\,358.97&0.00401&-1\,206.92&0.0504\\
&-0.9&-1\,291.19&0.0059&-1\,168.48& 0.0262\\
\hline \hline
\end{tabular}
}
\caption*{Note: Using $N=10\,000$ particles, results are obtained from independent 50 independent estimations.}\label{sim:table1}
\end{table}

\section{Figures}

\setcounter{figure}{0}
\begin{figure}[H]
\begin{tabular}{ccc}
\multicolumn{2}{c}{$\mathcal{ESN}_1$ distribution \eqref{sim:eq:ESN1}}&$\mathcal{ESN}_1$ distribution \eqref{sim:eq:ESN2}\\
n=1\,000&n=10\,000&n=1\,000\\
\includegraphics[scale=0.25]{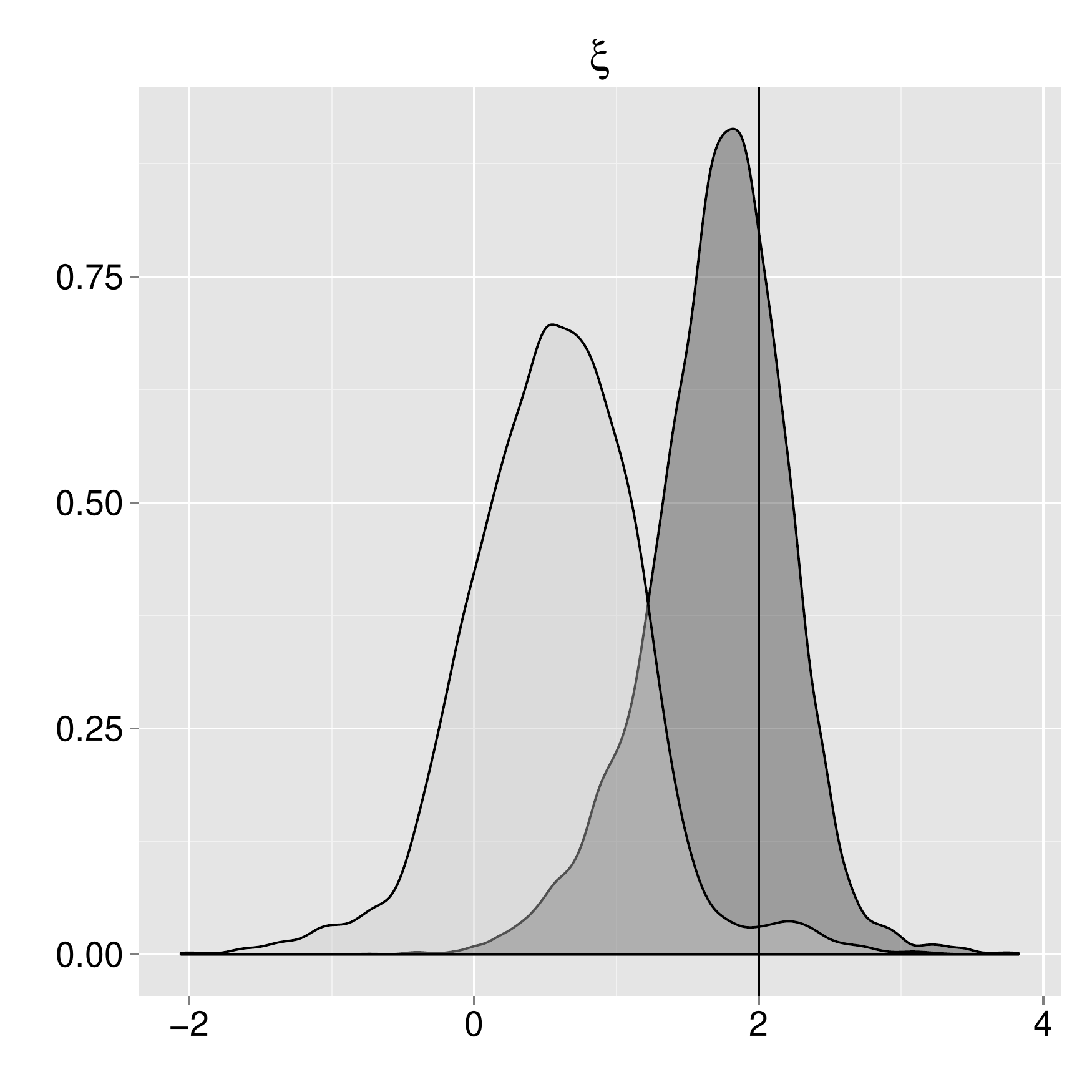}&\includegraphics[scale=0.25]{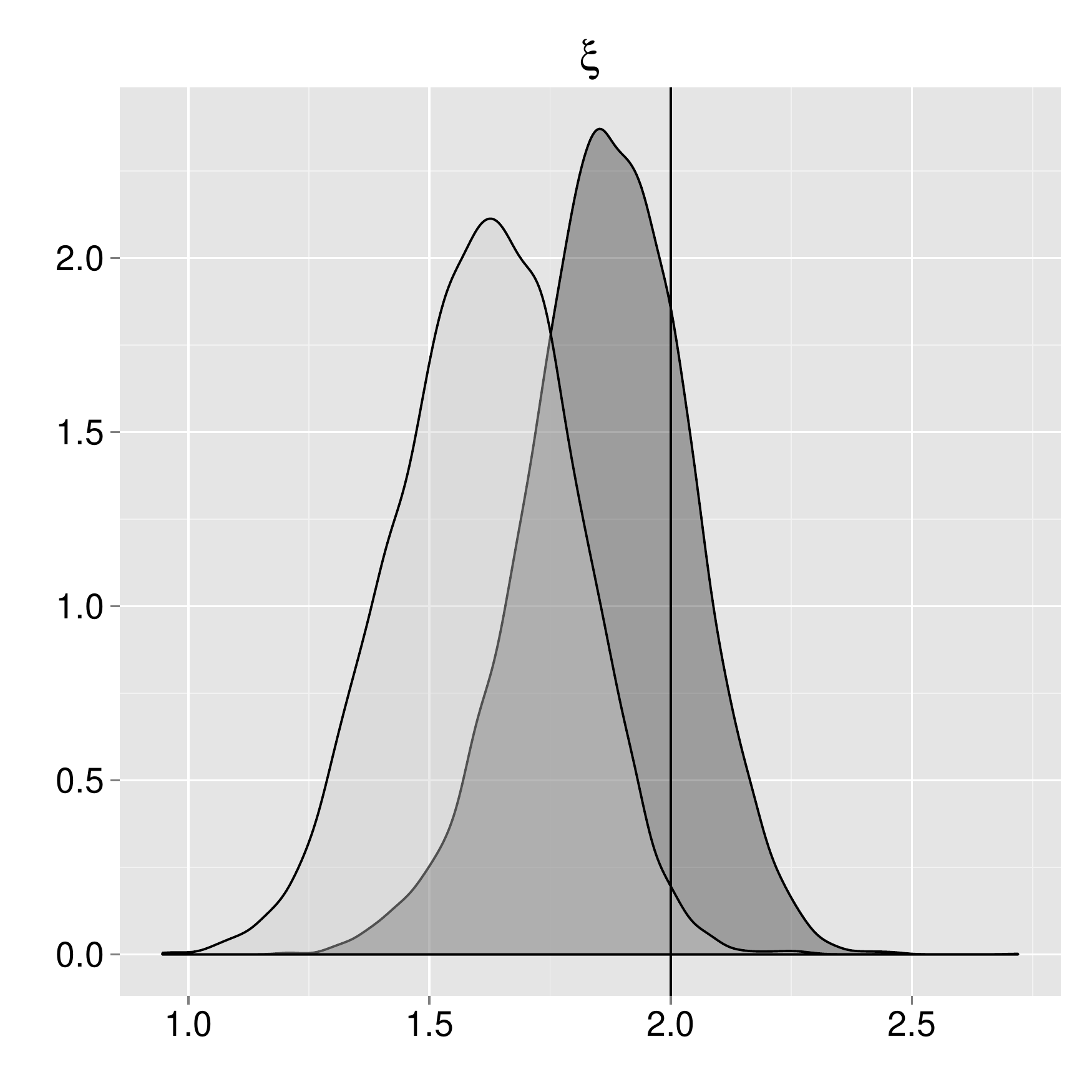}&\includegraphics[scale=0.25]{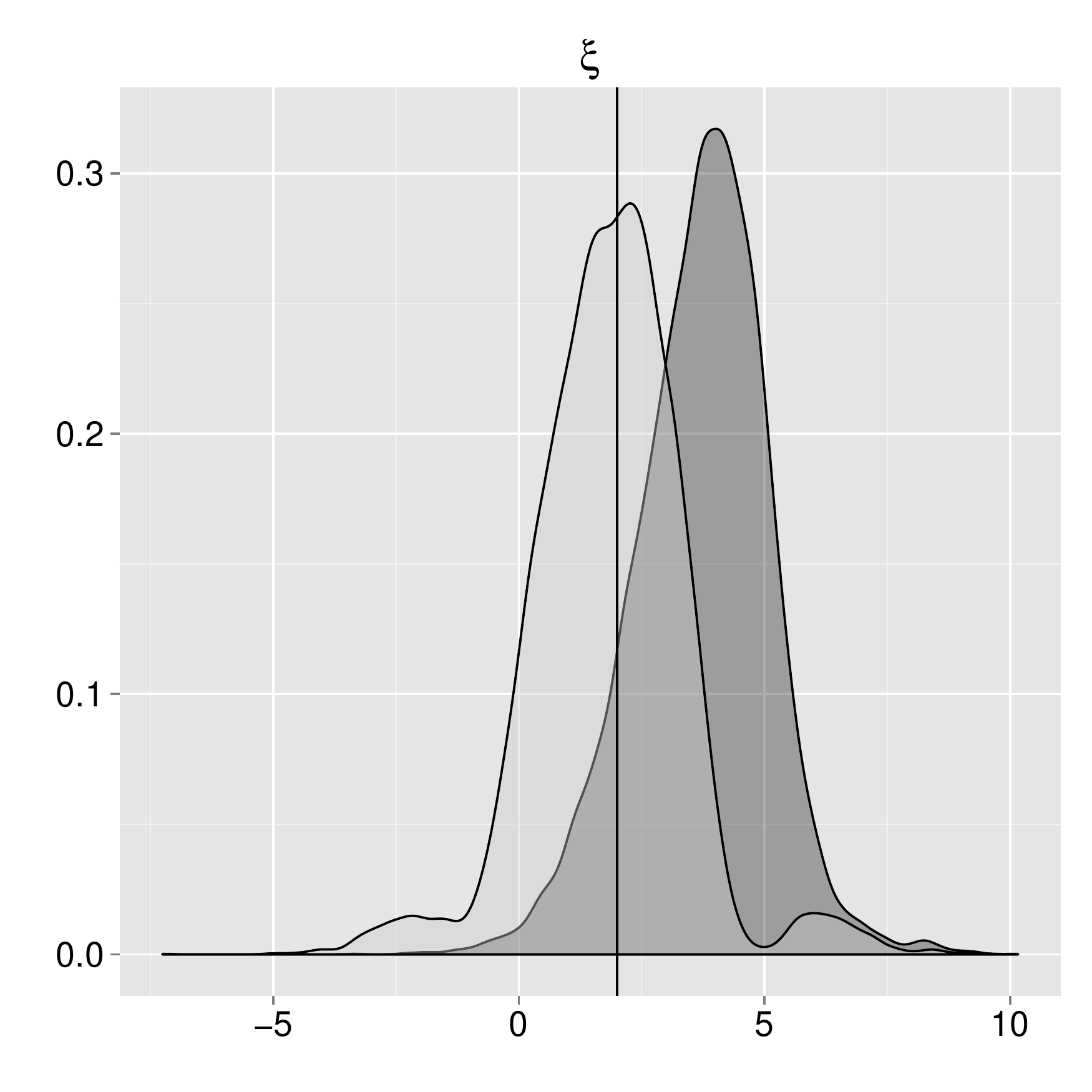}\\
\includegraphics[scale=0.25]{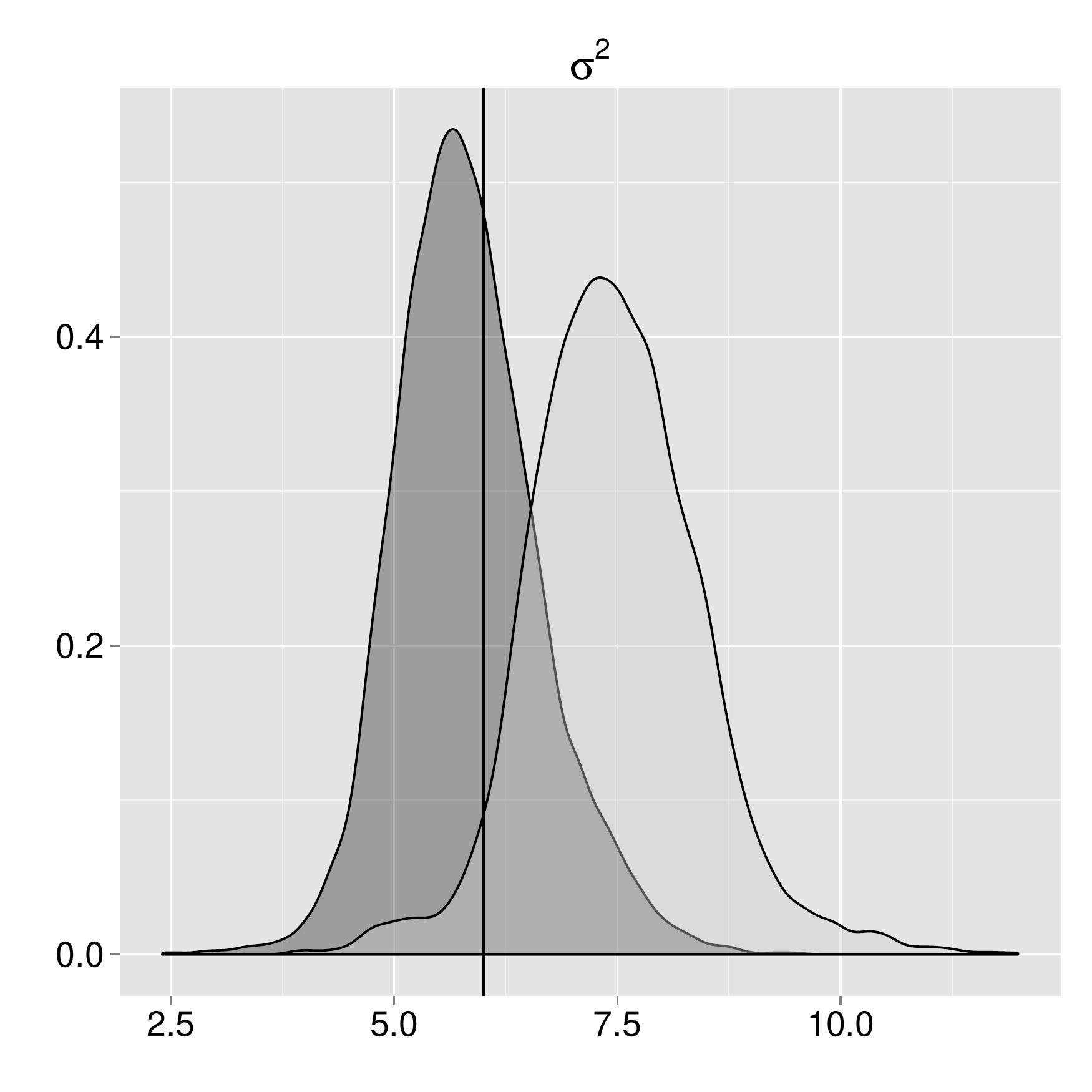}&\includegraphics[scale=0.25]{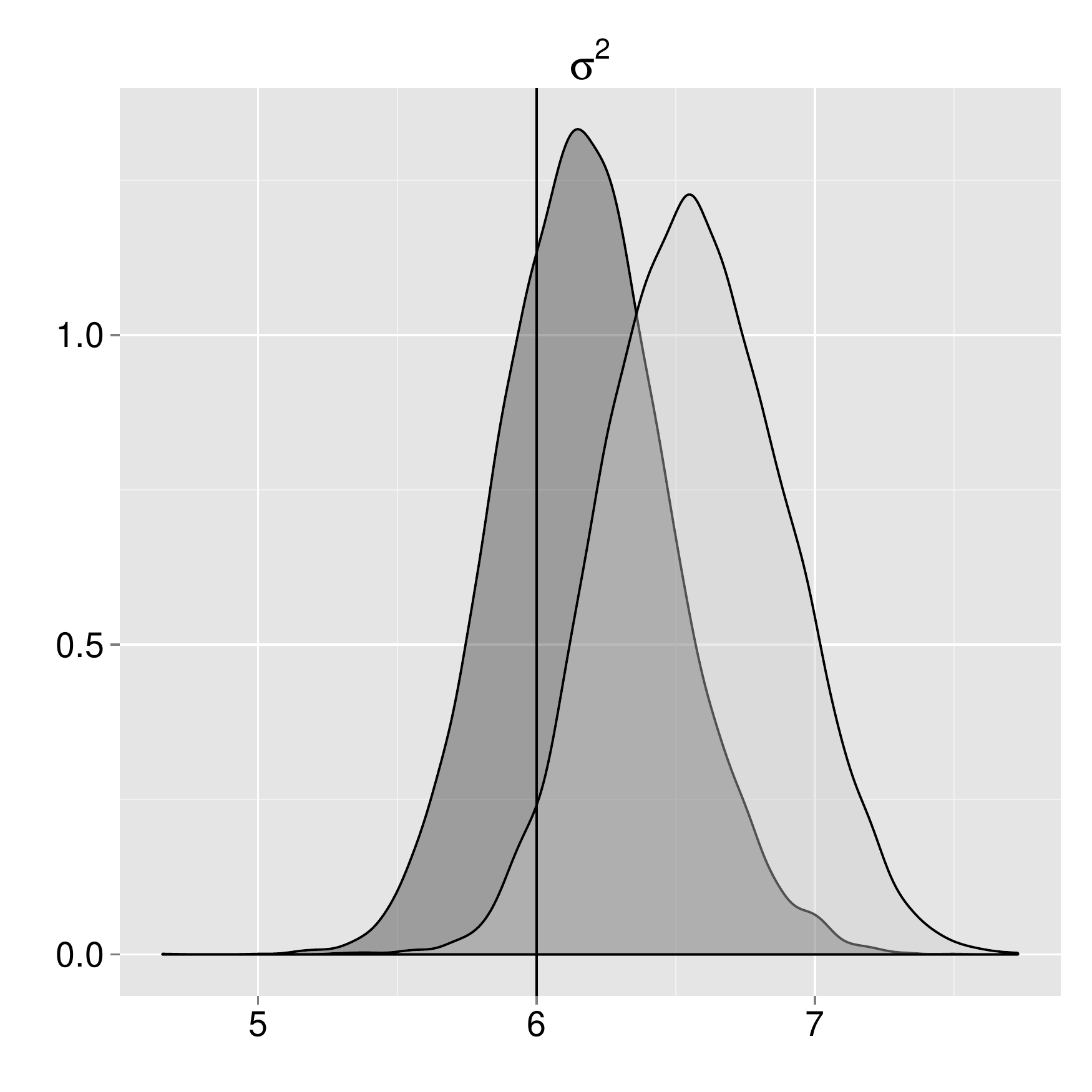}&\includegraphics[scale=0.25]{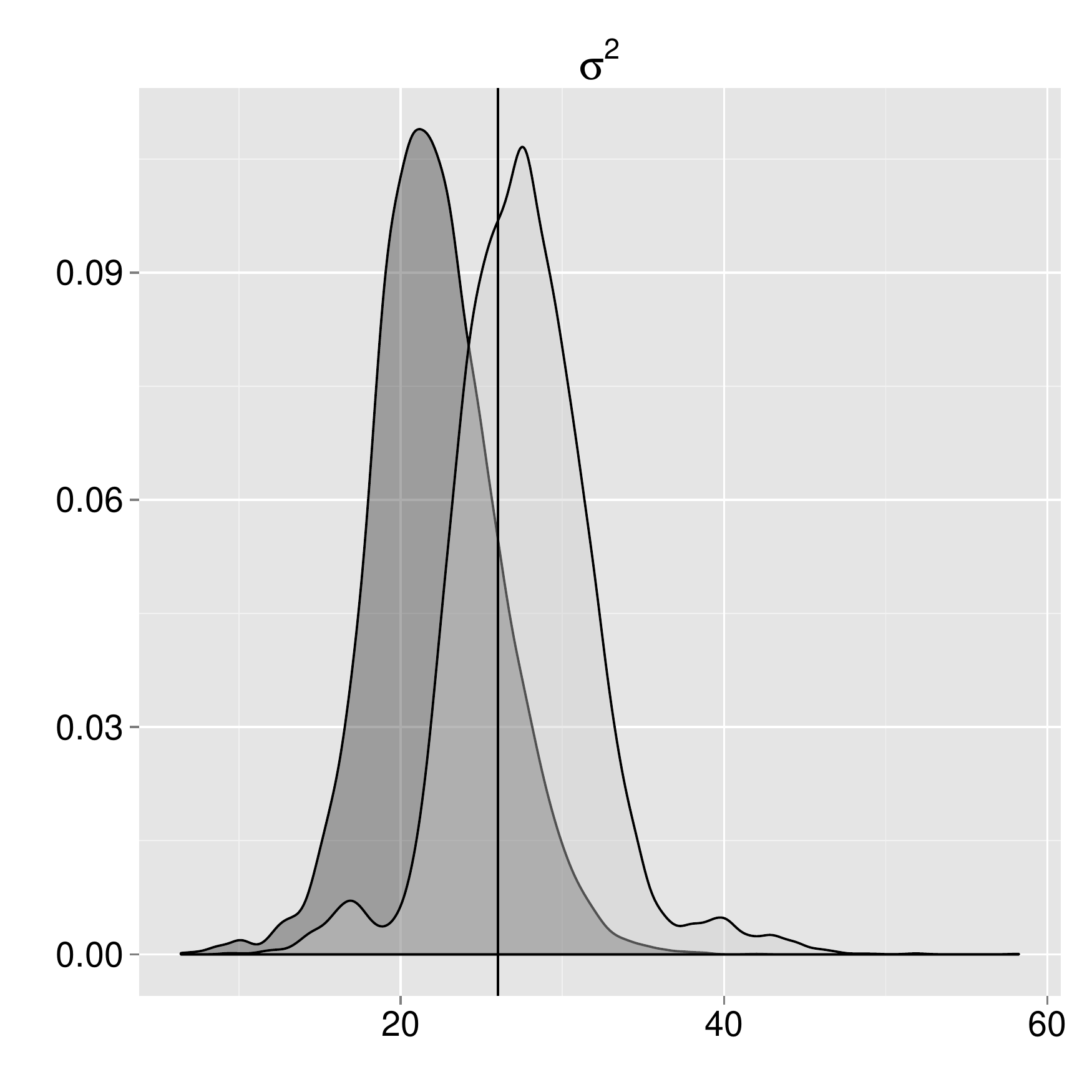}\\
\includegraphics[scale=0.25]{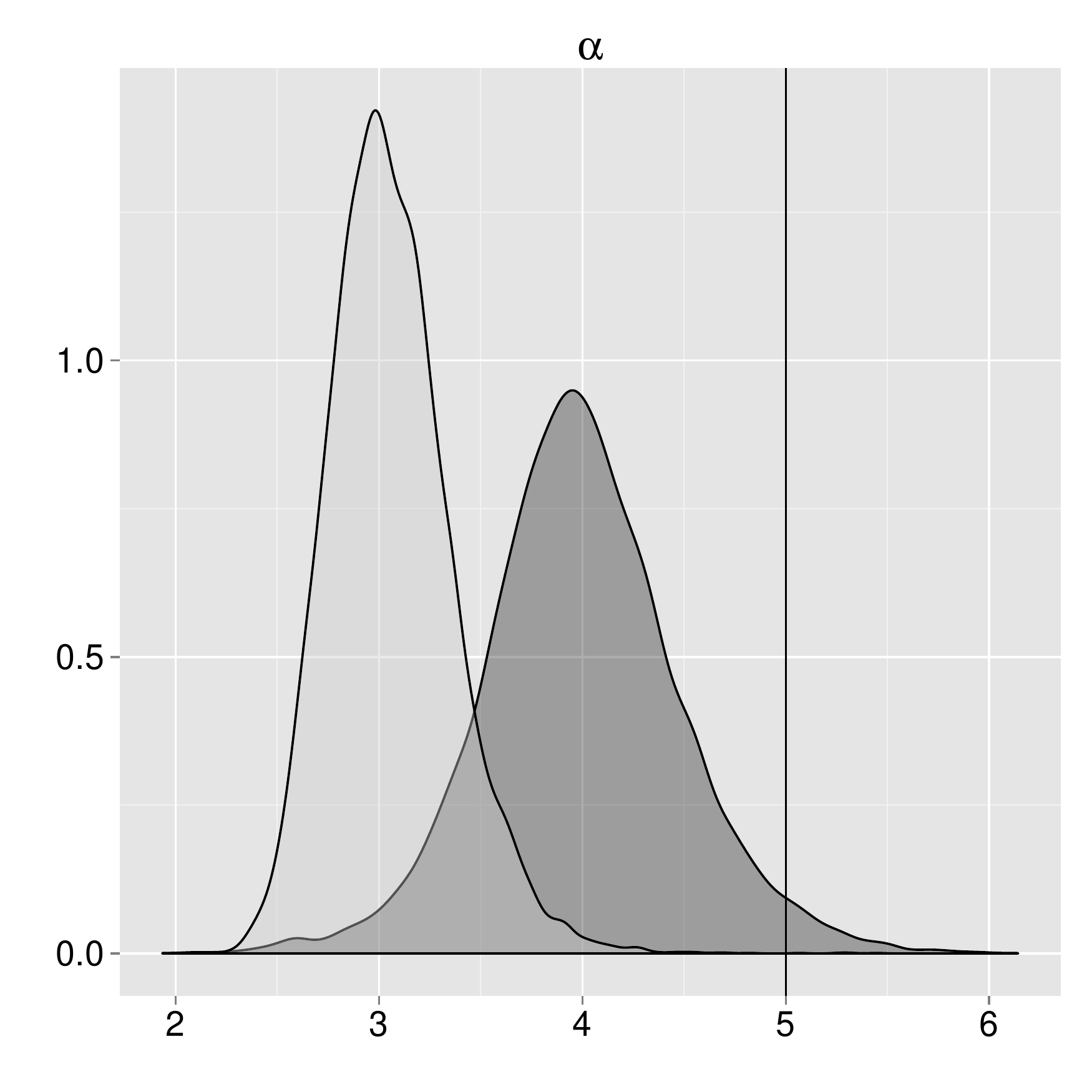}&\includegraphics[scale=0.25]{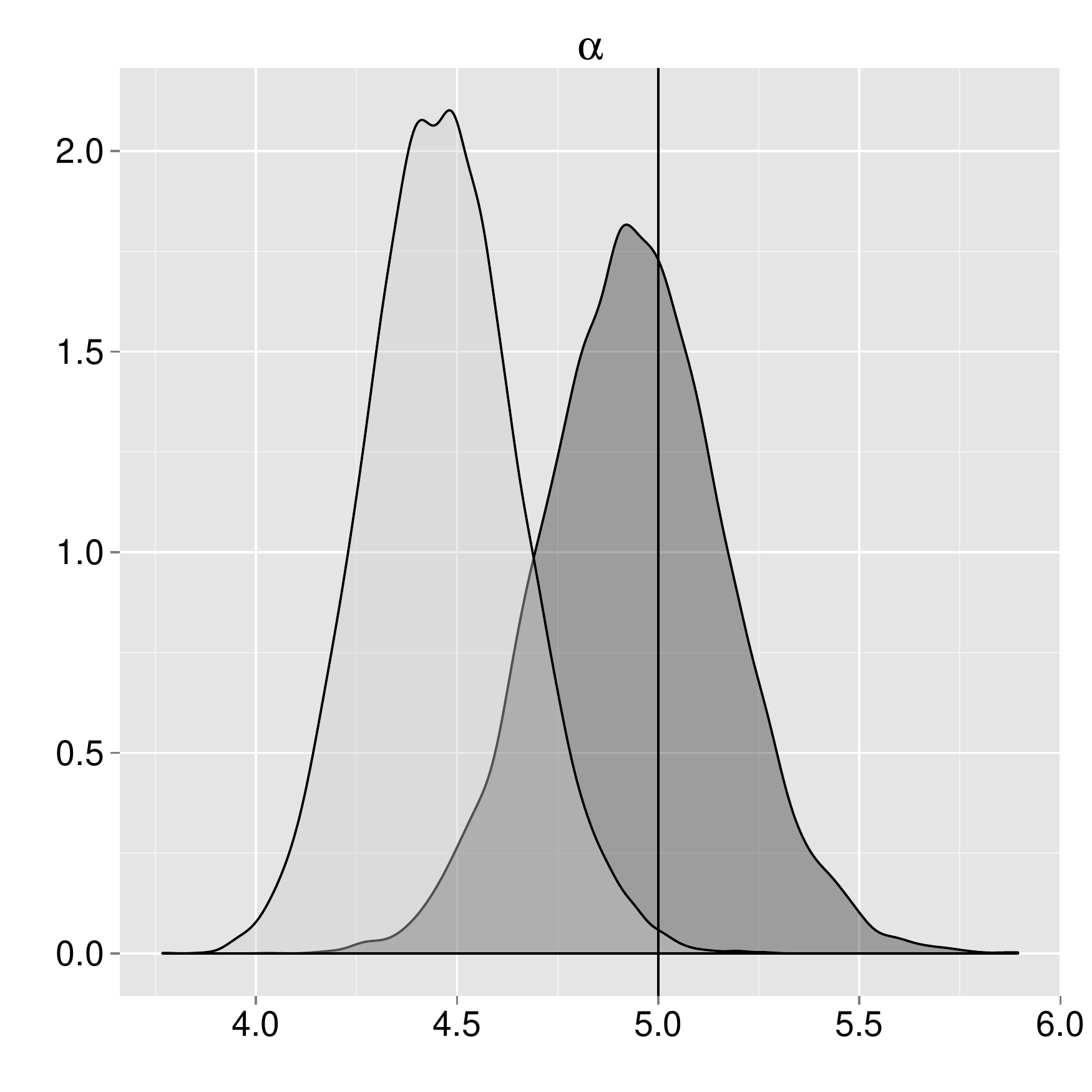}&\includegraphics[scale=0.25]{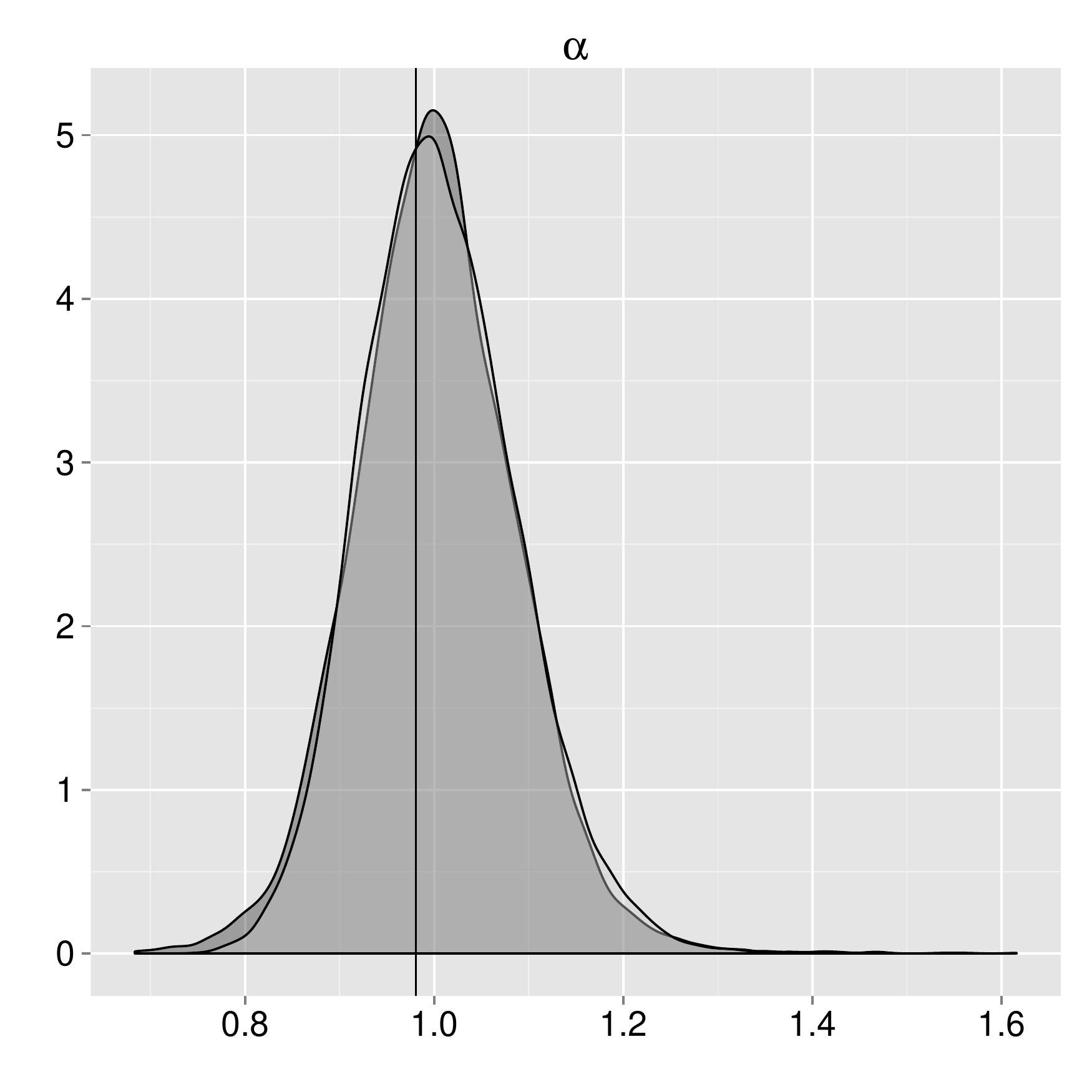}\\
\includegraphics[scale=0.25]{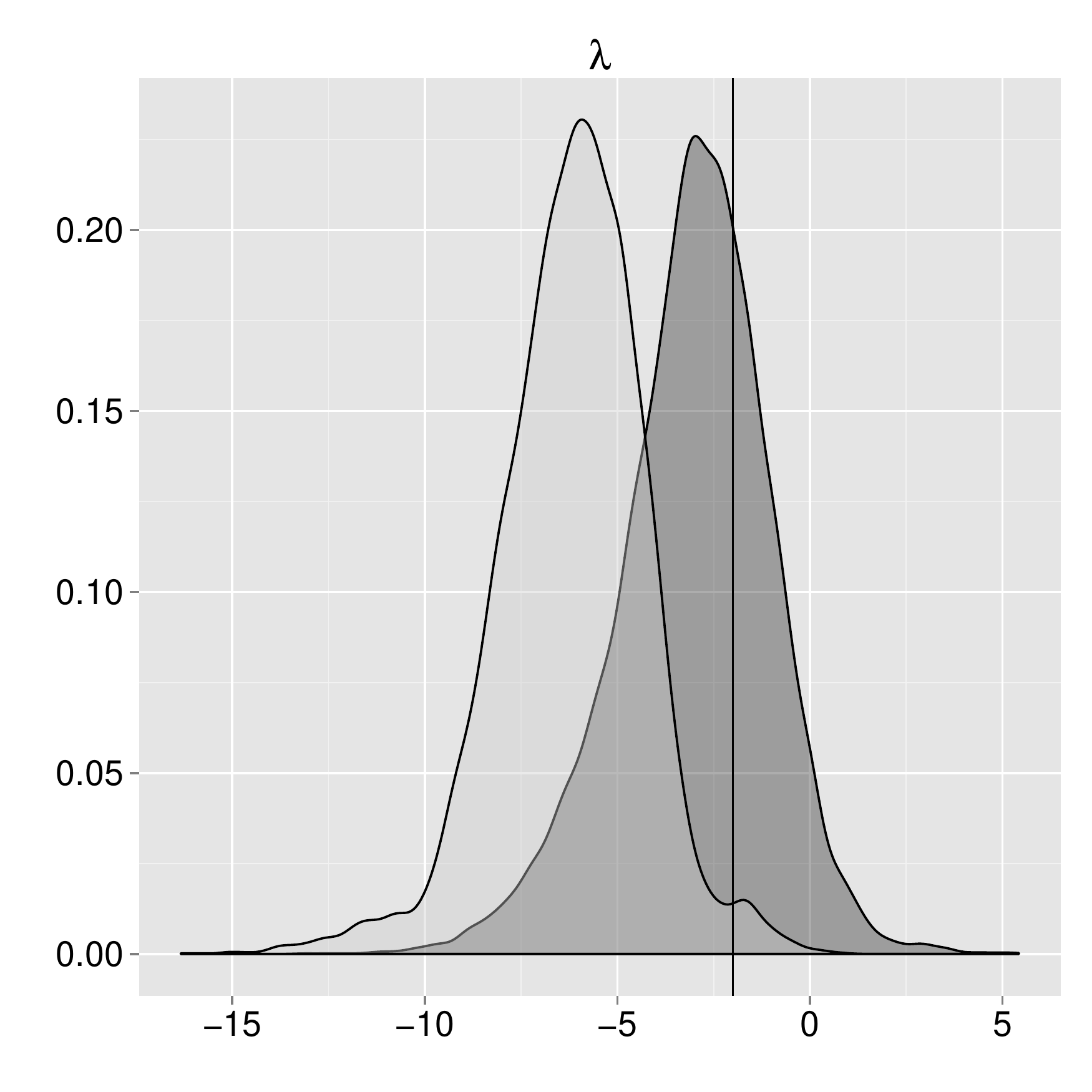}&\includegraphics[scale=0.25]{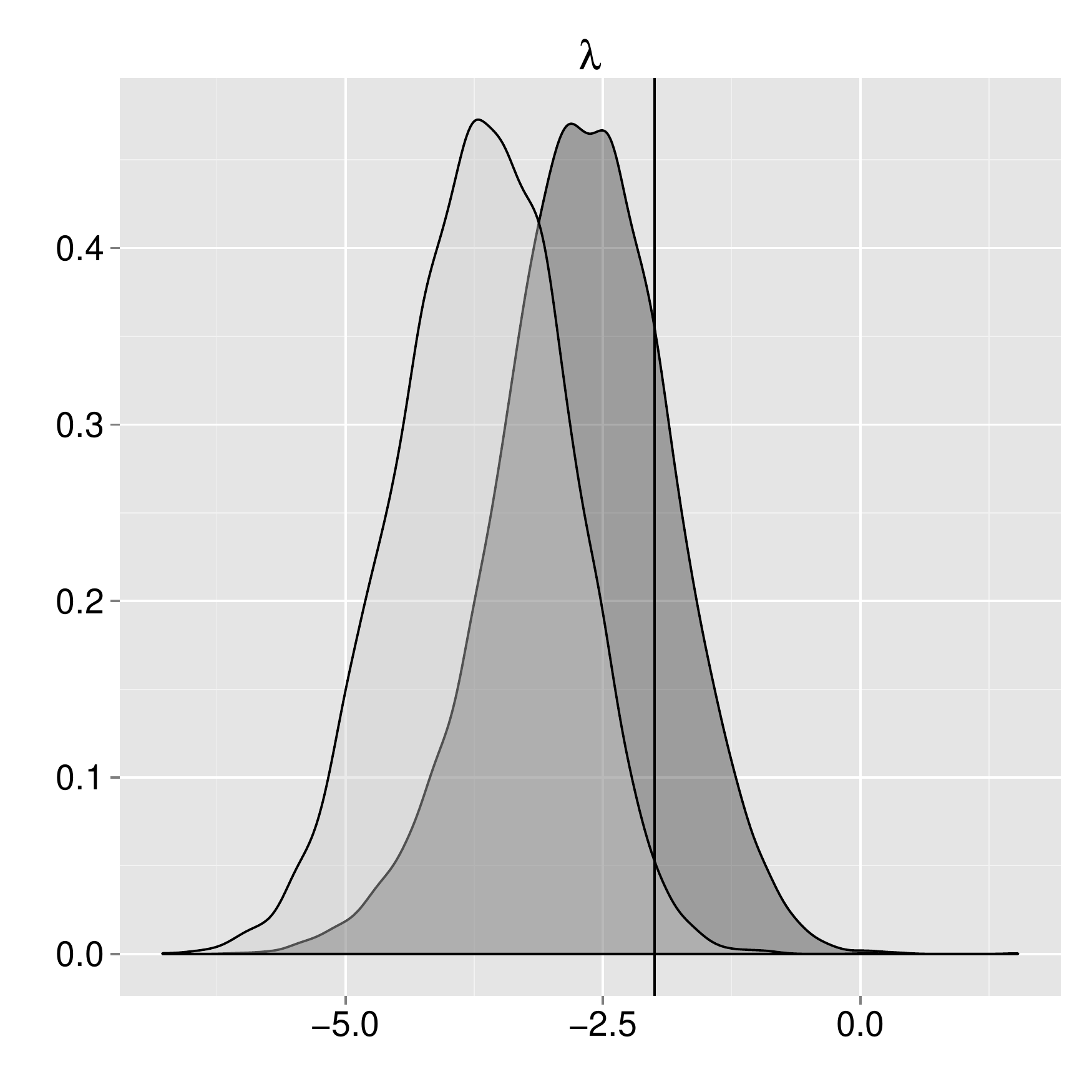}&\includegraphics[scale=0.25]{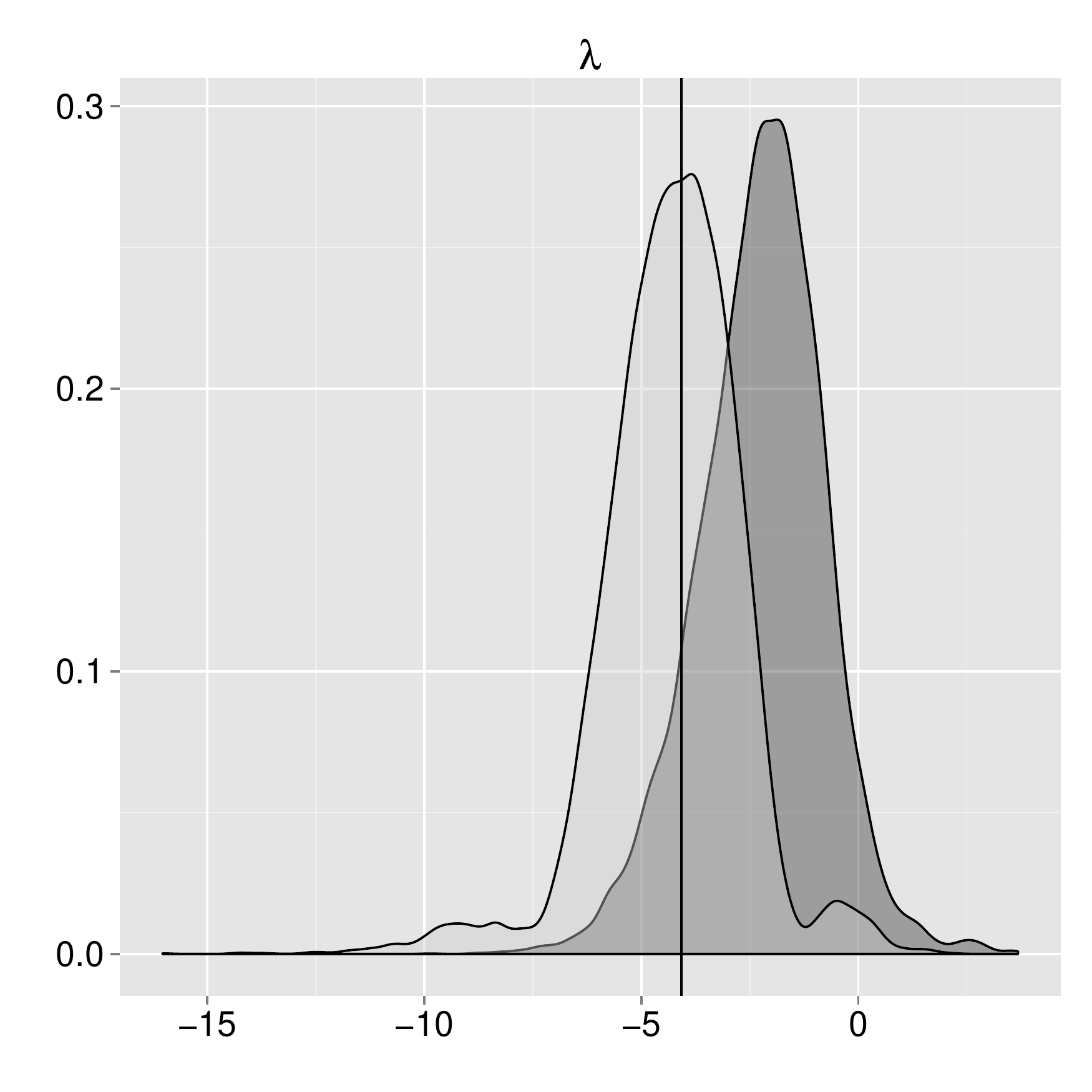}\\
\end{tabular}
\caption{Marginal posterior distributions for the parameters of the $\mathcal{ESN}_1$ distributions \eqref{sim:eq:ESN1}  and  \eqref{sim:eq:ESN2}. The results for P1 (respectively, P2) are in dark (respectively, in grey) and are obtained with $N=10\,000$ particles.\label{Fig:UnivPost}}
\end{figure}

\begin{figure}[H]
\centering
\includegraphics[scale=0.26]{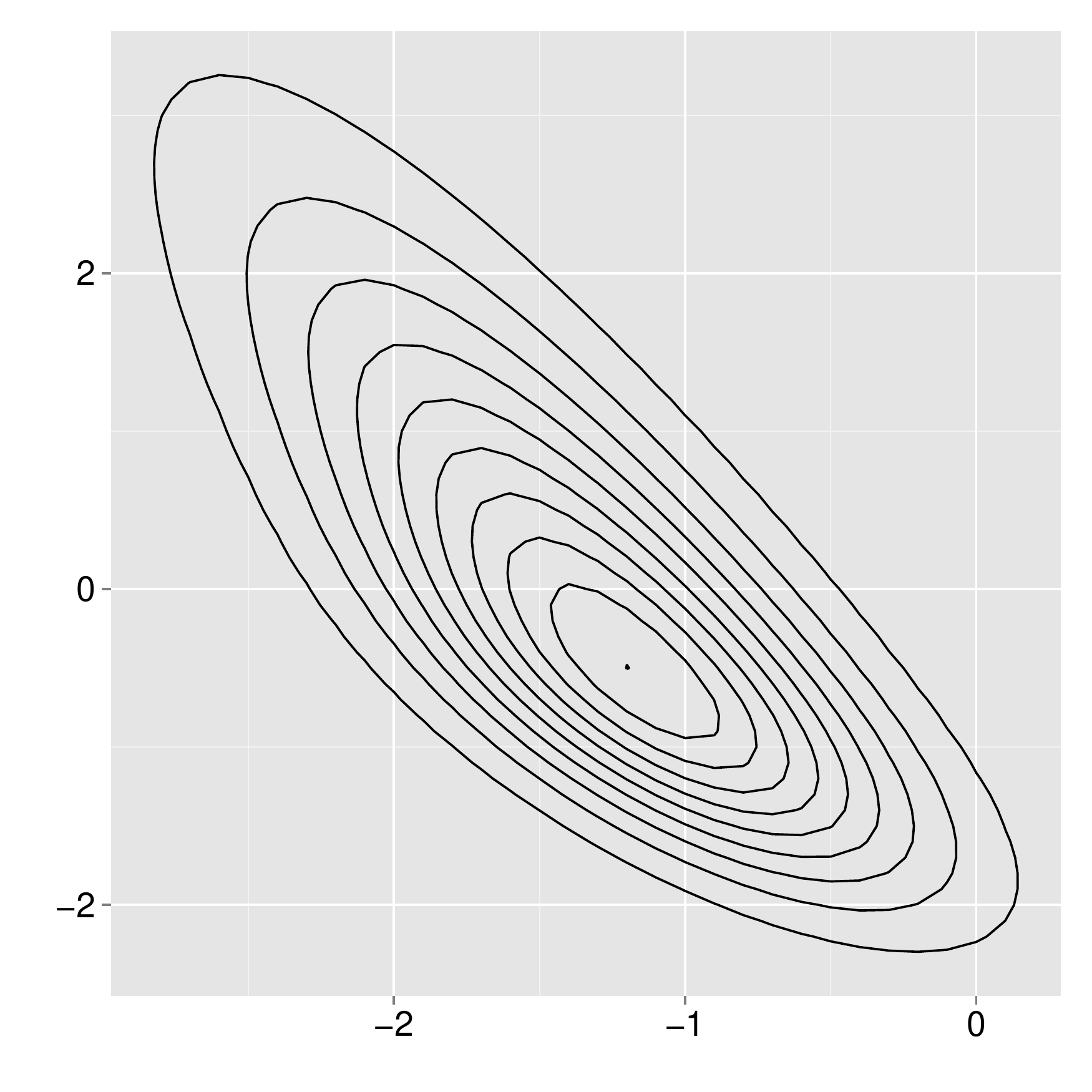}\includegraphics[scale=0.26]{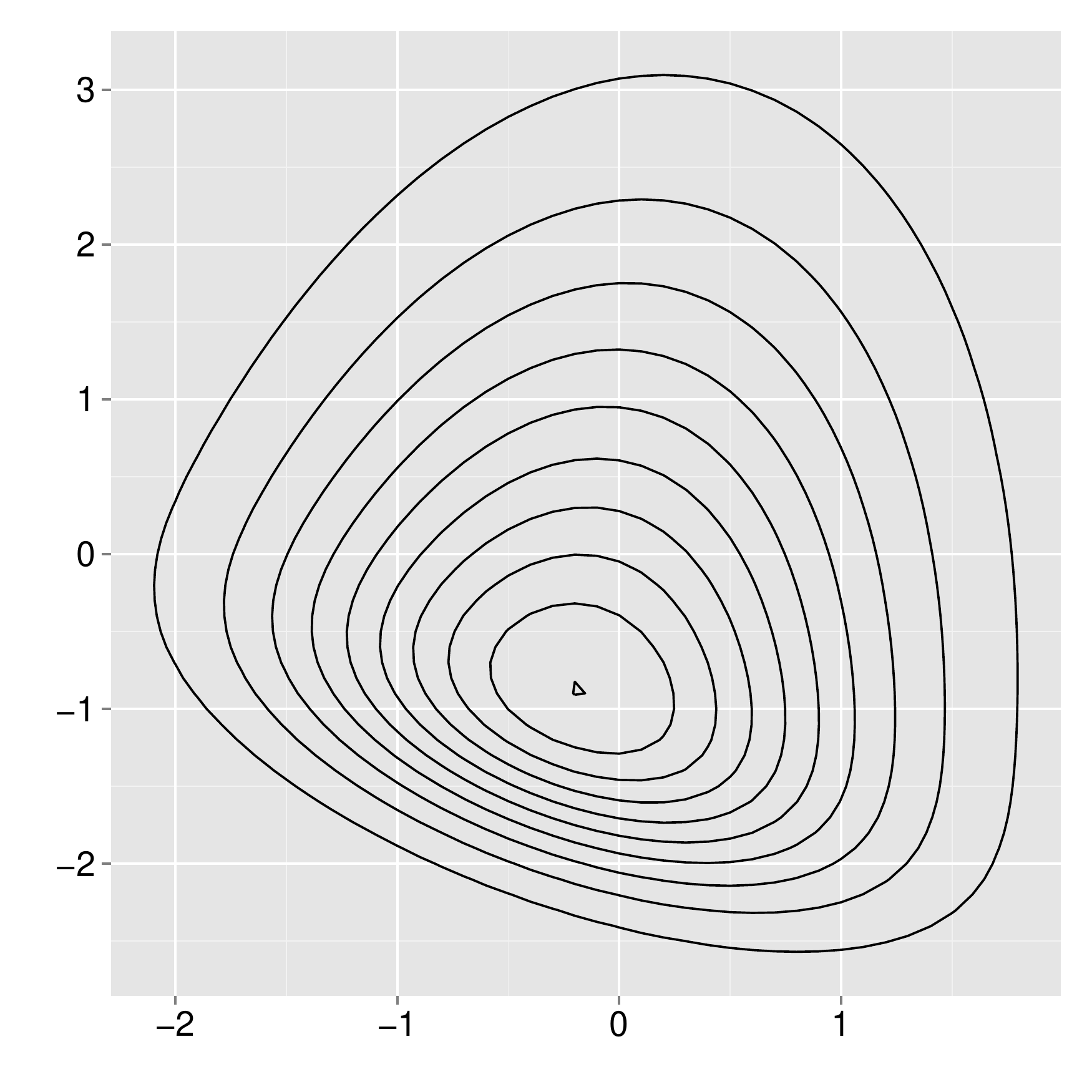}\includegraphics[scale=0.26]{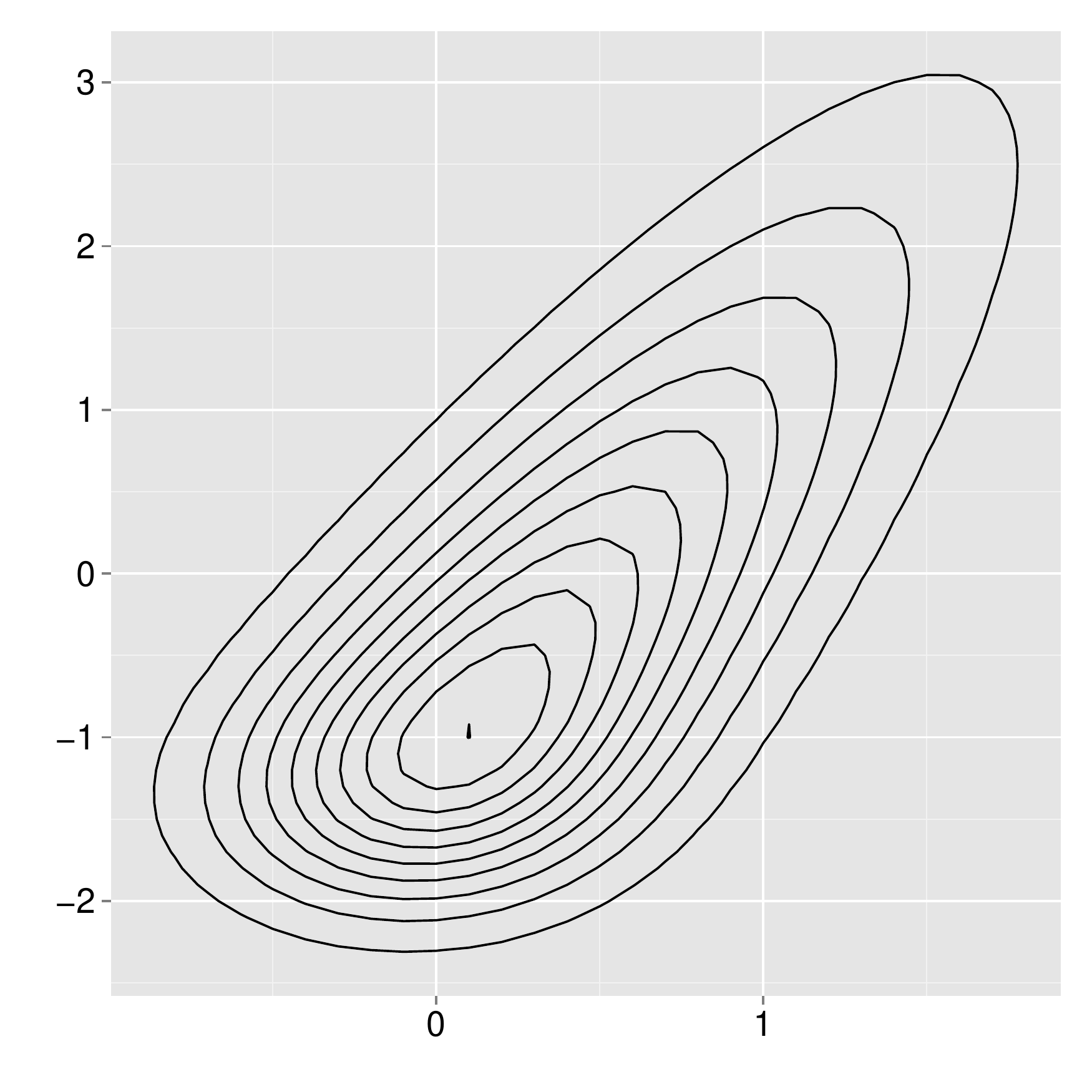}
\caption{Contour plots of the zero mean  $\mathcal{ESN}_2$ distribution \eqref{sim:eq:resESN}. The left, middle and right panels represent the contour plots of the zero mean $\mathcal{ESN}_2$ distribution when the correlation parameter $rho$ is respectively given by -0.9, 0.3, and 0.9.}\label{sim:fig:shape}
\end{figure}

\begin{figure}[H]
\begin{subfigure}{0.46\textwidth}
\centering
\includegraphics[scale=0.29, trim=0 0 0 0.7cm]{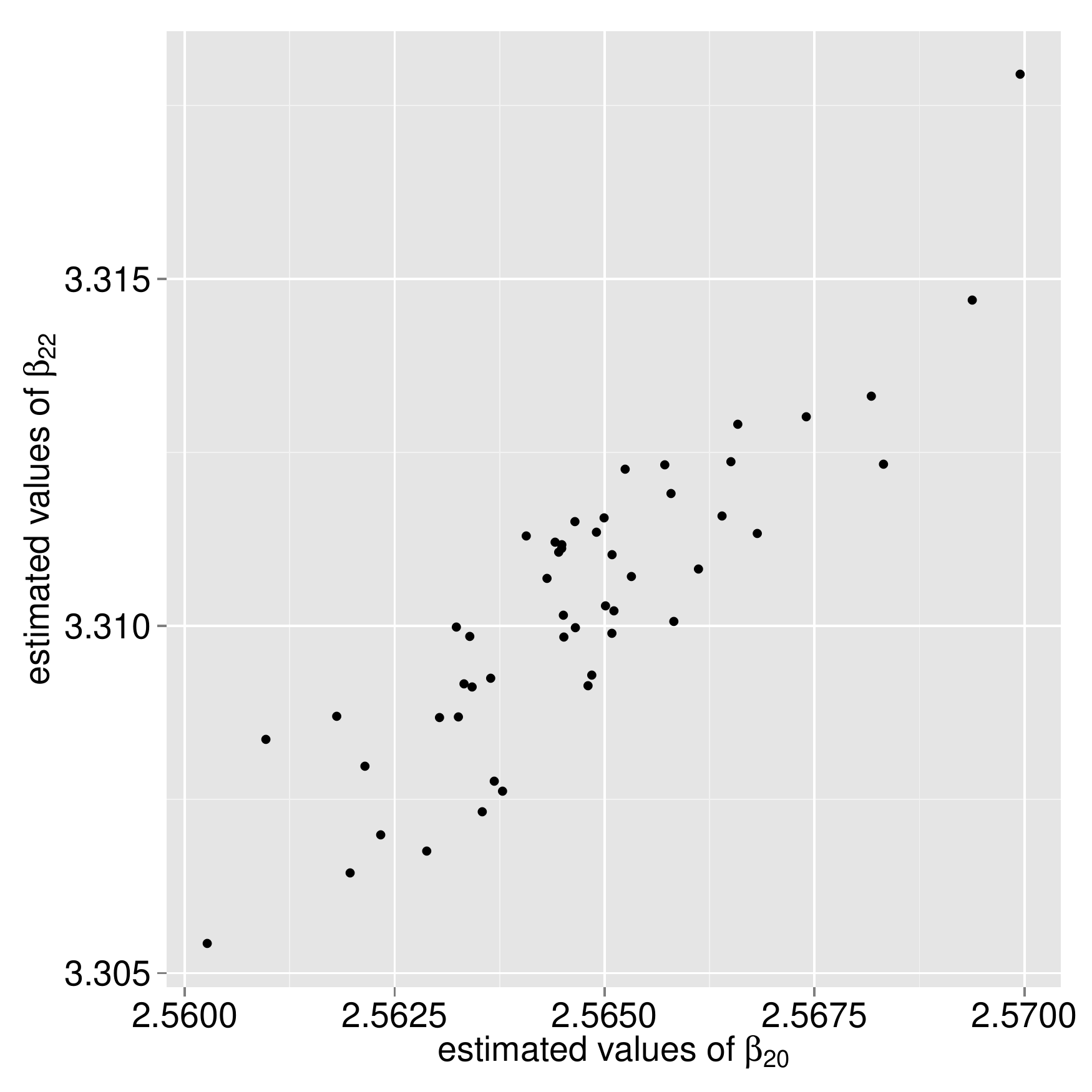}
\caption{\footnotesize{The results are obtained using $N=10\,000$ particles, 50 independent estimates of the selection parameters $(\beta_{20},\beta_{22})$ are reported when the true parameter vector is $(\beta_{20},\beta_{22})=(1.5,2)$ and $\rho=0.9$.}\label{sim:fig:Biais}}
\end{subfigure}
\hspace{0.2cm}
\begin{subfigure}{0.46\textwidth}
\centering
\includegraphics[scale=0.29]{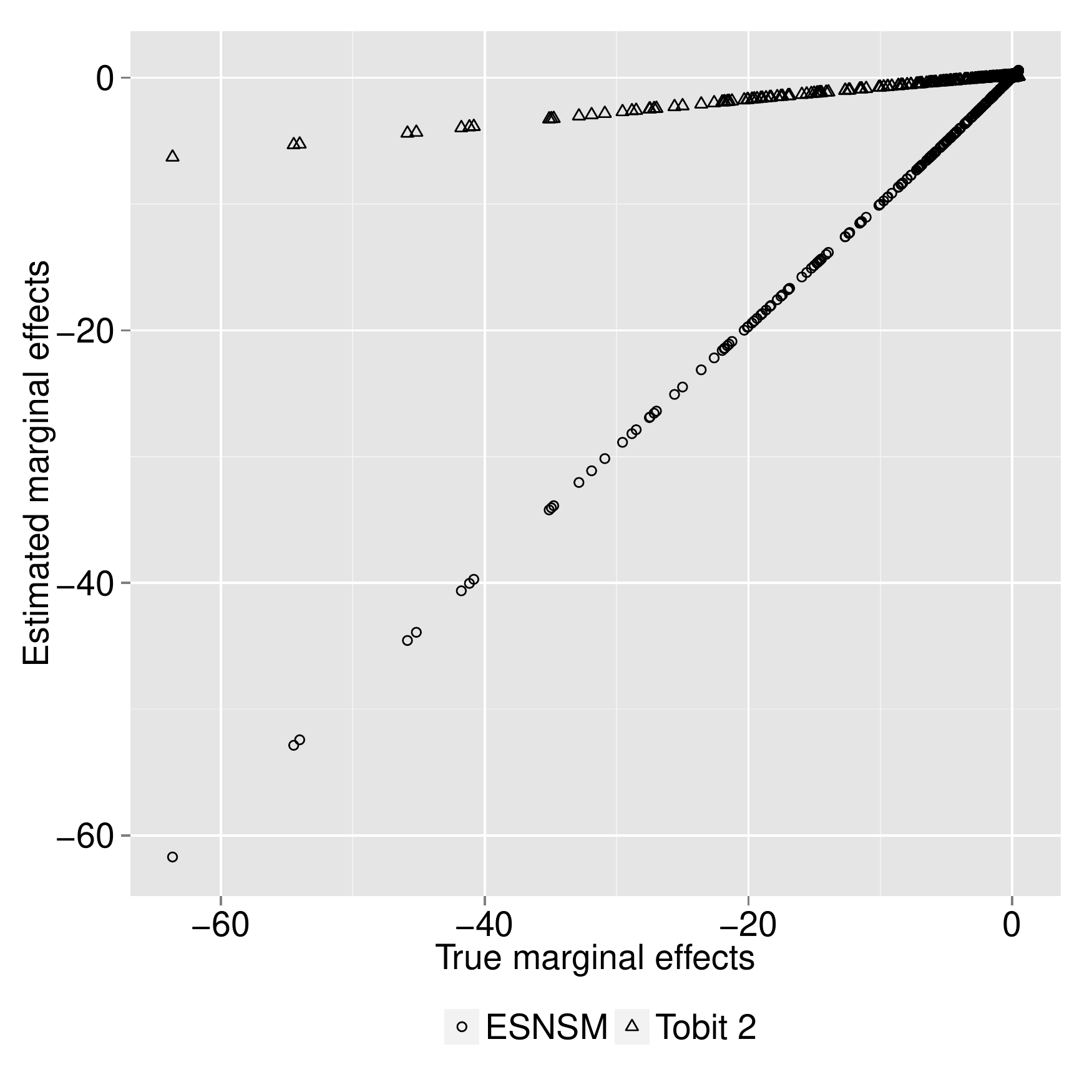}
\caption{\footnotesize{The results are obtained  with  $N=50\,000$ particle and $\rho=0.3$. For each individual, we report the marginal effect estimate using either the Tobit type-2 model (triangular markers) or the ESNSM model (circular markers) against the true marginal effect.}\label{sim:ME1}}
\end{subfigure}
\caption{Bias for selection coefficients of a Tobit type-2 model (Figure \ref{sim:fig:Biais}) and marginal effects for the ESNSM  \eqref{sim:eq:model}-\eqref{sim:eq:resESN} (Figure \ref{sim:fig:posterior}).\label{Fig:6}}
\end{figure}

\begin{figure}[H]
\centering
\includegraphics[scale=0.8]{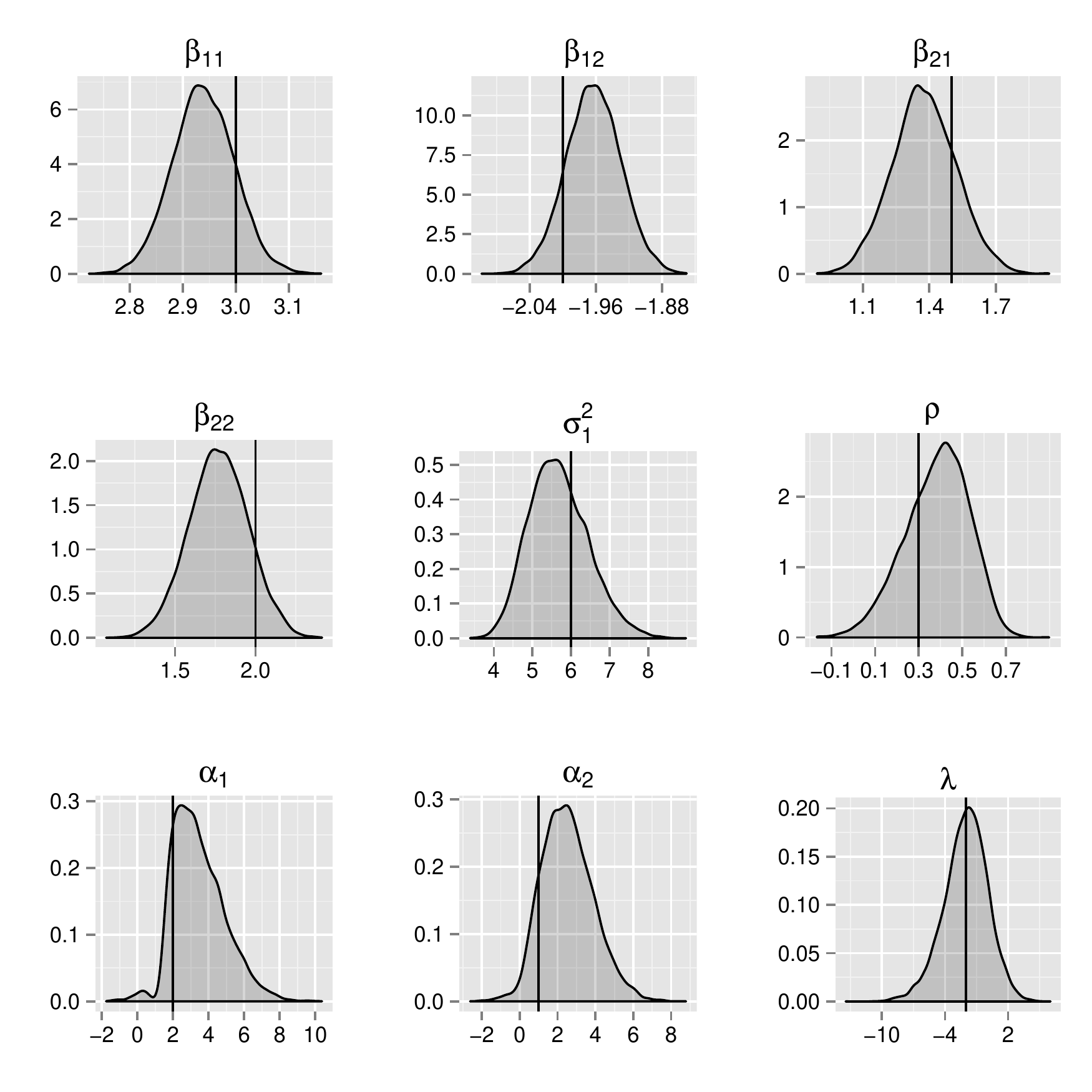}
\caption{Marginal Posterior distribution of the ESNSM \eqref{sim:eq:model}-\eqref{sim:eq:resESN}, evaluated with 50\,000 particles when $\rho=0.3$.}\label{sim:fig:posterior}
\end{figure}

\begin{figure}[H]
\centering
\includegraphics[scale=0.22]{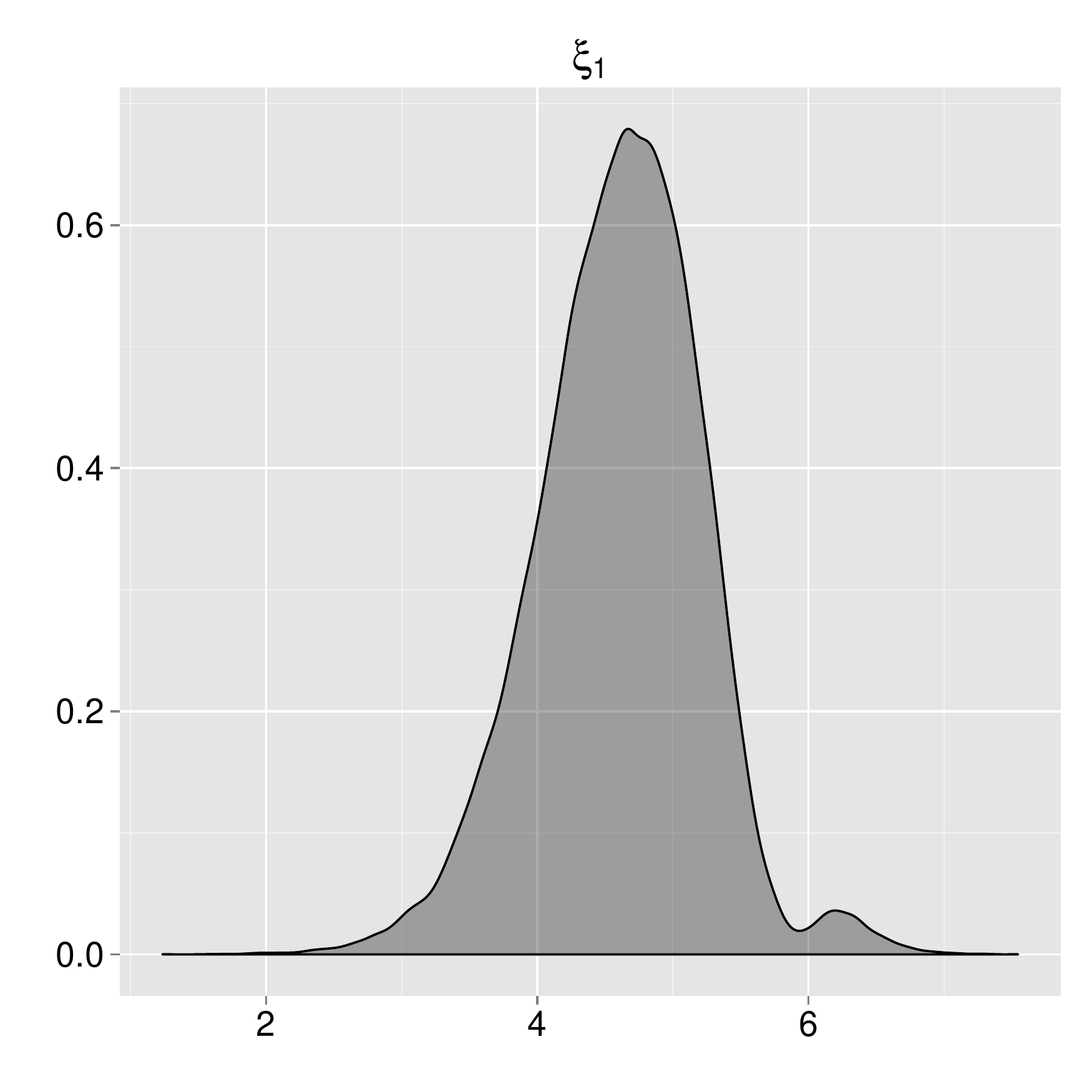}\includegraphics[scale=0.22]{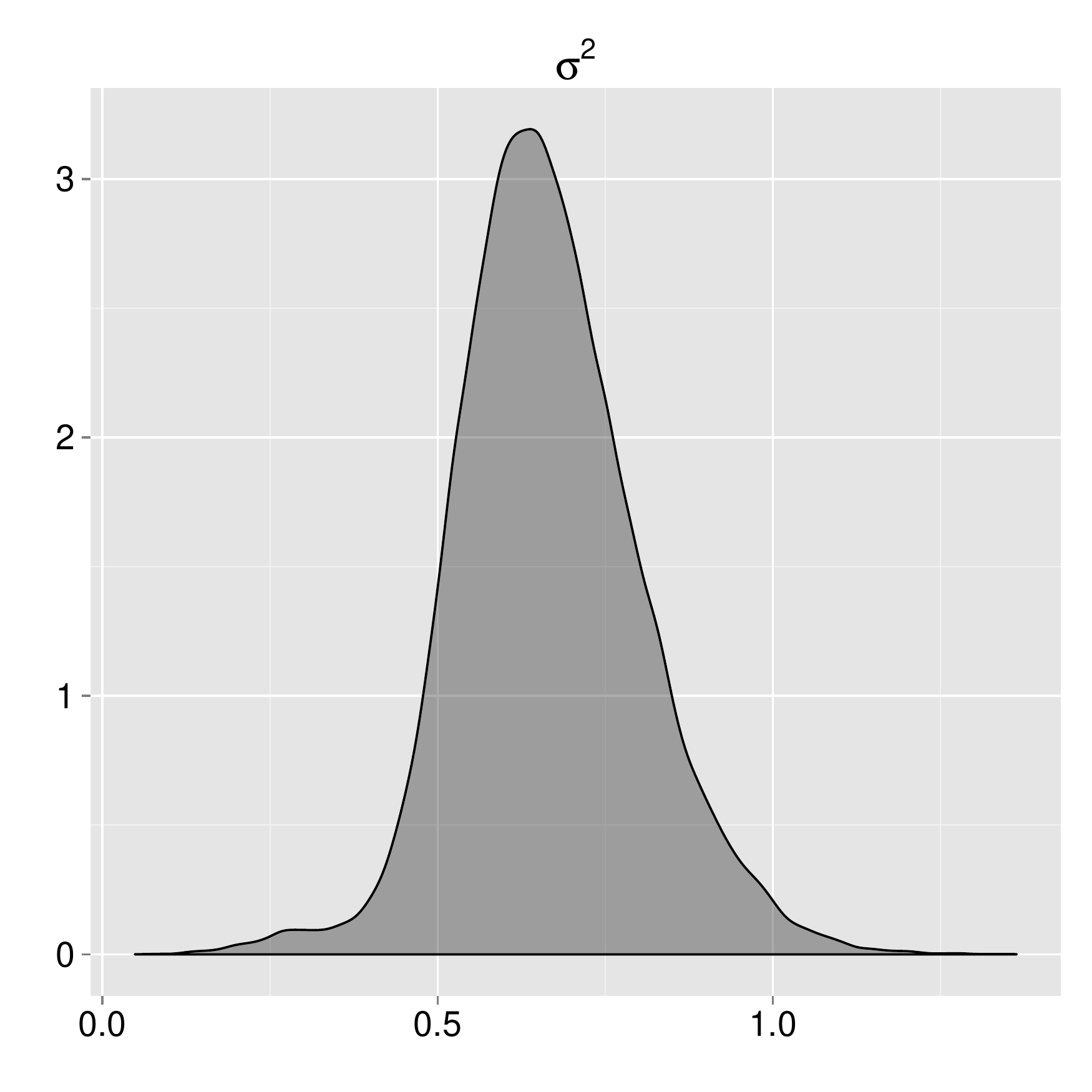}\includegraphics[scale=0.22]{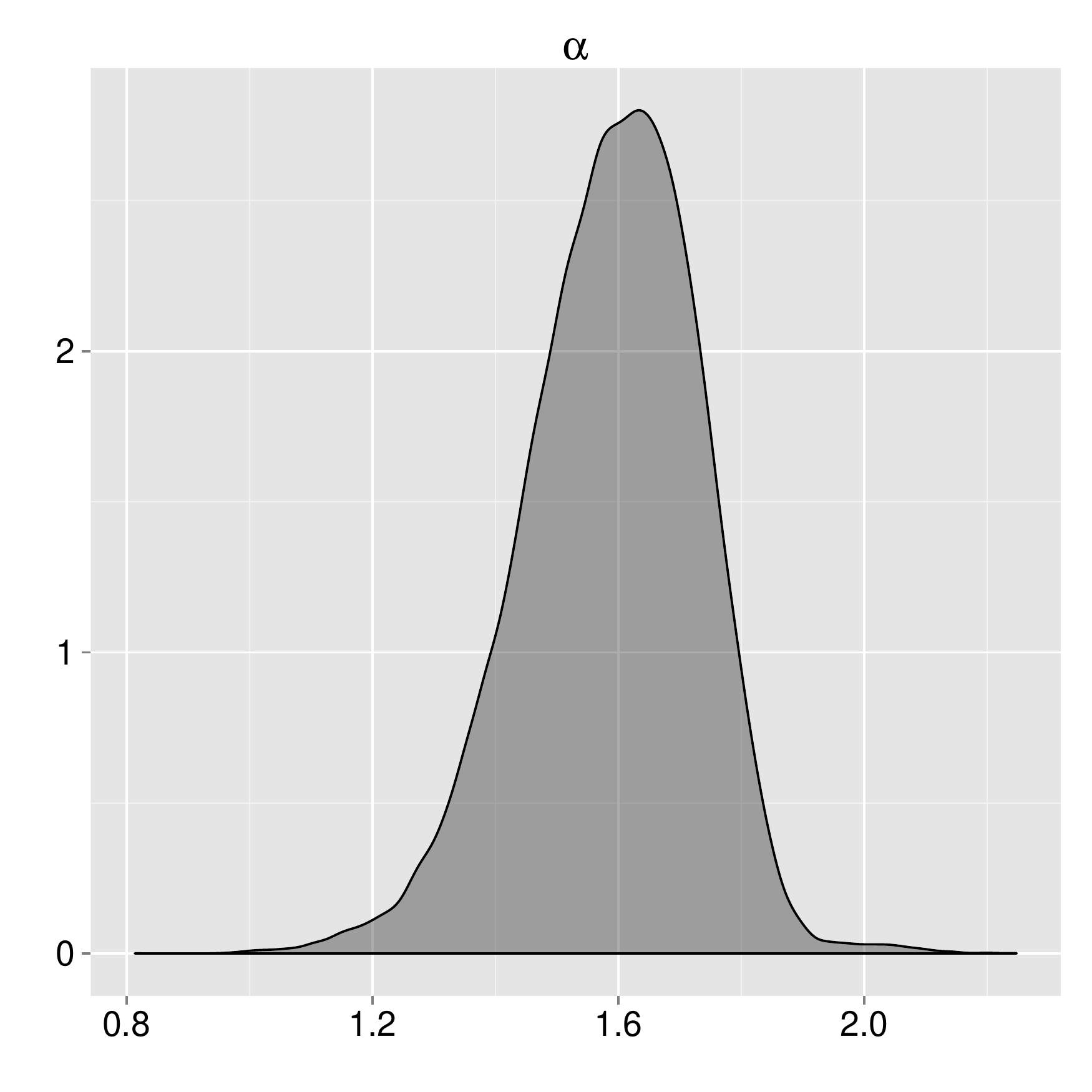}\includegraphics[scale=0.22]{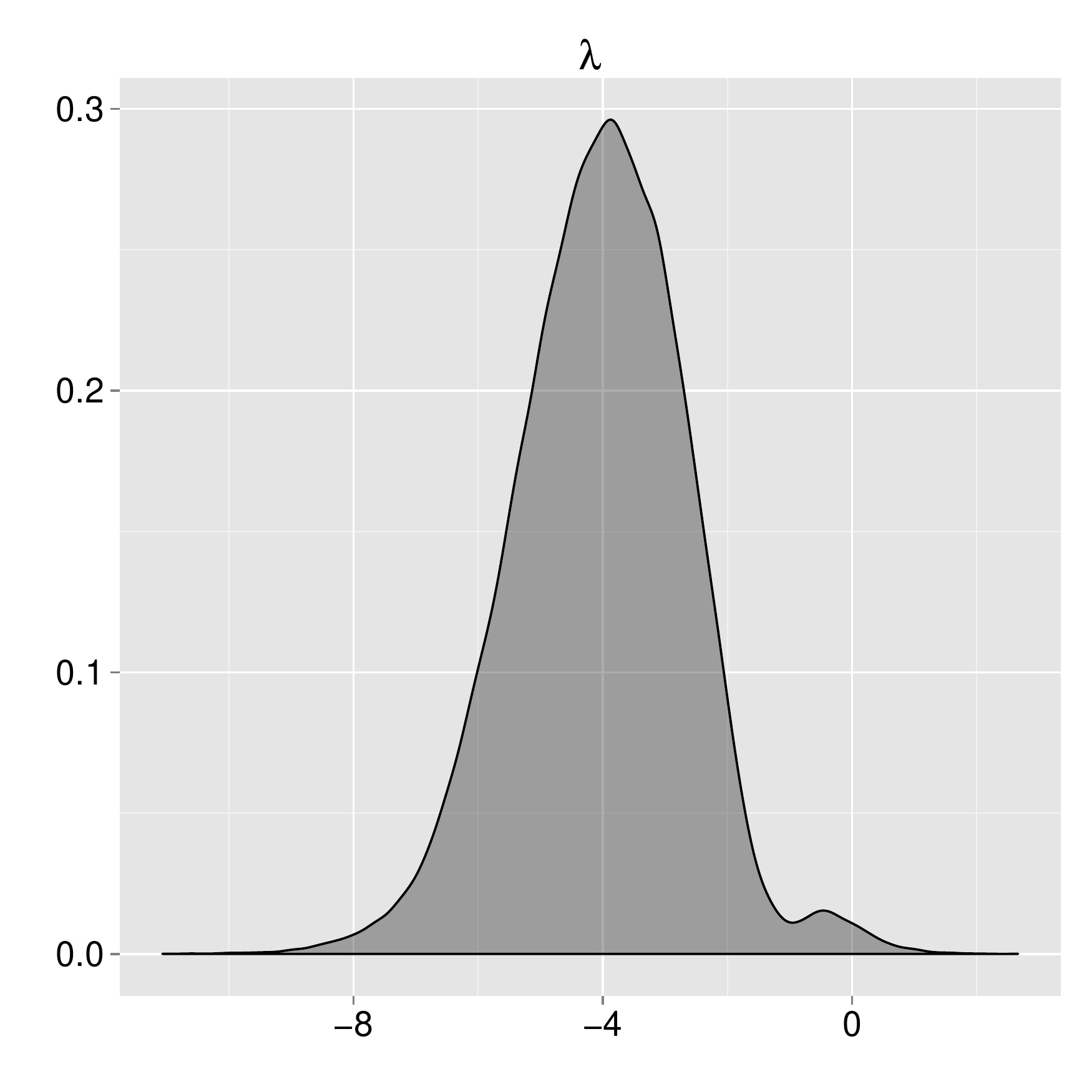}
\caption{Transfer fees of soccer players and marginal posterior distributions of the SN parameters. The results are obtained with $N=50\,000$  particles.}  \label{Fig:UnivPostFootESN}
\end{figure}

\begin{figure}[H]
\centering
\includegraphics[scale=0.22]{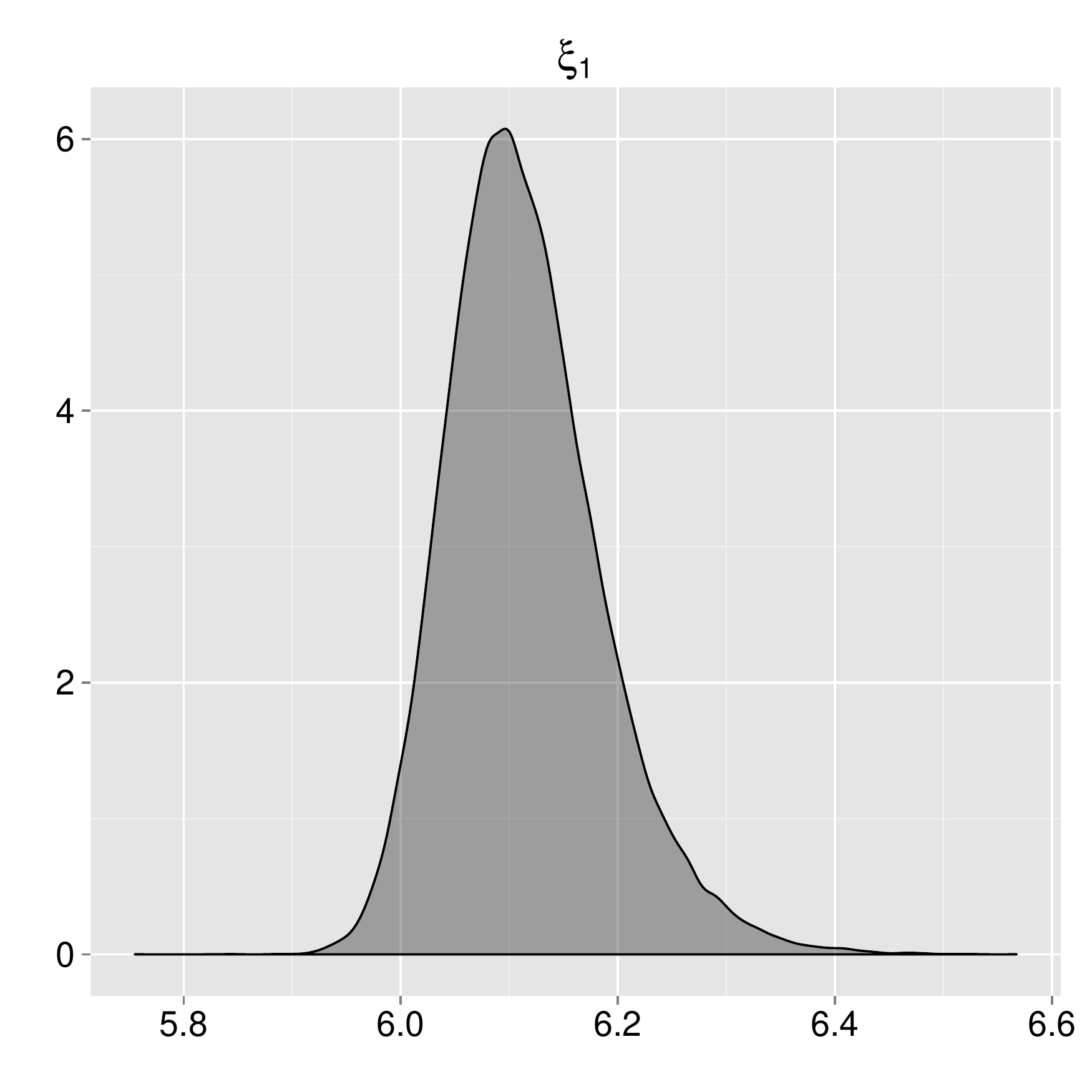}\includegraphics[scale=0.22]{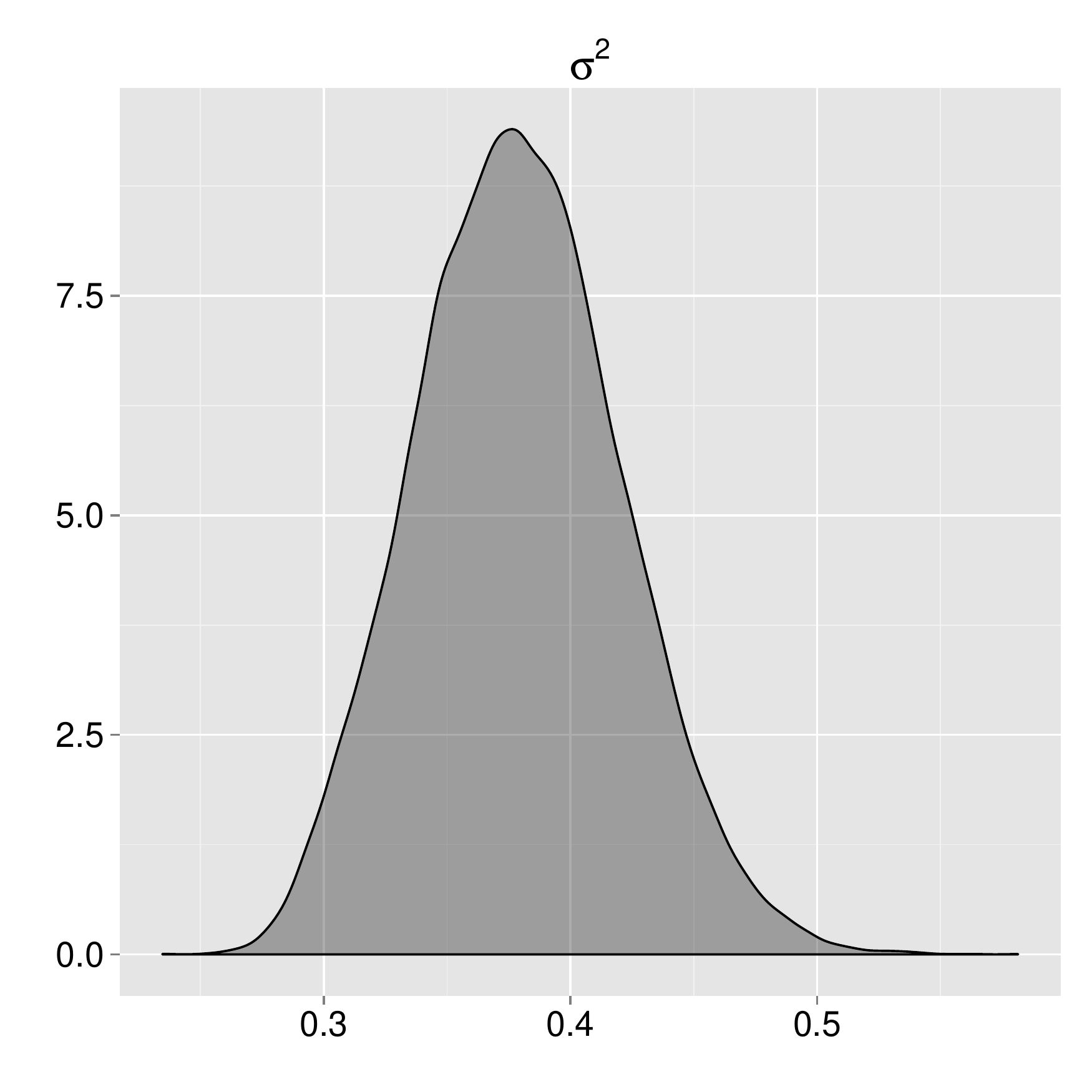}\includegraphics[scale=0.22]{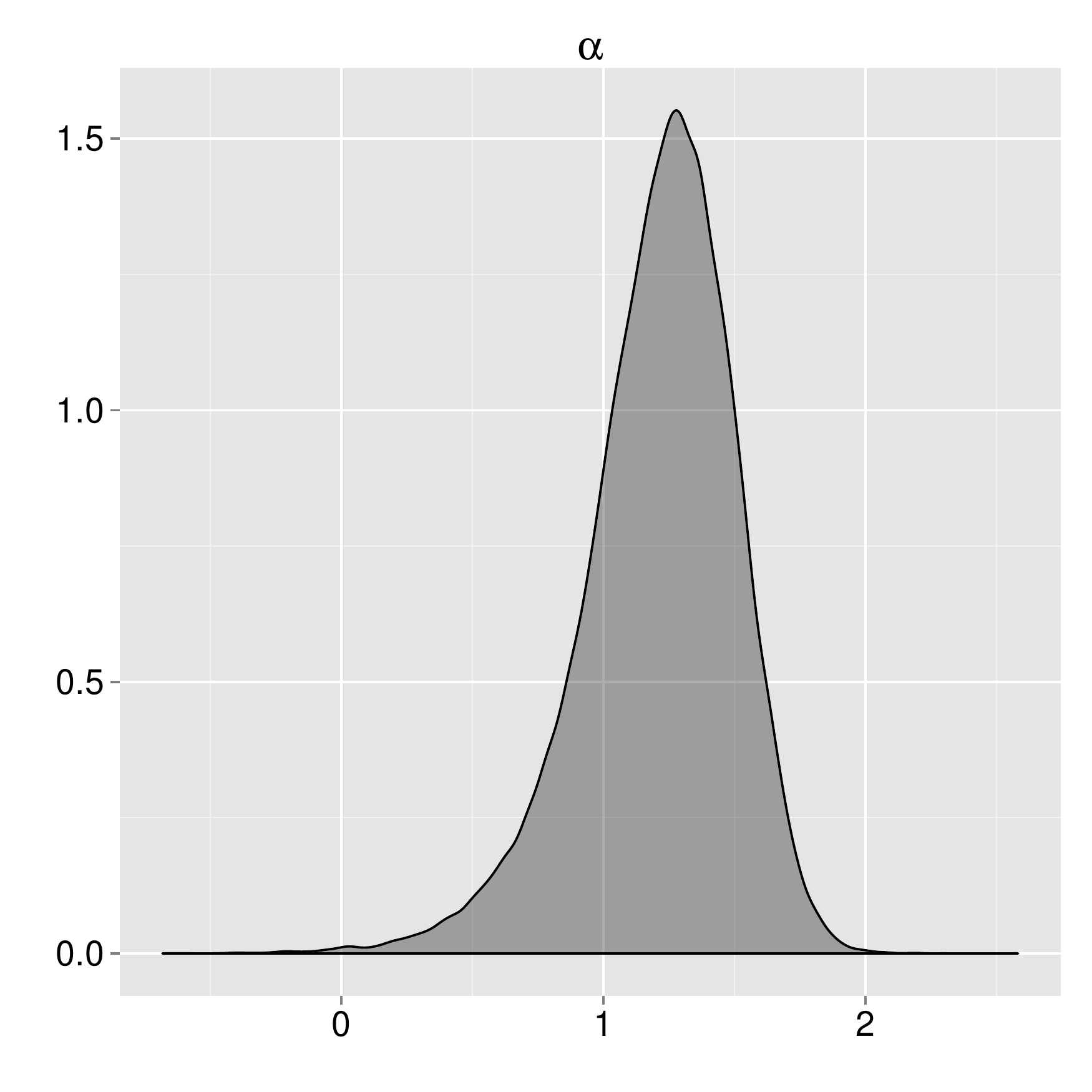}
\caption{Transfer fees of soccer players and marginal posterior distributions of the SN parameters. The results are obtained with $N=50\,000$  particles.}  \label{Fig:UnivPostFootSN}
\end{figure}

\begin{figure}[H]
\centering
\includegraphics[scale=0.3]{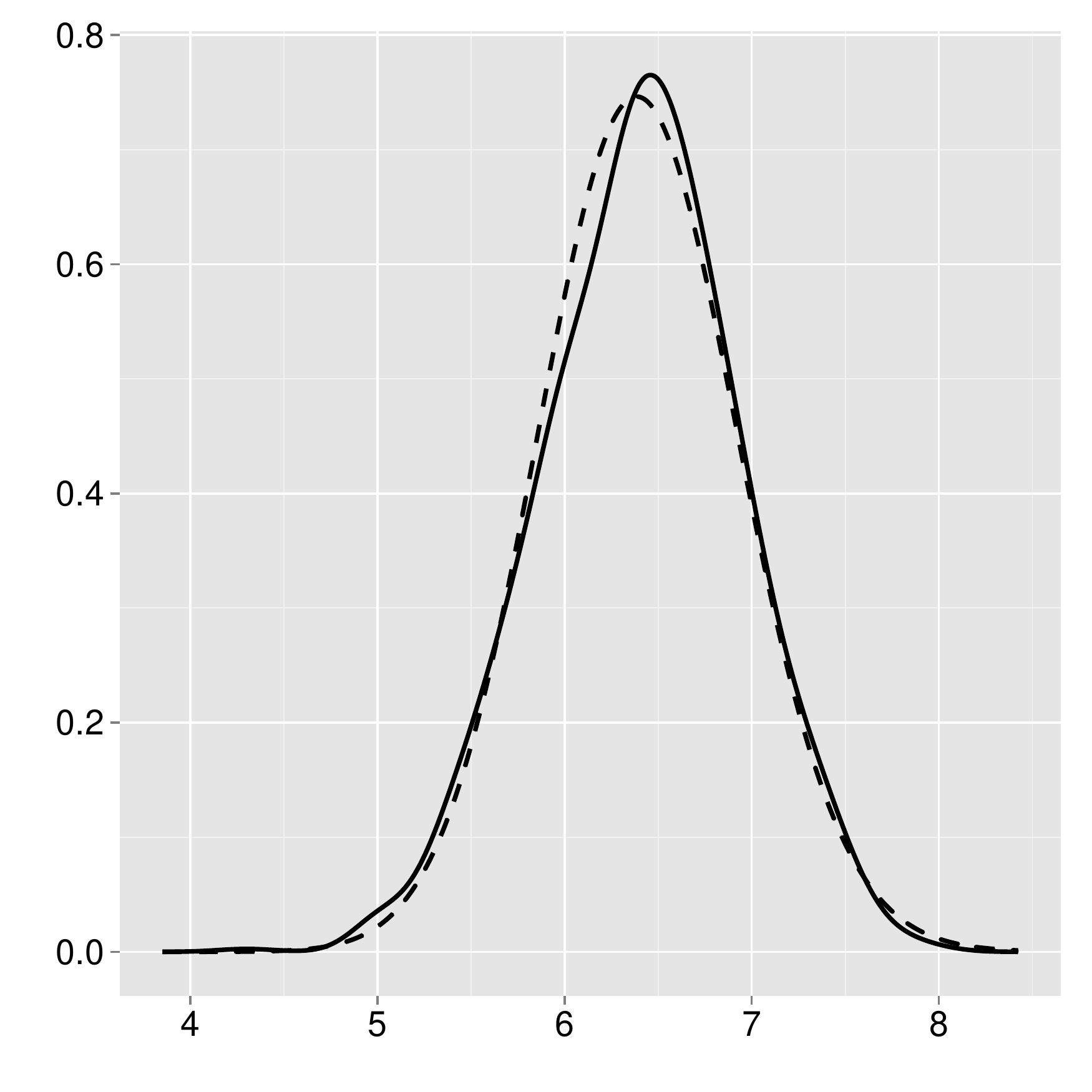}
\caption{Estimates of the Non-parametric and ESN transfer fees distribution. The dashed line (respectively, solid line) represents the estimate of the ESN (respectively, non-parametric) transfer fees distribution. The ESN estimate is obtained with $N=50\,000$  particles.} \label{Fig:densityFoot}
\end{figure}

\begin{figure}[H]
\centering
\includegraphics[scale=0.22]{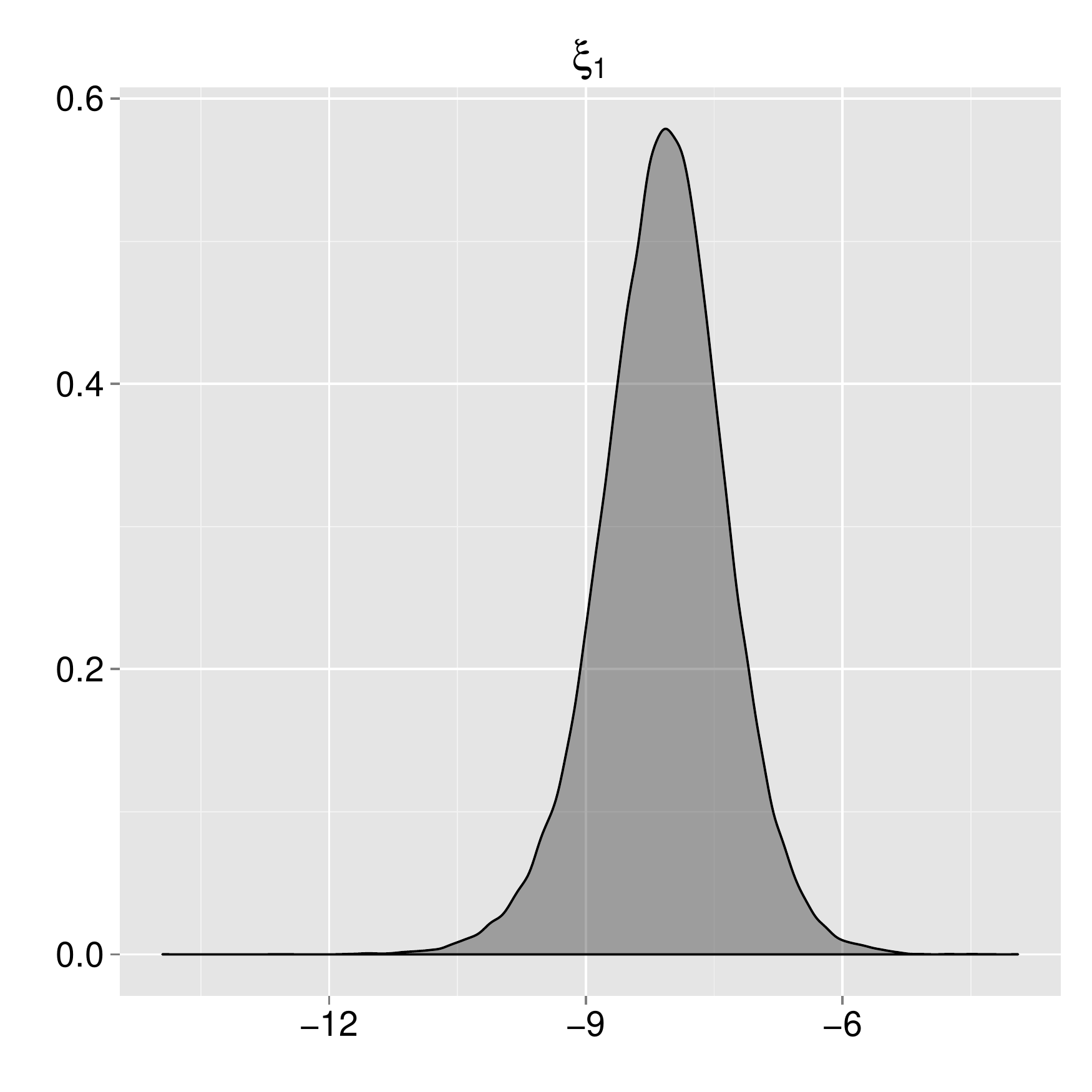}\includegraphics[scale=0.22]{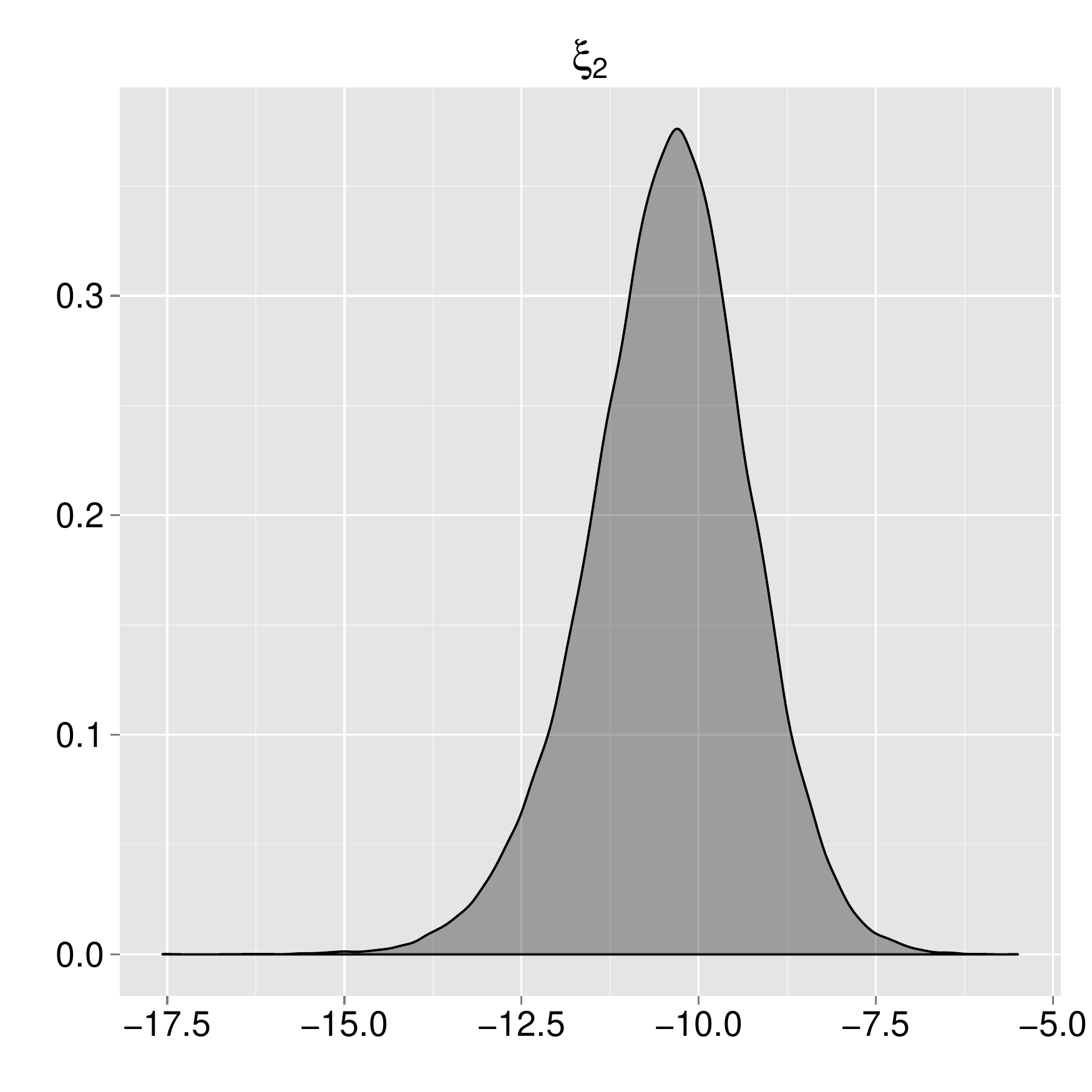}\includegraphics[scale=0.22]{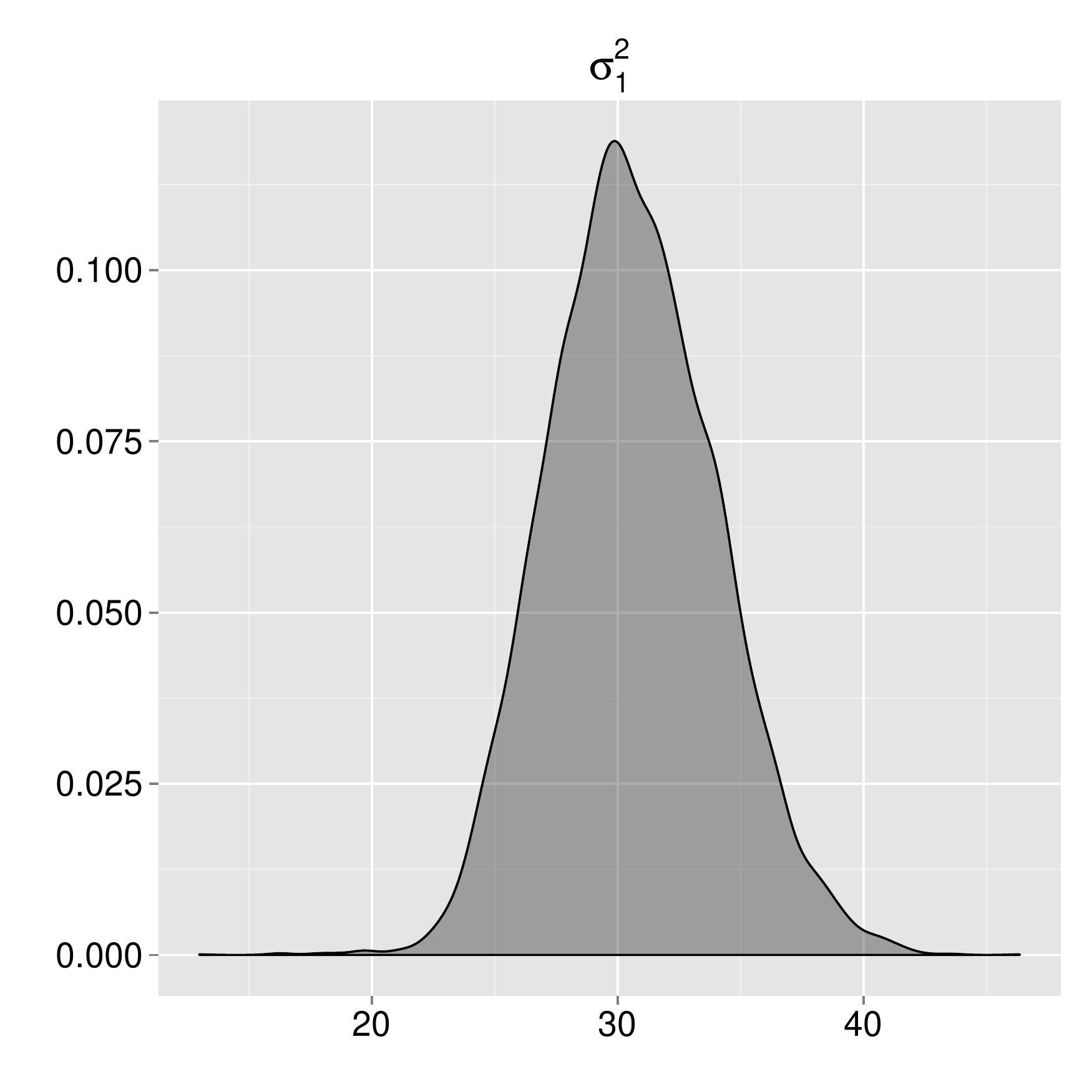}\includegraphics[scale=0.22]{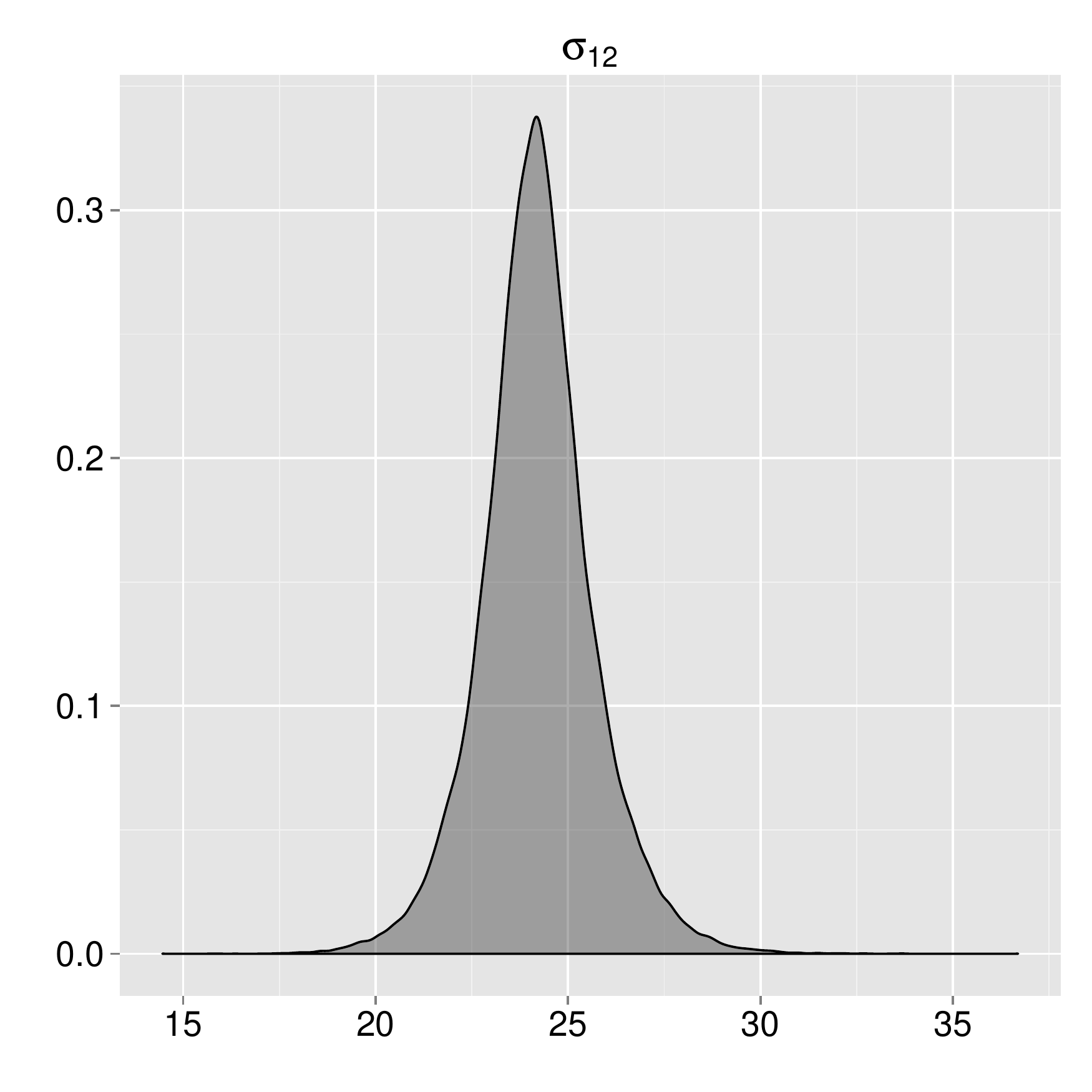}

\includegraphics[scale=0.22]{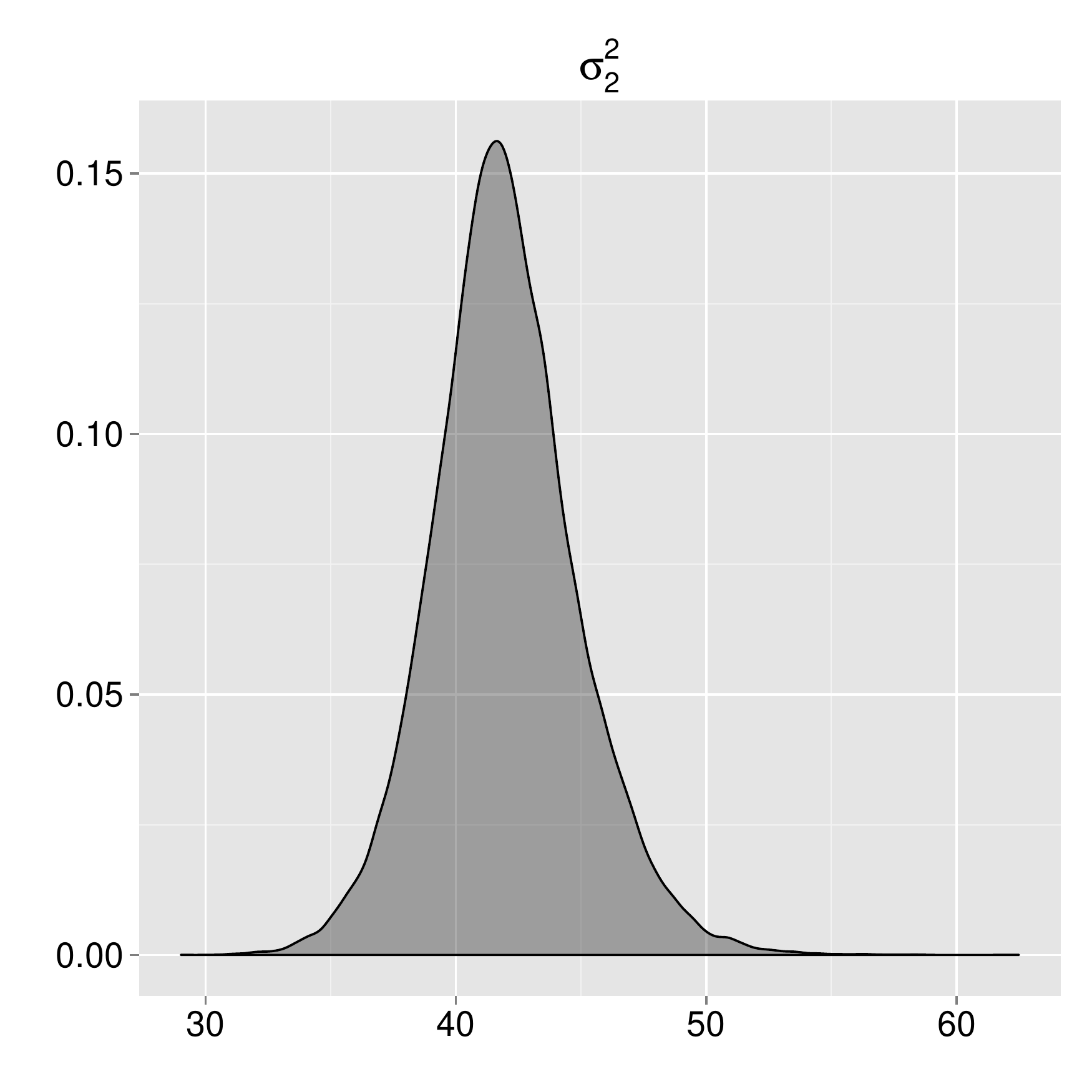}\includegraphics[scale=0.22]{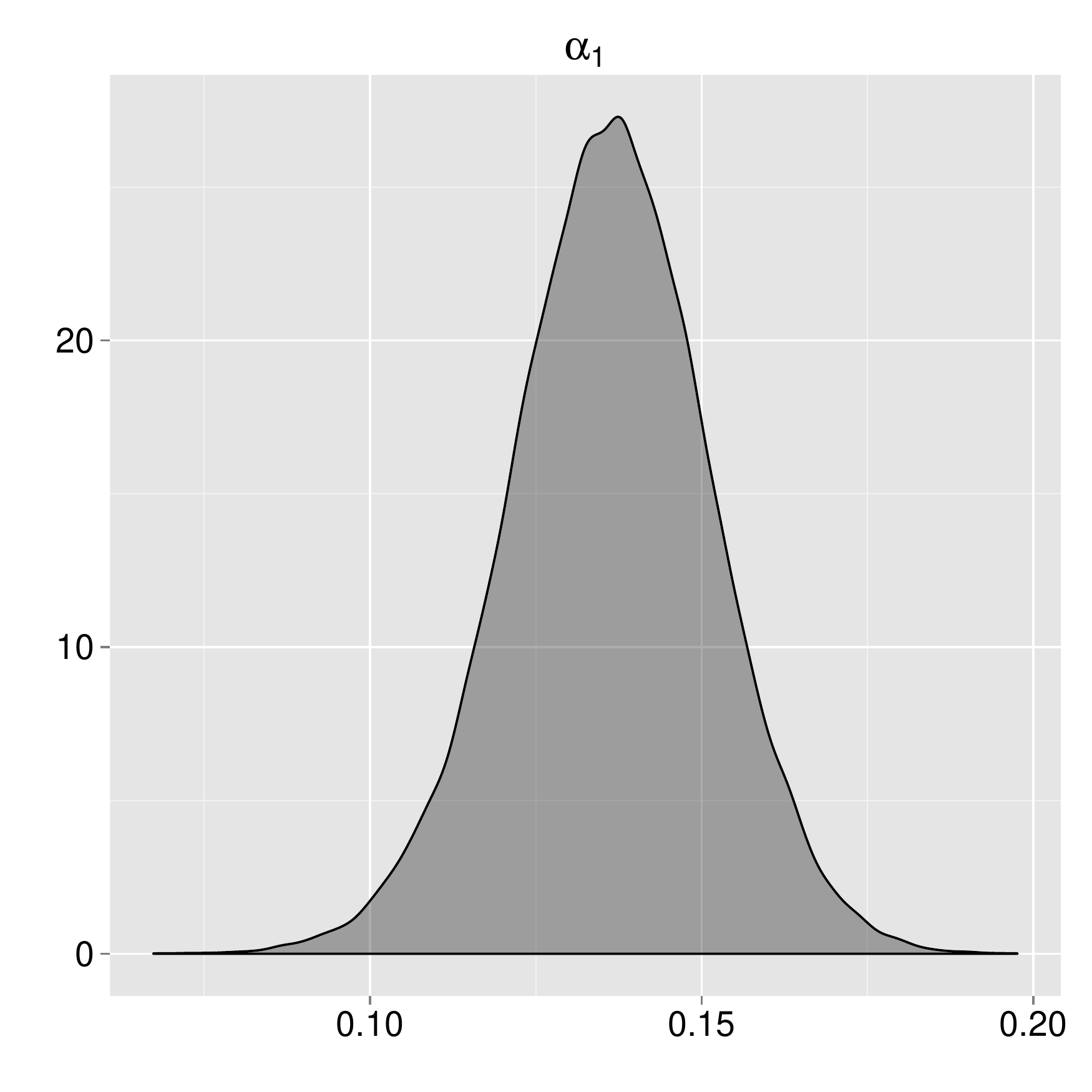}\includegraphics[scale=0.22]{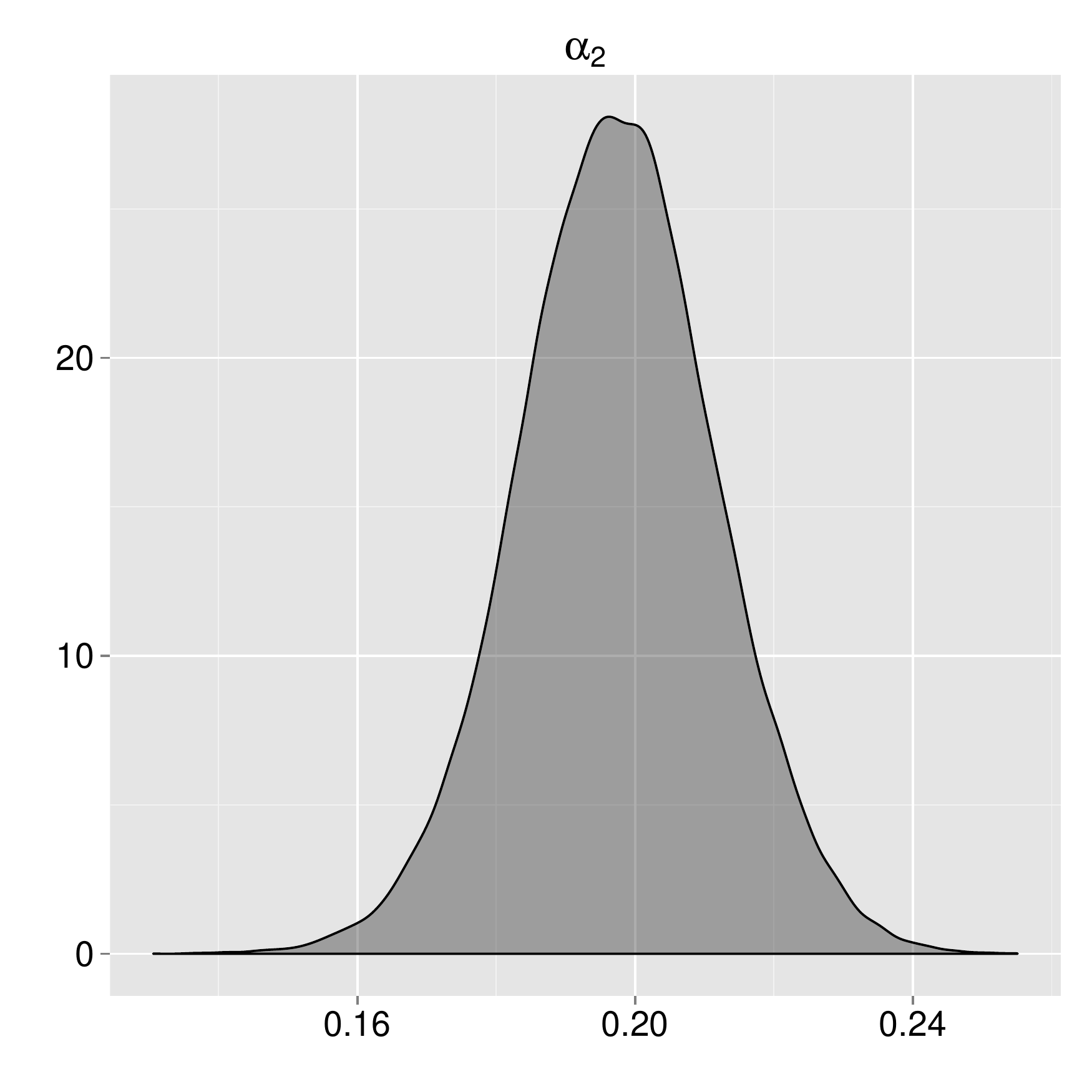}\includegraphics[scale=0.22]{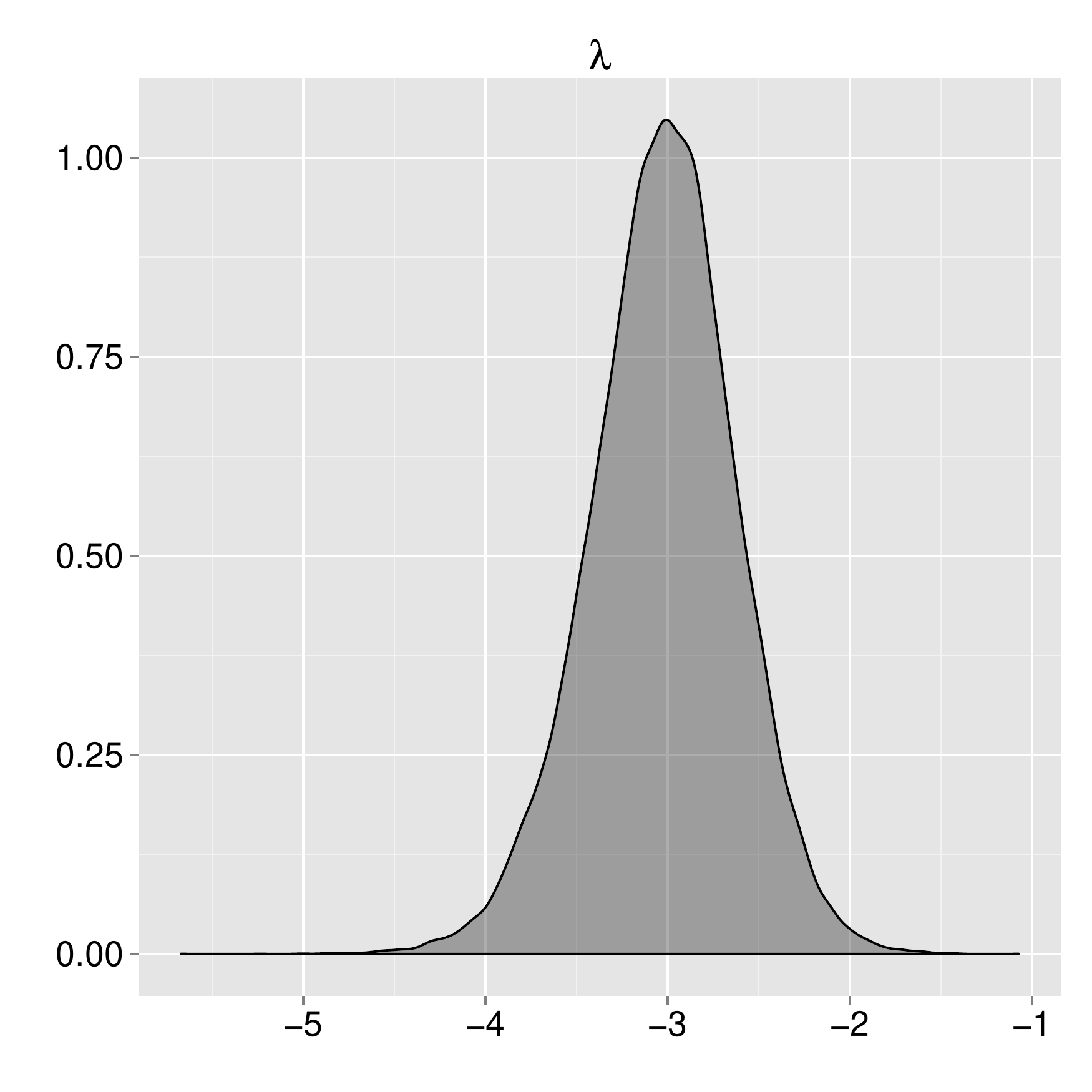}
\caption{Marginal Posterior distributions for the financial data under the ESN assumption. The results are obtained with $N=10\,000$  particles.  Evidence (in log) is  -8\,631.379.} \label{Fig:UnivPostReal}
\end{figure}

\begin{figure}[H]
\centering
\includegraphics[scale=0.22]{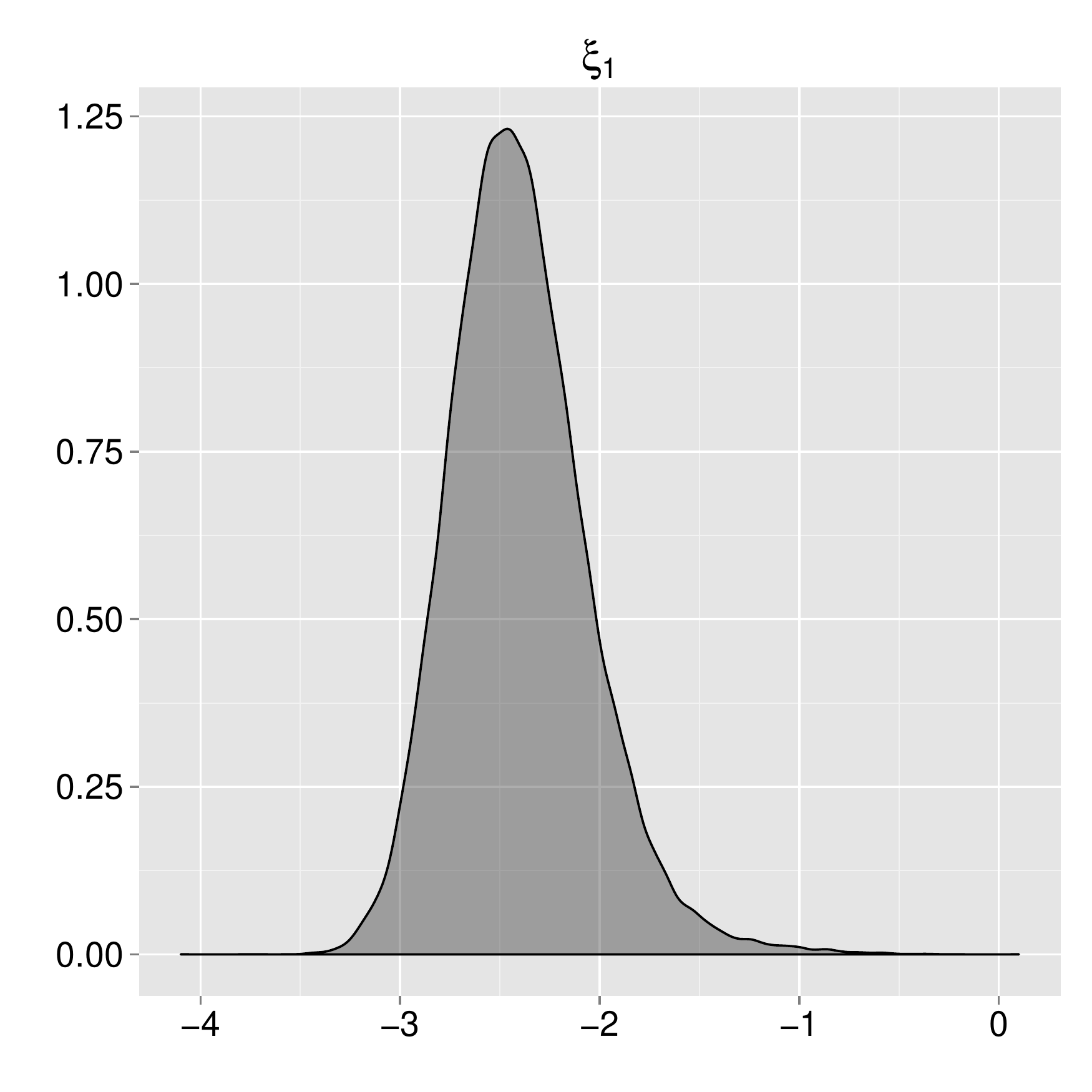}\includegraphics[scale=0.22]{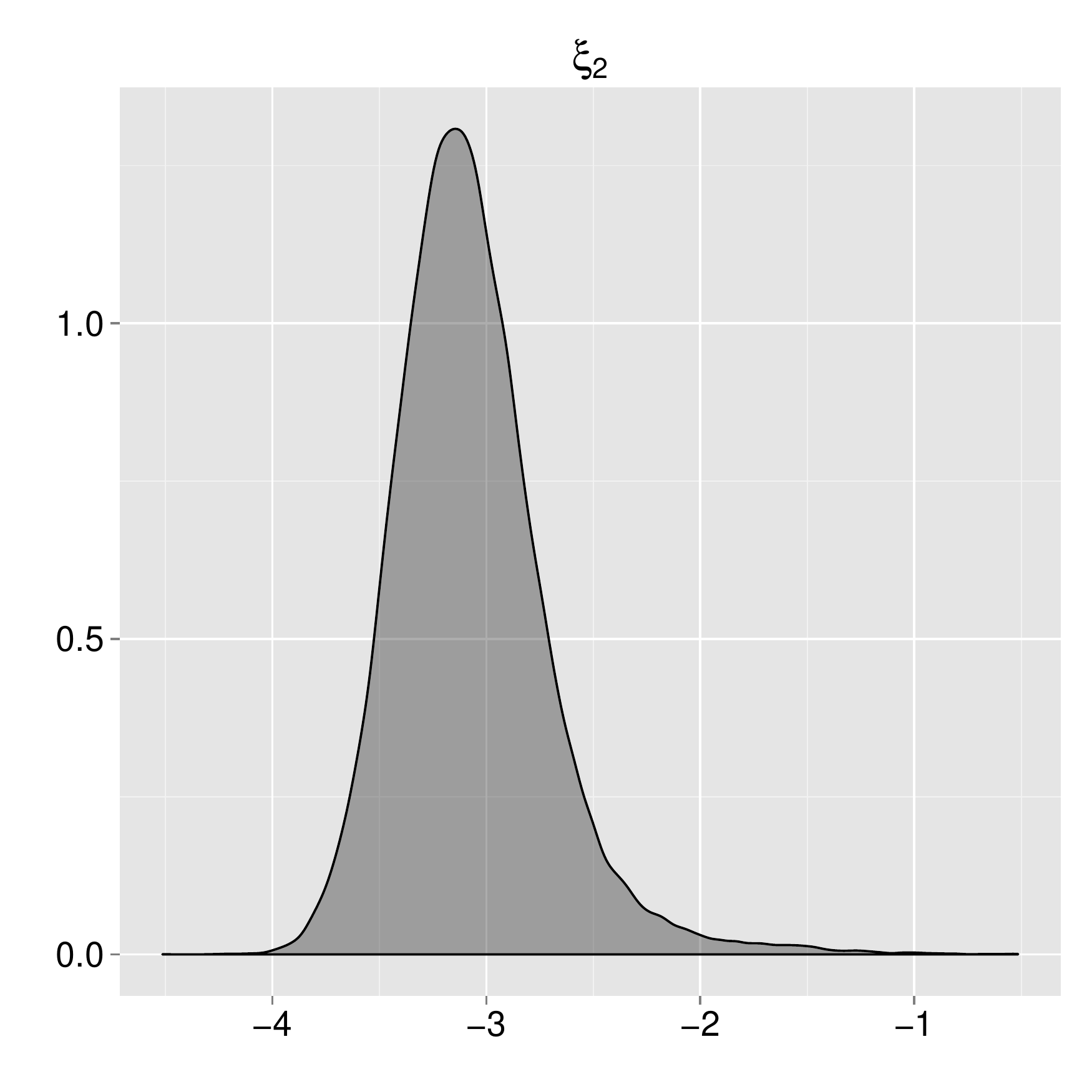}\includegraphics[scale=0.22]{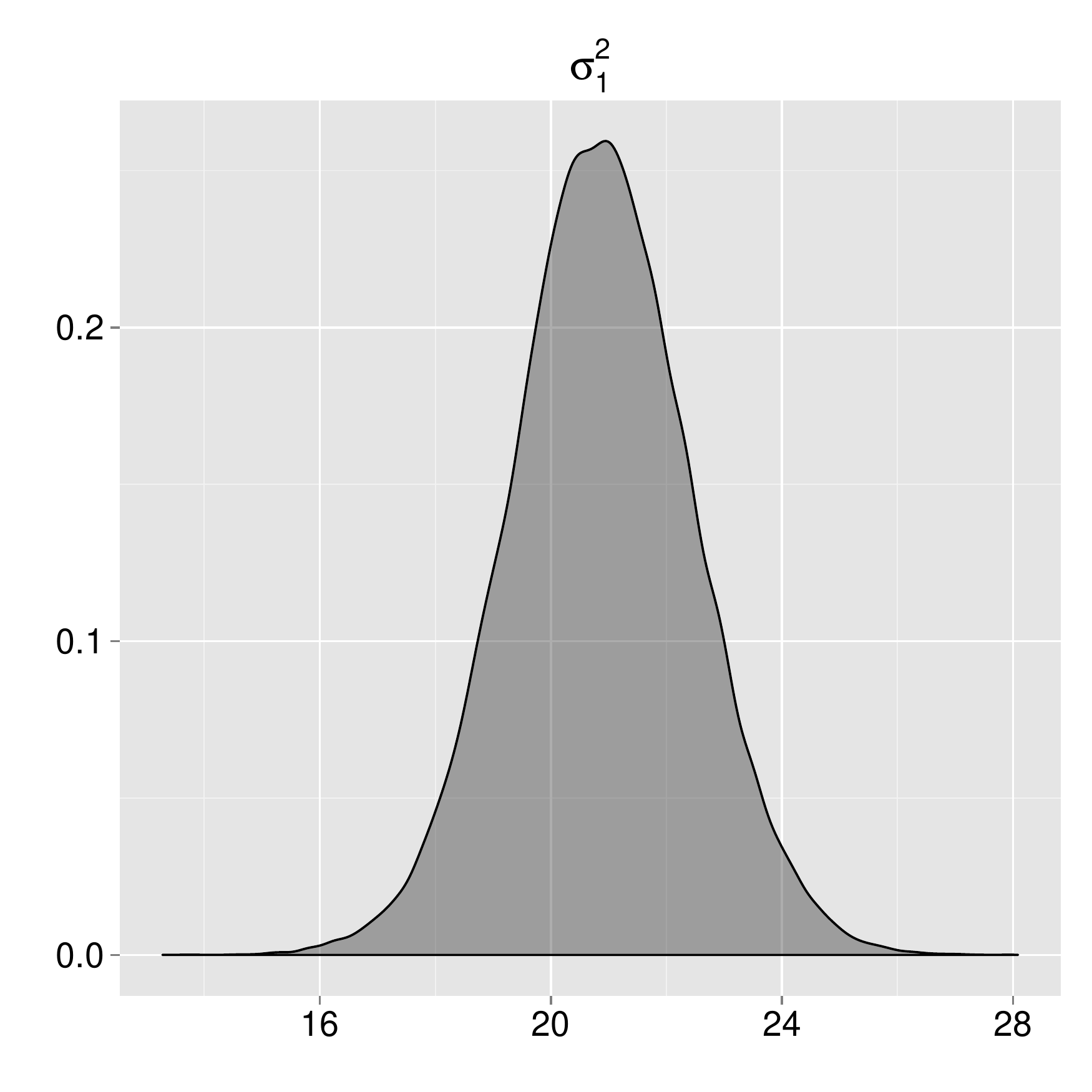}\includegraphics[scale=0.22]{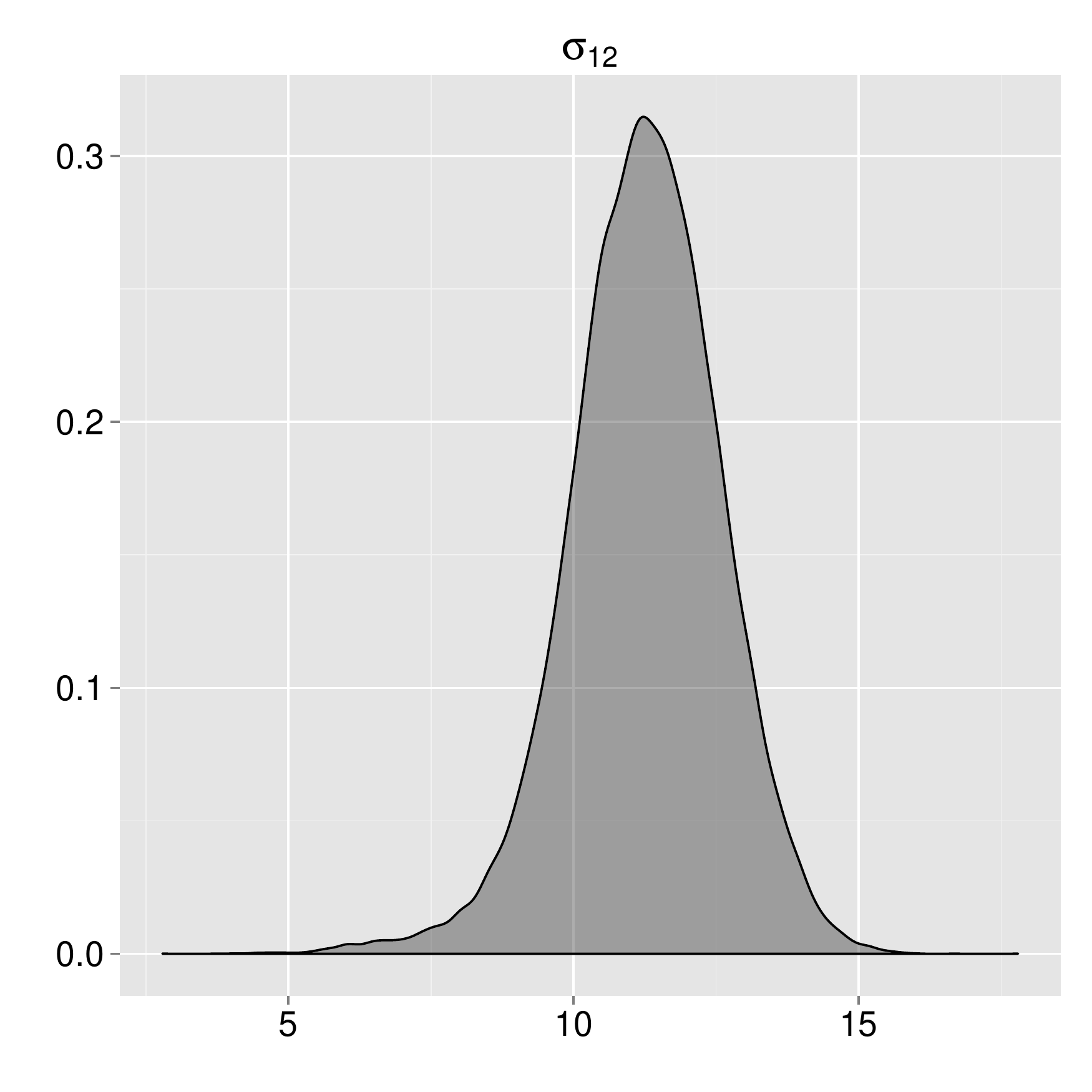}

\includegraphics[scale=0.22]{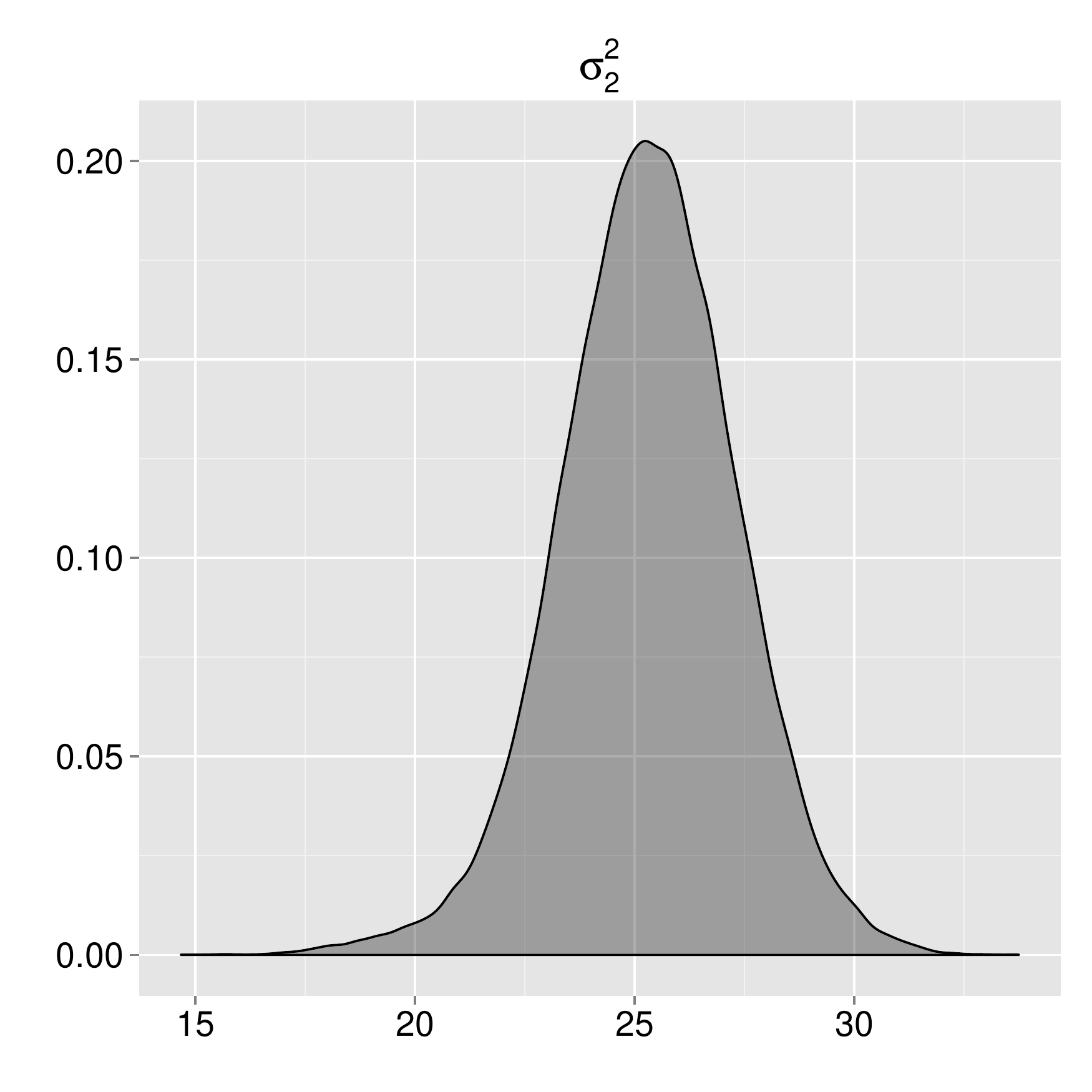}\includegraphics[scale=0.22]{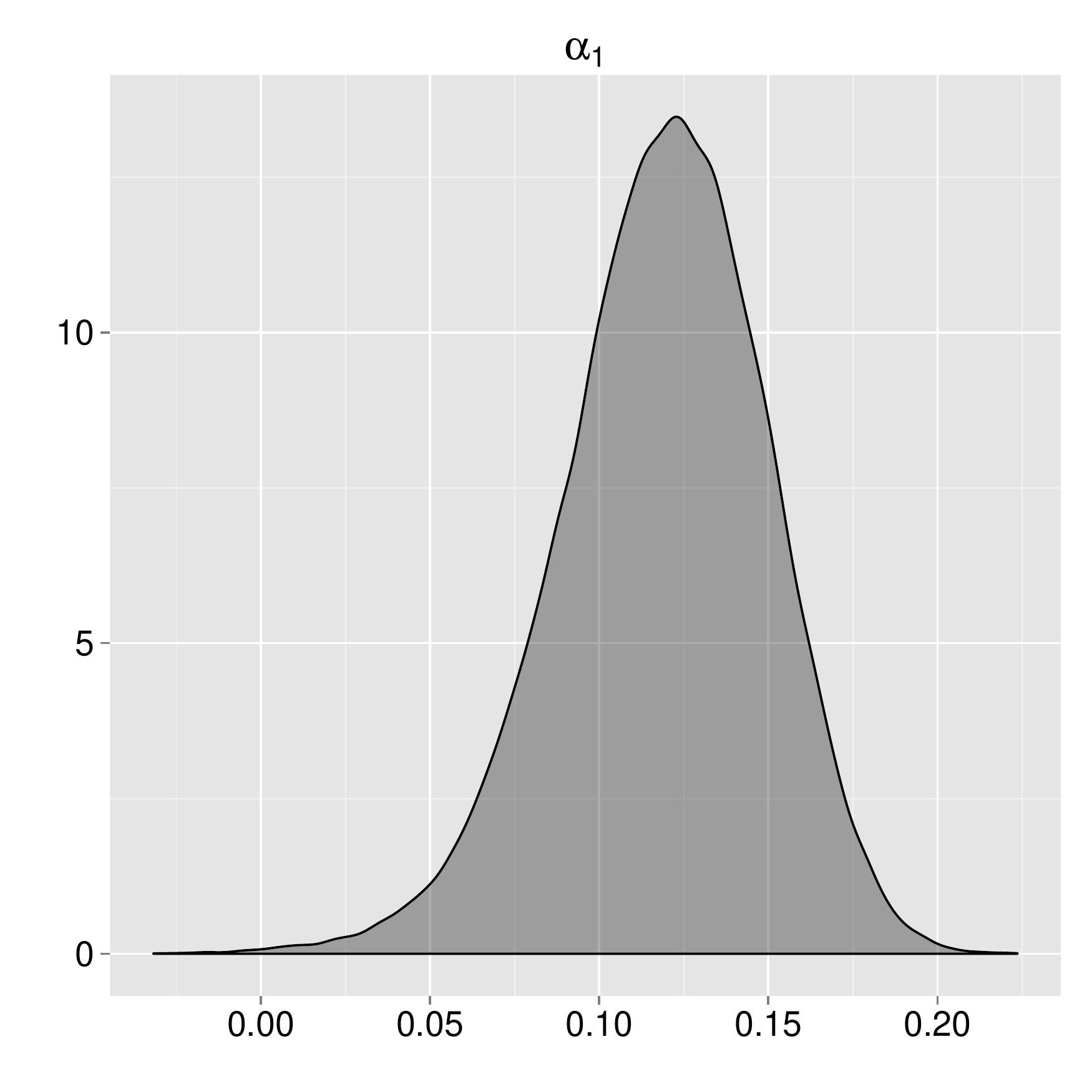}\includegraphics[scale=0.22]{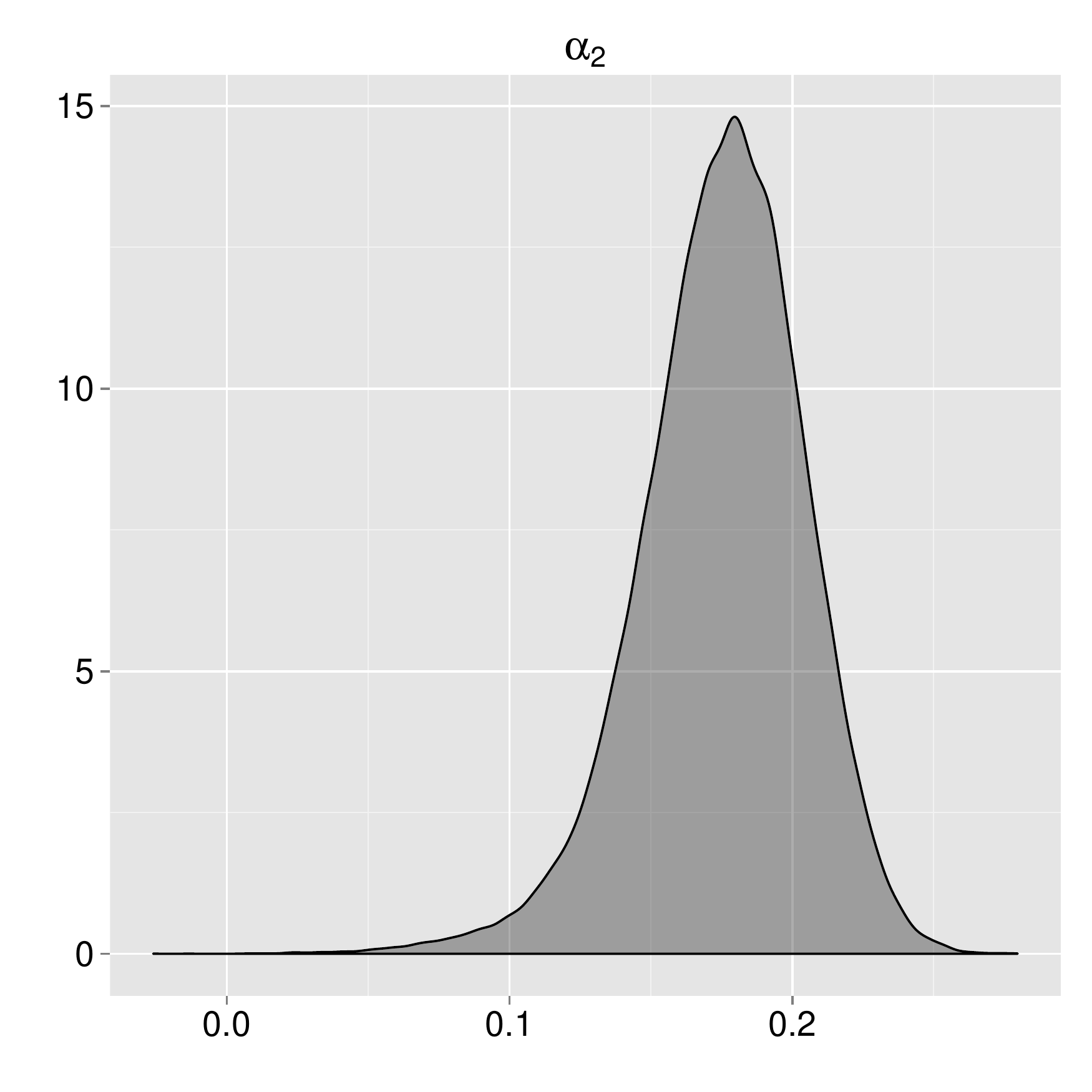}

\caption{Marginal Posterior distributions for the financial data under the SN assumption. The results are obtained for $N=50\,000$. Evidence (in log) is -8\,647.239.} \label{Fig:UnivPostRealSN}
\end{figure}

\bibliographystyle{apalike}
\bibliography{bibcsn}

\end{document}